\documentclass{article}

\usepackage{arxiv}

\usepackage[utf8]{inputenc} 
\usepackage[T1]{fontenc}    
\usepackage{hyperref}       
\usepackage{url}            
\usepackage{booktabs}       
\usepackage{amsfonts}       
\usepackage{nicefrac}       
\usepackage{microtype}      
\usepackage{graphicx}

\usepackage{amssymb}
\usepackage{latexsym}
\usepackage{amsmath}
\usepackage{physics}
\usepackage{enumitem}
\usepackage{caption}
\usepackage{tabu}
\usepackage{stackengine}
\usepackage{lineno}
\usepackage{floatrow}
\usepackage{subfig}
\usepackage{ulem}  
\usepackage{cancel} 

\usepackage{url}
\usepackage[svgnames]{xcolor}
\xdefinecolor{dgreen}{named}{DarkOliveGreen}
\definecolor{newcolor}{rgb}{.8,.349,.1}

\usepackage{tikz}
\usetikzlibrary{patterns}
\usetikzlibrary{decorations}
\usetikzlibrary{decorations.pathmorphing}
\usetikzlibrary{decorations.pathreplacing}
\usetikzlibrary{decorations.shapes}
\usetikzlibrary{decorations.text}
\usetikzlibrary{decorations.markings}
\usetikzlibrary{decorations.fractals}
\usetikzlibrary{decorations.footprints}
\usetikzlibrary{arrows,calc}
\usetikzlibrary{shapes}

\tikzset{base/.style={draw=gray, align=center, semithick,minimum height=4ex,scale=0.3},
         test1/.style={base, diamond, aspect=3, text width=0.1em, inner sep=1.pt},
        }

\title{Analysis and reduction of spurious noise generated at grid refinement interfaces with the lattice Boltzmann method}

\author{{\hspace{1mm}Thomas Astoul}\\
	CERFACS, 42 Avenue G. Coriolis,\\
	31057 Toulouse Cedex, France\\
	Airbus Operations,316 Route de\\
	Bayonne, 31300 Toulouse, France\\
	\texttt{tastoul@cerfacs.fr}\\
	\And{\hspace{1mm}Gauthier Wissocq} \\
	CERFACS, 42 Avenue G. Coriolis,\\
	31057 Toulouse Cedex, France\\
	Safran Aircraft Engines,\\
	77550 Moissy-Cramayel, France\\
	\And{\hspace{1mm}Jean-Fran{\c{c}}ois Boussuge} \\
	CERFACS, 42 Avenue G. Coriolis,\\
	31057 Toulouse Cedex, France\\
	\And{\hspace{1mm}Alois Sengissen} \\
	Airbus Operations,316 Route de\\
	Bayonne, 31300 Toulouse, France\\
	\And{\hspace{1mm}Pierre Sagaut} \\
	Aix Marseille Univ, CNRS.\\
	Centrale Marseille, M2P2 UMR 7340,\\
	13451 Marseille, France
}

\renewcommand{\headeright}{}

\begin{document}

\newcommand{\reels}{\mathbb{R}}
\newcommand{\red}{\textcolor{red}}
\newcommand{\orange}{\textcolor{orange}}
\newcommand{\green}{\textcolor{LimeGreen}}
\newcommand{\blue}{\textcolor{blue}}
\newcommand{\purple}{\textcolor{purple}}

\newcommand{\herm}{\mathcal{H}}
\newcommand{\TF}{\begin{tikzpicture} \draw node[circle,draw,thick,fill=gray, scale=0.7]{}; \end{tikzpicture}}
\newcommand{\IE}{\begin{tikzpicture} \draw node[circle,draw,thick,fill=white, scale=0.7]{}; \end{tikzpicture}}
\newcommand{\TN}{\begin{tikzpicture} \draw node[rectangle,draw,thick,fill=gray, scale=0.9]{}; \end{tikzpicture}}
\newcommand{\redC}{\begin{tikzpicture}[scale=1] \draw (-0.1,0) node{} ; \draw (0.1,0) node{} ; \draw [thick,draw=red, opacity=1] (-0.2,0) -- (0.2,0); \end{tikzpicture}}
\newcommand{\grayC}{\begin{tikzpicture}[scale=1] \draw (-0.1,0) node{} ; \draw (0.1,0) node{} ; \draw [thick,draw=gray, opacity=1] (-0.2,0) -- (0.2,0); \end{tikzpicture}}
\newcommand{\blackC}{\begin{tikzpicture}[scale=1] \draw (-0.1,0) node{} ; \draw (0.1,0) node{} ; \draw [thick,draw=black, opacity=1] (-0.2,0) -- (0.2,0); \end{tikzpicture}}
\newcommand{\grayCO}{\begin{tikzpicture}[scale=1] \draw (-0.1,0) node{} ; \draw (0.1,0) node{} ; \draw [thick,draw=gray!70, opacity=1] (-0.25,0) -- (0.25,0); 
\draw node[circle,draw=gray!70,thick,fill=white, scale=0.4]{}; 
\end{tikzpicture}}
\newcommand{\greenC}{\begin{tikzpicture}[scale=1] \\draw (-0.1,0) node{} ; \draw (0.1,0) node{} ; \draw [thick,draw=Green] (-0.25,0) -- (0.25,0);\end{tikzpicture}}

\newcommand{\PSDgrayA}{\begin{tikzpicture}[scale=1] \\draw (-0.1,0) node{} ; \draw (0.1,0) node{} ; \draw [thick,draw=gray!100] (-0.2,0) -- (0.2,0);\end{tikzpicture}}
\newcommand{\PSDgrayB}{\begin{tikzpicture}[scale=1] \\draw (-0.1,0) node{} ; \draw (0.1,0) node{} ; \draw [thick,draw=gray!70] (-0.2,0) -- (0.2,0);\end{tikzpicture}}
\newcommand{\PSDgrayC}{\begin{tikzpicture}[scale=1] \\draw (-0.1,0) node{} ; \draw (0.1,0) node{} ; \draw [thick,draw=gray!45] (-0.2,0) -- (0.2,0);\end{tikzpicture}}
\newcommand{\PSDred}{\begin{tikzpicture}[scale=1] \\draw (-0.1,0) node{} ; \draw (0.1,0) node{} ; \draw [thick,draw=FireBrick,densely dashed] (-0.25,0) -- (0.25,0);\end{tikzpicture}}
\newcommand{\grayStar}{\begin{tikzpicture}
\draw (-0.1,0) node{} ; \draw (0.1,0) node{} ; \draw [very thick,draw=gray, opacity=1,densely dashed] (-0.25,0) -- (0.25,0);
\draw (0,0) node[star, star points=5,draw,scale=0.3,fill=white,thick,draw=gray] {};
\end{tikzpicture}}
\newcommand{\blackStar}{\begin{tikzpicture}
\draw (-0.1,0) node{} ; \draw (0.1,0) node{} ; \draw [very thick,draw=black, opacity=1,densely dashed] (-0.25,0) -- (0.25,0);
\draw (0,0) node[star, star points=5,draw,scale=0.3,fill=white,thick,draw=black] {};
\end{tikzpicture}}
\newcommand{\grayPolyg}{\begin{tikzpicture}
\draw (-0.1,0) node{} ; \draw (0.1,0) node{} ; \draw [very thick,draw=gray, opacity=1] (-0.25,0) -- (0.25,0);
\draw (0,0) node[regular polygon,regular polygon sides=5,draw,scale=0.35,fill=white,thick,draw=gray] {};
\end{tikzpicture}}
\newcommand{\blackPolyg}{\begin{tikzpicture}
\draw (-0.1,0) node{} ; \draw (0.1,0) node{} ; \draw [very thick,draw=black, opacity=1] (-0.25,0) -- (0.25,0);
\draw (0,0) node[regular polygon,regular polygon sides=5,draw,scale=0.35,fill=white,thick,draw=black] {};
\end{tikzpicture}}
\newcommand{\grayCdash}{\begin{tikzpicture}[scale=1] \\draw (-0.1,0) node{} ; \draw (0.1,0) node{} ; \draw [thick,draw=gray, densely dashed] (-0.25,0) -- (0.2,0);\end{tikzpicture}}

\newcommand{\redSquare}{\begin{tikzpicture}
\draw (-0.1,0) node{} ; \draw (0.1,0) node{} ; \draw [very thick,draw=FireBrick, opacity=1] (-0.25,0) -- (0.25,0);
\draw (0,0) node[rectangle,draw,scale=0.45,fill=white,thin,draw=FireBrick] {};
\end{tikzpicture}}

\newcommand{\blueStar}{\begin{tikzpicture}
\draw (-0.1,0) node{} ; \draw (0.1,0) node{} ; \draw [very thick,draw=MidnightBlue!70!RoyalBlue, opacity=1] (-0.25,0) -- (0.25,0);
\draw (0,0) node[star, star points=5,draw,scale=0.3,fill=white,thin,draw=MidnightBlue!70!RoyalBlue] {};
\end{tikzpicture}}

\newcommand{\shearP}{\begin{tikzpicture} \draw node[circle,fill=red!, scale=0.6]{}; \end{tikzpicture}}
\newcommand{\numS}{\begin{tikzpicture} \fill node[circle,draw=red,fill=white,very thick, scale=0.5]{}; \end{tikzpicture}}
\newcommand{\numB}{\begin{tikzpicture} \fill node[rectangle,draw=black,fill=white,thick, scale=0.7]{}; \end{tikzpicture}}
\newcommand{\acousP}{\begin{tikzpicture} \draw[rotate=-90,scale=0.75] (0,0) -- (0.23,0) -- (1/2*0.23, {sqrt(3)/2*0.23}) -- cycle[draw=blue,fill=blue]{};\end{tikzpicture}}
\newcommand{\acousN}{\begin{tikzpicture} \draw[rotate=90,scale=0.75] (0,0) -- (0.23,0) -- (1/2*0.23, {sqrt(3)/2*0.23}) -- cycle[draw=dgreen,fill=dgreen]{};\end{tikzpicture}}
\newcommand{\numG}{\begin{tikzpicture} \node[test1,scale=0.9] at (0,0) {}; \end{tikzpicture}}

\newenvironment{Hfigure}{\setcaptiontype{figure}%
  \vskip\textfloatsep\begin{minipage}{\columnwidth}}%
  {\end{minipage}\vskip\textfloatsep\noindent}

\maketitle

\begin{abstract}
The present study focuses on the unphysical effects induced by the use of non-uniform grids in the lattice Boltzmann method. In particular, the convection of vortical structures across a grid refinement interface is likely to generate spurious noise that may impact the whole computation domain. This issue becomes critical in the case of aeroacoustic simulations, where accurate pressure estimations are of paramount importance. The purpose of this article is to identify the issues occurring at the interface and to propose possible solutions yielding significant improvements for aeroacoustic simulations. More specifically, this study highlights the critical involvement of non-physical modes in the generation of spurious vorticity and acoustics. The identification of these modes is made possible thanks to linear stability analyses performed in the fluid core, and non-hydrodynamic sensors specifically developed to systematically emphasize them during a simulation. Investigations seeking pure acoustic waves and sheared flows allow for isolating the contribution of each mode. An important result is that spurious wave generation is intrinsically due to the change in the grid resolution (i.e. aliasing) independently of the details of the grid transition algorithm. Finally, the solution proposed to minimize spurious wave amplitude consists of choosing an appropriate collision model in the fluid core so as to cancel the non-hydrodynamic mode contribution regardless the grid coupling algorithm. Results are validated on a convected vortex and on a turbulent flow around a cylinder where a huge reduction of both spurious noise and vorticity are obtained.

\end{abstract}


\newpage
\section{Introduction}
\label{sec:Intro}

The lattice Boltzmann method (LBM) has emerged as a very efficient approach for computational fluid dynamics over the last two decades. Its high degree of versatility makes it applicable to a large variety of highly complex physical phenomena, such as turbulence~\cite{Yu2005,Sagaut2010}, multiphase flows~\cite{Shan1993,Luo1998}, porous media~\cite{Bernsdorf2000} or even hemodynamics~\cite{Ye2015}, and it has increasingly interested both industrial and academic actors. Its main advantage are, \textit{inter alia}, a very simple and weakly dissipative numerical scheme representing weakly compressible flows which makes the LBM suitable for aeroacoustic simulations. Furthermore, the well known collide \& stream algorithm requires a Cartesian grid that allows for a seamless way to handle complex geometries~\cite{Ravetta2018} through automated octree meshes and immersed boundary conditions.

Current challenges faced by industrial companies require large scale problems to be simulated with a high degree of accuracy. This issue involves the use of non-uniform grids to reduce computational costs and focus the mesh refinements on regions of interest determined by the physical phenomena at stake. Unfortunately, such grid topologies are likely to generate spurious vorticity or acoustic disturbances that may pollute aeroacoustic simulations for which acoustic pressure fluctuations are much lower than the aerodynamic ones. In such situations, avoiding parasitic sources is of paramount importance. For now on, and considering the very enlightening bibliographic review of Gendre \textit{et al.}~\cite{Gendre2017}, a few aeroacoustic studies in the presence of grid refinements can be found in the literature but these are limited to pure acoustic propagation problems~\cite{Marie2008,Hasert2014}. The first study with acoustic validation in the presence of vortices that cross a refinement interface has been performed by Gendre \textit{et al.} using the standard BGK collision model with an increased kinematic viscosity for stability purposes. Despite the fact that vortices are brought to cross interfaces in many aeroacoustic applications such as turbulent jet noise~\cite{Brogi2017}, landing gear noise~\cite{Sengissen2015} or cavity noise~\cite{Coreixas2015}, it is interesting to wonder why this \textit{a priori}, simple case is almost never studied in the LBM literature. 

Furthermore, the very same undesirable phenomenon occurs in the Navier-Stokes (NS) framework~\cite{Vanharen2017}. A first indication can be found in the PhD Thesis of Hasert~\cite{Hasert2014}, in which a turbulent flow crosses grid refinements. In this simulation, spurious vortices and pressure spots appear at grid interfaces, but also far from high fluctuation hydrodynamics areas. These spurious artifacts disappear, decreasing the order of the spatial interpolation scheme used at the interface to reconstruct missing data, which increases the dissipation in this region. Obviously, using a first-order interpolation scheme is not a reliable solution as the accuracy is decreased~\cite{Lagrava2012}. Another enlightening point can be found in~\cite{Dorschner2016}, where a turbulent channel is simulated using an entropic lattice Boltzmann model~\cite{Karlin2014}. Even if acoustic phenomena are not primarily investigated, the entropic stabilizer varies considerably close to the grid interface. Since the latter is related to interactions between stress variables and ghost ones~\cite{Benzi1990,Benzi1992} which are closely linked with non-hydrodynamic modes (also referred to as ghost modes~\cite{Adhikari2008}) activity, a close relationship between grid refinement and these modes activity can be supposed. 

Therefore, as these modes have an expected non-hydrodynamic contribution, they may be further investigated thanks to a von Neumann analysis~\cite{Neumann1950}. This method is widely used to understand and predict the stability of a numerical scheme. It consists in evaluating the response of a system, described by a given set of equations, to linear perturbations. Sterling and Chen~\cite{Sterling1996a} were among the firsts to apply this method to the LBM scheme. Subsequently, this technique has allowed to exhibit the coexistence and the spectral properties of both hydrodynamic and non-hydrodynamic modes in a LBM scheme~\cite{Adhikari2008}. Moreover, in a recent study~\cite{Wissocq2019}, an extended spectral analysis was proposed, consisting in an investigation of the eigenvector's macroscopic content of the linearized lattice Boltzmann scheme. This last reveals that non physical modes can have macroscopic contributions, and allows for their identification in term of acoustics or shear contribution. 

This article aims at improving the analysis of spurious wave generation at grid interface, and proposing efficient solutions to damp them. Such an improvement could lead to the design of Lattice Boltzmann Methods well suited for aeroacoustic simulations on non-uniform grids. Until now, the origin of spurious phenomena has not been convincingly explained, which makes the resolution of this issue difficult. In this context, it is proposed (1) to understand and explain the spurious phenomena that occur at mesh refinement interface during the crossing of acoustics and vortices, by studying in particular the non-hydrodynamic modes involvement, and (2) to propose a collision model to avoid both spurious emissions and vorticity. It is worth emphasizing that the methodology introduced in the following is general and does not depend on the grid refinement algorithm.

The paper is organized as follows. First of all, the context of aeroacoustic simulations on non-uniform grids is introduced and the issues that may appear at grid interfaces are highlighted in Sec.~\ref{sec:acous_context}. Next, key features of the lattice Boltzmann method are briefly summarized and three collision models are considered specifically for their interesting spectral properties regarding non-hydrodynamic modes in Sec.~\ref{sec:LBM}. Then, a grid refinement algorithm is described, along with a suggested modification to enable a proper velocity gradient computation which is required for the collision model studied in Sec.~\ref{sec:refinement}. Spectral analysis tools are subsequently presented in Sec.~\ref{sec:LSA} to highlight the specific behavior of each collision model, especially with regard to the treatment of non-hydrodynamic modes. The effect of the resolution change on modes is then investigated in Sec.~\ref{sec:modal_interactions} by performing the spectral analysis of the projection of a fine mode onto a coarser resolution. It is important to notice that this projection is intrinsically tied to the use of different grid resolution, independently of the details of the grid transition algorithm. It is therefore a universal phenomenon, shared by all grid transition methods. Subsequently, in Sec.~\ref{sec:NHsensor}, sensors are developed so that non-physical modes can be identified in a systematic way during simulations. This step is necessary to make the link between spectral analysis observations, carried out in the very particular case of plane monochromatic waves in the linear approximation, and realistic LBM simulations. Afterwards, non-hydrodynamic effects on grid refinement will be meticulously studied in Sec.~\ref{sec:NHeffect} for both pure acoustic propagation and shear flows on acoustic and transversal shear waves. Finally, a proposition for solving these spurious artifacts is proposed and validated on a convected vortex in Sec.~\ref{sec:improvement} and on a highly turbulent flow around a cylinder in Sec.~\ref{sec:cylinder}.

\section{Aeroacoustic context}
\label{sec:acous_context}

This section aims at introducing specific concerns that may appear in the aeroacoustic framework on non-uniform grids.
When dealing with non-uniform grids in standard LBM, two main categories of algorithms exist in the literature: on one hand, the so-called ``cell-vertex'' formulation~\cite{Filippova1998,DupuisChopard2003,Lagrava2012} and on the other hand, the one referred to as ``cell-centered''~\cite{Rohde2006,Chen2006}. These algorithms are highly dissimilar at the first glance, as in the cell-centered case, fine and coarse nodes are never co-located. Thus, the latter requires specific steps of \textit{Coalescence} during the fine to coarse transfer and \textit{Explosion} for the reverse transfer.

The following paragraph briefly and qualitatively compares the behavior of two examples of cell-vertex / cell-centered algorithms on a convected vortex that crosses a fine to coarse transition:
\begin{itemize}
    \item the cell-vertex algorithm considered as detailed in Sec.~\ref{sec:refinement},
    \item the cell-centered algorithm as detailed in~\cite{Staubach2013}, which uses a uniform coalescence procedure and a linear explosion.
\end{itemize}

With both algorithms, the collision model used is the recursive regularized collision model from Malaspinas~\cite{Malaspinas2015}. This model is adopted here for its stability properties on this kind of application compared to the standard BGK collision model. It will be more specifically detailed and studied in the following.

\begin{figure}[H]
	\begin{center}
		\includegraphics[scale=0.6]{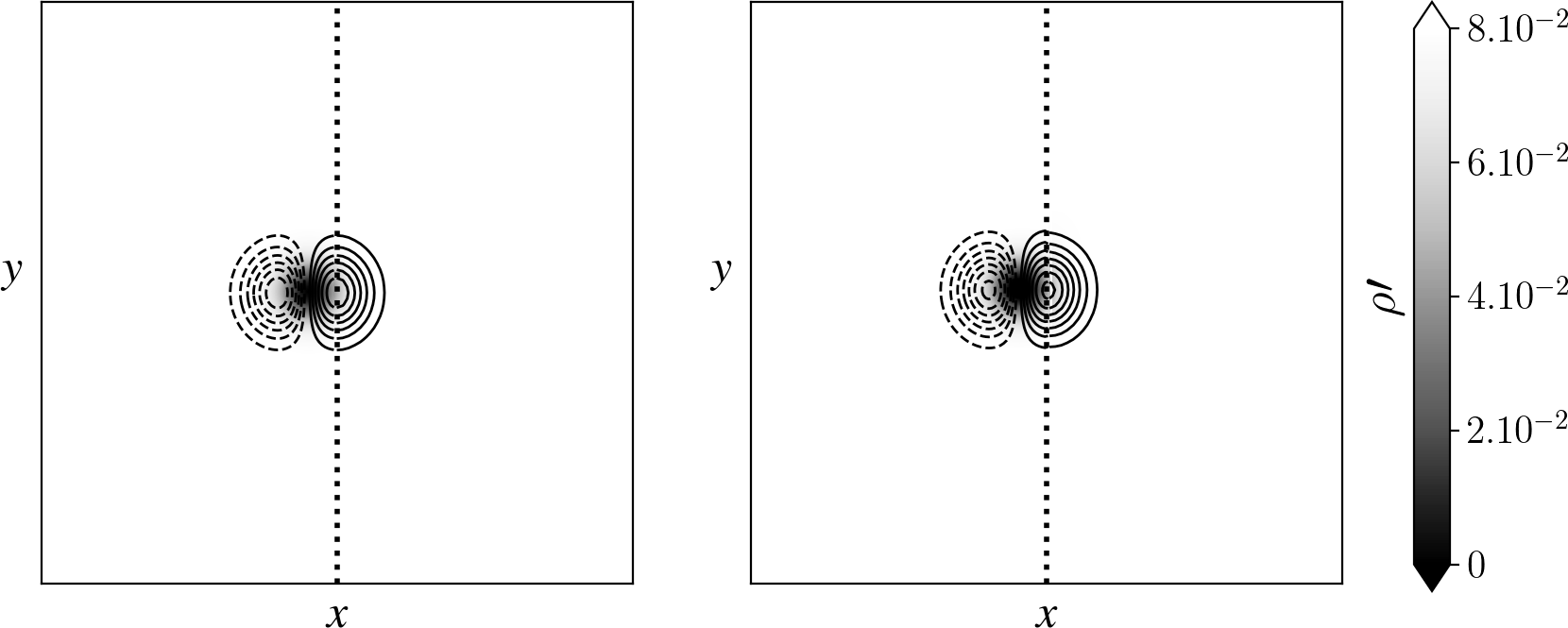}
		\caption{\label{fig:COVO_aero} Convected vortex that crosses a fine to coarse grid interface with a colormap of density scaled on hydrodynamic fluctuations and isocontours of transversal velocity. Left: Cell-vertex algorithm. Right: Cell-centered algorithm.}
		\end{center}
\end{figure}

\begin{figure}[H]
	\begin{center}
		\includegraphics[scale=0.6]{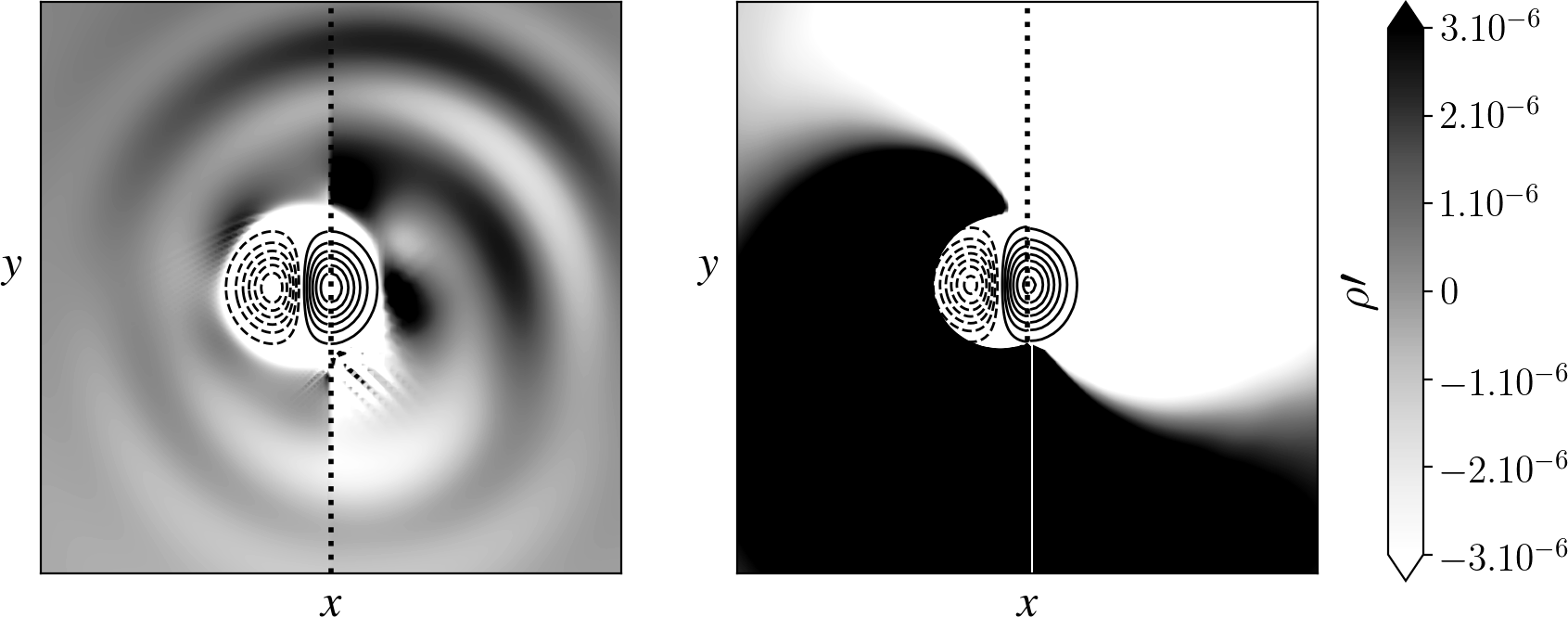}
		\caption{\label{fig:COVO_acous} Convected vortex that crosses a fine to coarse grid interface with a colormap of density scaled on acoustic fluctuations and isocontours of transversal velocity. Left: Cell-vertex algorithm. Right: Cell-centered algorithm.}
		\end{center}
\end{figure}

From an aerodynamic point of view, the convected vortex is qualitatively very well transmitted from one mesh to the other one whatever the algorithm used. 
No spurious emissions are apparent at this scale and both algorithms, with the collision model used here, seem satisfactory. 
However, this paper focuses on the acoustic emissions induced by distorted vortices. 
The related fluctuations at stake are several orders of magnitude lower than pure aerodynamic fluctuations, so it is essential to be extremely accurate when transferring information from one mesh to another.
Therefore, displaying on Fig.~\ref{fig:COVO_acous} the same figure as Fig.~\ref{fig:COVO_aero} with a much tighter colormap (pressure scales $10^4$ times lower) is required to evidence the emitted spurious acoustic fluctuations. 
This time, one can observe in both cases a significant spurious acoustic emission, as well as unwanted oscillations on the pressure fields. 

It has been recently shown by Gendre \textit{et al.}~\cite{Gendre2017} that the quality of the algorithm remains essential to reduce spurious acoustic waves. However, and without making any statement on the quality of the numerous cell-vertex or cell-centered algorithms that exist in the literature, the vortex test case of Fig.~\ref{fig:COVO_acous} exhibits the evidence of spurious acoustics for many very different mesh transition algorithms. It therefore seems that such an issue remains regardless of the local algorithm adopted at the interface.

In this regard, the objective of the present study is to identify the origin of these phenomena, that seems independent of the mesh transition algorithm. 
To this end, an original investigation of the LB model in the fluid core will be proposed by focusing on the effects of non-hydrodynamic modes crossing the interface. 
The latter will be highlighted thanks to linear stability analyses and specifically developed mode sensors. As the mesh transition algorithm is not the purpose of the present work, the classical cell-vertex algorithm will be adopted for the rest of the study, as it emits less spurious acoustics on Fig.~\ref{fig:COVO_acous}.

The next section is introduces the LB method with the classical BGK collision operator. Even if it suffers from known stability issues, it is interesting from a pedagogical point of view to understand the spurious behavior that occurs at grid interface on such mainstream collision scheme.

\section{Lattice Boltzmann method}
\label{sec:LBM}

\subsection{The lattice Boltzmann method with the BGK collision operator}
\label{subsec:LBM}

The lattice Boltzmann method describes the time and space evolution of the discrete particle distribution functions $f_i(\mathbf{x},t)$, which can be viewed as the probability density of finding fictive particles at position $\mathbf{x}$, at time $t$ and advected at discrete velocities $\boldsymbol{\xi} _i$.
With no body-force term, it can be expressed as 
\begin{equation} \label{eq:LBMequation}
f_i \left(\mathbf{x} + \boldsymbol{\xi} _i , t + 1\right) - f_i\left(\mathbf{x},t\right) = \Omega_i(\mathbf{x},t).
\end{equation}

The collision operator $\Omega_i(\mathbf{x},t)$ can be approximated by the single-relaxation time Bhatnagar-Gross-Krook (BGK) model~\cite{Bhatnagar1954}
\begin{equation} \label{eq:BGKmodel}
\Omega^{BGK} _i(\mathbf{x},t) = -\frac{1}{\tau}\bigg(f_i(\mathbf{x},t)-f_i^{(0)}(\mathbf{x},t)\bigg),
\end{equation}

\noindent where $\tau$ is the relaxation time towards an equilibrium distribution function $f_i^{(0)}$. The latter is usually approximated using an expansion in Hermite polynomials $\herm_i^{(n)}$ up to an order $N$~\cite{Shan2006,Philippi2006a}

\begin{equation} \label{eq:fEqH}
  f_i^{(0)} = w_i \sum_{n=0}^{N}\frac{1}{c_s^{2n}n!} \textit{\textbf{a}}_{0}^{n}:\herm^{(n)}_i,
\end{equation}

\noindent where the constant $c_s$ and the Gaussian weights $w_i$ are characteristic of the lattice of velocities $\boldsymbol{\xi}_i$, and $\textit{\textbf{a}}^{(n)}_{0}$ are equilibrium coefficients obtained by a projection of the Maxwell-Boltzmann distribution function onto the Hermite polynomials $\herm_i^{(n)}$ defined as
\begin{equation}
\mathbf{\herm}_i ^{(n)} = \frac{\left(-c_s^2\right)^n}{w\left(\boldsymbol{\xi}_i\right)} \nabla _{\boldsymbol{\xi}} ^n w\left(\boldsymbol{\xi}_i\right), \qquad \text{where} \qquad w\left(\boldsymbol{\xi}\right) = \frac{1}{\left(2\pi c_s^2\right)^{D/2}} \exp \left( -\frac{||\overline{\xi}|| ^2}{2c_s^2} \right).
\end{equation}

The D2Q9 lattice \cite{Qian1992a} is used throughout the rest of the present paper for clarity sake. The lattice constant $c_s=1/\sqrt{3}$, the discrete velocities $\boldsymbol{\xi} _i$ are

\begin{equation}
\label{eq:xi_i}
	\boldsymbol{\xi}_i = \left.
	\begin{cases}
	\quad	{0\ ,\ 1,\ 0,-1,\ \ 0,\ 1,-1,-1,\ \ 1}\\
	\quad 	{0\ ,\ 0,\ 1,\ \ 0,-1,\ 1,\ \ 1,-1,-1}\\
    \end{cases}
    \right\}, 
 \end{equation}
 
\noindent and the associated Gaussian weights are

\begin{equation}
\label{eq:w_i}
   	w_i = 
    \begin{cases}
    	4/9,  &  i=0,\\
    	1/9,  &  i=1,2,3,4,\\
    	1/36, &  i=5,6,7,8.
    \end{cases}
\end{equation}

The equilibrium distribution function is truncated at the third-order to improve the stability of the method~\cite{Wissocq2019}, especially for advanced collision models introduced subsequently~\cite{phDCoreixas2018}. For the D2Q9 lattice, it reads

\begin{equation} \label{eq:fEqO3}
	 f_i^{(0)} = w_i \rho \left[  1 + \frac{\boldsymbol{\xi}_i \cdot \boldsymbol{u}}{c_s ^2} + \frac{1}{2} \left(\frac{\boldsymbol{\xi_i} \cdot \boldsymbol{u}}{c_s ^2} \right)^2 - \frac{\boldsymbol{u}^2}{2 c_s^2}  + \frac{u_x u_y}{2 c_s^2} \left( \frac{\xi_{i,x} \xi_{i,y} \left( \boldsymbol{\xi}_i \cdot \boldsymbol{u} \right)}{c_s^4} - \frac{\xi_{i,x}u_y + \xi_{i,y}u_x}{c_s^2}  \right) \right].
\end{equation}

The macroscopic quantities of interest in the case of an isothermal flow (\textit{i.e.} the density $\rho$, and the velocity $\boldsymbol{u}[u_x,u_y]$), are given by the following moments of the distribution function

\begin{equation} \label{eq:moment_rho}
	\rho = \sum_{i} f_i,
\end{equation}
\begin{equation} \label{eq:moment_rhoU}
	\rho \boldsymbol{u} = \sum_{i} \boldsymbol{\xi}_i \, f_i.
\end{equation}

Finally, a Chapman-Enskog expansion~\cite{Chapman1990} provides a systematic link between the relaxation time $\tau$ and the dimensionless kinetic viscosity $\nu$. For the BGK collision operator, it gives

\begin{equation} \label{eq:linkTauNu}
	\nu = c_s^2 \left(\tau - \frac{1}{2} \right).\\
\end{equation}


As it is well known in the literature~\cite{DHumieres2002a}, the BGK collision model (\ref{eq:BGKmodel}) suffers from severe stability issues, especially due to non-hydrodynamic mode contributions for under resolved simulations~\cite{Dellar2002}. 
This model will be studied in Sec.~\ref{subsec:LSA_BGK}. To improve the LBM stability, many advanced models have been developed. Some very common examples are based on multiple relaxation times (MRT)~\cite{DHumieres1992a,DHumieres2002a} whose purpose is to relax each moment of $f$ at an intended relaxation rate. These models increase the LBM stability, by relaxing the ``ghost variables''~\cite{Adhikari2008} at a certain rate, and allow modifying the bulk viscosity~\cite{phDGendre2018}.\\

A well-known family of advanced collision models constructed in order to cancel high-order contributions are the regularized collision models. 
The standard regularization procedure can be interpreted as a MRT model in the Hermite basis~\cite{phDLatt2007} where ``ghost variables'' are relaxed at a specific value $\tau_g=1$. 
Previous studies have highlighted the capability of regularized collision models to completely filter some non-hydrodynamic modes out of a computation. For instance, the recursive regularized model allows for  reducing the number of modes of the D2Q9 lattice to six modes instead of nine~\cite{Malaspinas2015}.
This mode filtering property of regularized operators will be particularly interesting to emphasize the effects of non-hydrodynamic modes crossing a mesh refinement interface. 
It is also worth noting that regularized collision models can be re-interpreted in the framework of entropic LBM \cite{kramer2019}.
This is why they will be adopted in the following.

\noindent In the next section, regularized collision models are then introduced.



\subsection{Recursive regularized collision model: RR}
\label{subsec:RR}

The regularization procedure consists in filtering out high-order moments of the off-equilibrium distribution function to enhance the stability of the LBM scheme. This is done by reconstructing the distribution function before the collision step as proposed by Latt and Chopard~\cite{Latt2006}. The reconstruction consists in projecting the off-equilibrium part of the distribution function onto the Hermite basis up to the second-order.

\begin{equation} \label{eq:reguProcedure}
	f_{i}^{reg} \equiv f_{i}^{(0)} + f_{i}^{(1),reg},
\end{equation}

\noindent with

\begin{equation} \label{eq:reguProceduref1N}
  f_i^{(1), reg} = w_i \, \frac{1}{2 c_s^4} \, \boldsymbol{a}_1^{(2)}:\herm_i^{(n)}, 
\end{equation}
where $\textit{\textbf{a}}^{(2)}_{1}$ are the off-equilibrium expansion coefficients defined as
\begin{equation} \label{eq:a1Proj}
	\textit{\textbf{a}}^{(2)}_{1} = \sum \mathbf{\herm}_i ^{(2)} \left( f_i-f_i^{(0)} \right).
\end{equation}

This is equivalent to relaxing off-equilibrium moments of order higher than two at equilibrium before the collision step. Furthermore, the regularization method aims at filtering out higher-order contributions in Knudsen number by imposing Eq.~(\ref{eq:reguProcedure})~\cite{Latt2006}. The LBM scheme can be modified to apply the regularized procedure to all the distribution functions $f_i$. Its becomes

\begin{equation} \label{eq:LBMscheme_reg}
f_i \left(\mathbf{x} + \boldsymbol{\xi} _i , t + 1\right)= f_i^{(0)} + \left(1 - \frac{1}{\tau}\right) f_i^{(1), reg}.
\end{equation}

It has been shown in~\cite{Malaspinas2015,phDCoreixas2018} that this model suffers from stability issues for relatively small Mach numbers which make it not well suited for many aeroacoustic applications. 
 
More recently, it has been proposed to enhance the content of $f_i^{(1),reg}$ by including some higher-order off-equilibrium expansion coefficients (identified as moments of $f^{(1),reg}$~\cite{Shan2006}) in the regularization procedure (Eq.~\ref{eq:reguProcedure}). It has been done through a recurrence formula~\cite{Malaspinas2015} obtained thanks to a Chapman-Enskog expansion. These models are referred to as recursive regularized collision models and have yielded greatly improved stability for the regularization procedure~\cite{Malaspinas2015,Coreixas2017,phDCoreixas2018}. For the D2Q9 lattice, the off-equilibrium distribution function $f_i^{(1),reg}$ up to the third-order can be expressed as:
\begin{equation} \label{eq:f1RR}
	 f_i^{(1),reg} = w_i \left( \frac{1}{2c_s ^4} \mathcal{H}_i^{(2)}:\textit{\textbf{a}}^{(2)}_{1} + \frac{1}{2 c_s ^6} \left( \mathcal{H}^{(3)}_{i,xxy} \textit{a}^{(3)}_{1,xxy} + \mathcal{H}^{(3)}_{i,xyy} \textit{a}^{(3)}_{1,xyy} \right)   \right),
\end{equation}

\noindent where

\begin{equation} \label{eq:a13RR}
	\textit{a}^{(3)}_{1,xxy} = 2u_x a_{1,xy}^{(2)}+u_y a_{1,xx}^{(2)}, \qquad \textit{a}^{(3)}_{1,xyy} = 2u_y a_{1,xy}^{(2)}+u_x a_{1,yy}^{(2)}.
\end{equation}

Note that the recursive regularization procedure can also include fourth-order terms in $f_i^{(1),reg}$. Yet, no difference was observed in this study when including these terms. Hence, the RR model will refer to the expression of Eq.~(\ref{eq:f1RR}) injected into Eq.~(\ref{eq:reguProceduref1N}).


\subsection{Hybrid-Recursive regularize collision model: H-RR}
\label{subsec:HRR}
The last collision model introduced is chosen for its highly relevant spectral properties that will be presented in Sec.~\ref{subsec:LSA_HRR}. This recent model~\cite{Jacob2018} is based on a hybridization of the recursive regularized collision model with a finite difference reconstruction of the viscous stress tensor. Indeed, a Chapman-Enskog expansion~\cite{Chapman1990} allows to link the off-equilibrium populations $f_i^{(1)}=f_i-f_i^{(0)}$ with the deviatoric tensor $S_{\alpha \beta} = 1/2\left(\nabla \boldsymbol{u} + \left(\nabla \boldsymbol{u} \right)^T\right)$ through the second-order moments of $f^{(1)}$

\begin{equation} \label{eq:stressT}
	\sum_{i}\xi_{i,\alpha} \xi_{i,\beta}f_i^{(1)} \simeq -2 \tau \rho c_s^2 S_{\alpha \beta}.
\end{equation}
It it then possible to hybridize the viscous stress tensor computation as follow

\begin{equation} \label{eq:a1ij_HRR}
  \textit{\textbf{a}}_1^{(2),HRR} = \sigma \textit{\textbf{a}}_1^{(2)} + \left(1-\sigma \right) \left(-2 \tau \rho c_s^2 S_{\alpha \beta}\right), \qquad  \left(0 < \sigma < 1 \right),
\end{equation}

\noindent where $\textit{\textbf{a}}_{1}^{(2)}$ are computed thanks to Eq.~(\ref{eq:a1Proj}) and $S_{\alpha \beta}$ is estimated with a second-order centered finite-difference scheme. Then, third-order off-equilibrium Hermite coefficients are computed recursively using the modified $\textit{\textbf{a}}_1^{(2),HRR}$ with Eq.~(\ref{eq:a13RR}). As in the RR collision model, $f_i^{(1)}$ is computed using Eq.~(\ref{eq:f1RR}) before the collision step, and injected into the LBM scheme (\ref{eq:LBMscheme_reg}).\\

The choice of the three collision models presented above is far from exhaustive, and has been motivated by their interesting spectral properties that will be highlighted in Sec.~\ref{sec:LSA}. The purpose of the present article is to exhibit the effects of these collision models on the spurious phenomena occurring at a mesh transition, and provide convincing explanations to them. Hence, the next section is devoted to the grid refinement algorithm that has been used in this work.


\section{Grid refinement algorithm}
\label{sec:refinement}
Before introducing a particular grid refinement algorithm, it is worth mentioning that the concepts presented in the following are independent of the grid refinement algorithm and have been validated for both cell-vertex and cell-centered algorithms in two and three dimensions. Since the aim of the article is not a comparison of grid coupling algorithms, the one from Lagrava \textit{et al.}~\cite{Lagrava2012} is chosen as it is one of the most popular and one of the simplest in term of implementation.\\
 This algorithm is based on the one from Dupuis \& Chopard~\cite{DupuisChopard2003} where the distribution functions are rescaled before the collision step in contrast to Filippova \textit{et al.}~\cite{Filippova1998}. An additional filtering procedure~\cite{Touil2014} is applied during the fine to coarse grid transfer to avoid aliasing effects.\\
Grid refinement algorithm details and conversion relations between fine and coarse quantities are described below.

\subsection{Rescaling of physical quantities}
\label{subsec:rescaling}
In the following, a plane transition separating a fine and a coarse grid resolution domain is considered. Since the dimensionless convention has been adopted, each resolution level possesses its own ``lattice units". This change of scale requires a rescaling of the physical quantities between grids.

\noindent Any quantity related to the fine or coarse domain is denoted by a superscript \textit{f} and \textit{c}, respectively. The coarse and fine mesh sizes are linked with each other as $\Delta x^c = 2 \Delta x^f$. In the case of an acoustic scaling, the timestep is imposed as $\Delta t^c = 2\Delta t^f $. In the following, the coarse scale is chosen to make  the space and time steps dimensionless.\\

\noindent The dimensionless viscosity must be rescaled in order to ensure the Reynolds number continuity~\cite{Filippova1998}
\begin{equation} \label{eq:rescNu}
	\nu ^f = \frac{\Delta x^c}{\Delta x^f} \nu ^c = 2 \nu ^c,
\end{equation}

\noindent leading to the following relation between the relaxation times

\begin{equation} \label{eq:rescTau}
	\tau ^f = 2 \tau ^c - \frac{1}{2}.
\end{equation}

Unlike the equilibrium part of the distribution function which depends only on macroscopic quantities that are continuous through the interface, the off-equilibrium part $f_i^{(1)}$ has to be rescaled since it depends on velocity gradients through Eq.~(\ref{eq:stressT}). By a combination of (\ref{eq:rescTau}) and (\ref{eq:stressT}), the relation between the off-equilibrium parts of the fine and coarse populations is
\begin{equation} \label{eq:fneqResc}
	f_i^{(1),c} = 2 \frac{\tau^c}{\tau^f} f_i^{(1),f}.
\end{equation}

It is worth noting that this relation allows to build distribution functions that are missing at the grid interface after a streaming step:

\begin{equation} \label{eq:rescff}
	f_i ^f = f_i ^{(0)} + \frac{\tau ^f}{2 \tau ^c} f_i ^{(1),c},
\end{equation}

\begin{equation} \label{eq:rescfc}
	f_i ^c = f_i ^{(0)} + \frac{2 \tau ^c}{\tau ^f} \tilde{f}_i ^{(1),f},
\end{equation}

\noindent where the superscript $\tilde{f}^{(1)}$ stands for the filtered value of the off-equilibrium distribution function. This filtering step is highly recommended for stability and accuracy reasons~\cite{Gendre2017,Lagrava2012}. It is also worth noting that cell-centered algorithms implicitly use a spatial filtering during the coalescence step \cite{phDGendre2018}. The filter used in the present work is the one proposed in~\cite{Touil2014}. For a D2Q9 lattice, it reads

\begin{equation} \label{eq:filterFneq}
\tilde{f}_i^{(1),f} \left(\mathbf{x},t\right) = \frac{1}{4} f_i^{(1),f}\left(\mathbf{x},t\right) + \frac{1}{8} \sum_{\alpha=1}^4 f_i^{(1),f}\left(\mathbf{x} + \boldsymbol{\xi}_\alpha,t \right) + \frac{1}{16} \sum_{\alpha=5}^8 f_i^{(1),f}\left(\mathbf{x} + \boldsymbol{\xi}_\alpha,t \right),
\end{equation}

\noindent where the three contributions correspond respectively to the \textit{center}, \textit{normal} and \textit{diagonal} directions over the neighboring nodes with the convention of lattice velocities $\boldsymbol{\xi}_i$ given in Sec.\ref{sec:LBM}.

The last quantity that needs to be rescaled, especially for the hybrid recursive regularized collision model (H-RR) introduced in Sec.~\ref{subsec:HRR}, is the strain rate tensor

\begin{equation} \label{eq:PijResc}
	S_{\alpha \beta}^{c} = 2 S_{\alpha \beta}^{f}.
\end{equation}

\subsection{Cell-vertex algorithm with overlapping area}
\label{subsec:algorithm}
In the standard collide \& stream algorithm, some distribution functions are missing at the interface after a streaming step. The use of an overlapping area to couple grids is a very common way to recover the missing populations. This strategy is adopted for most of both cell-centered~\cite{Rohde2006,Chen2006} or cell-vertex~\cite{DupuisChopard2003,Filippova1998} algorithms.

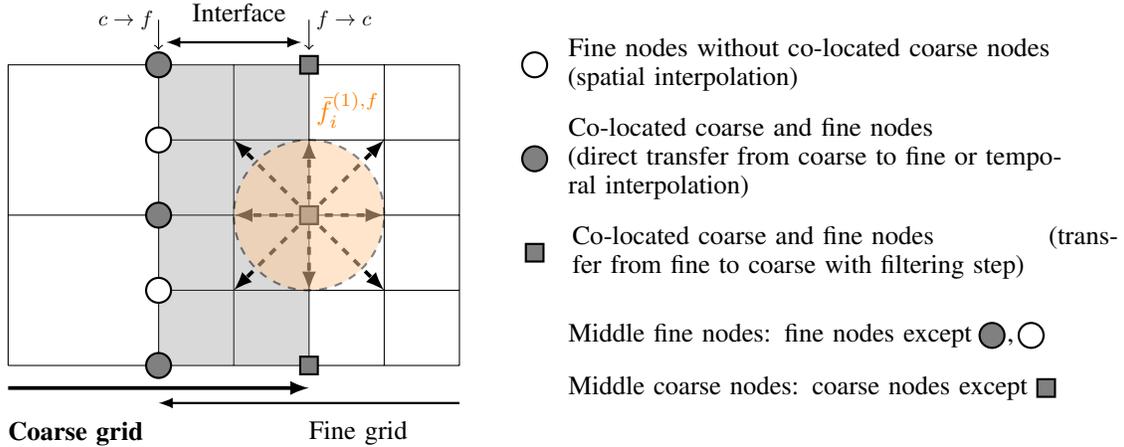
\begin{figure}[H]
		\begin{center}
		\begin{tikzpicture}[scale=1.]
			\draw (0,0) grid[step=2.] (2,4);
			\draw (2,0) grid[step=1.] (6,4);
			\fill[gray,opacity=0.3] (2,0) rectangle (4,4) ;
			\tikzstyle{TC}=[circle,draw,thick,fill=gray]
			\draw (2,0) node[TC]{};
			\draw (2,2) node[TC]{};
			\draw (2,4) node[TC]{};
			\tikzstyle{TN}=[draw,rectangle,thick,fill=gray]
			\draw (4,0) node[TN]{};
			\draw (4,2) node[TN]{};
			\draw (4,4) node[TN]{};
			\tikzstyle{IE}=[circle,draw,thick,fill=white]
			\draw (2,1) node[IE]{};
			\draw (2,3) node[IE]{};
			\tikzstyle{IF}=[draw,rectangle,thick,fill=white]
			\draw [<-> , >=latex , line width=0.3mm] (2.1,4.3) -- (3.9,4.3);
			\node[text width=3.5cm] at(4.2,4.7){Interface};
			\draw [->] (2,4.6) -- (2,4.2);	
			\node[text width=3.5cm,scale=0.8] at(2.6,4.6){$c \rightarrow f$};	
			\draw [->] (4,4.6) -- (4,4.2);		
			\node[text width=3.5cm,scale=0.8] at(5.5,4.6){$f \rightarrow c$};		
			\draw [-> , >=latex , line width=0.5mm] (0,-0.3) -- (4,-0.3);
			\node[text width=3cm] at(1.5,-0.9){\textbf{Coarse grid}};		
			\draw [<- , >=latex , line width=0.3mm] (2,-0.5) -- (6,-0.5);
			\node[text width=4cm] at(6,-0.9){Fine  grid};	
			
			\draw [-> , >=latex ,line width=0.5mm,dashed] (4,2) -- (5,3);		
			\draw [-> , >=latex ,line width=0.5mm,dashed] (4,2) -- (5,2);
			\draw [-> , >=latex ,line width=0.5mm,dashed] (4,2) -- (5,1);
			\draw [-> , >=latex ,line width=0.5mm,dashed] (4,2) -- (4,1);
			\draw [-> , >=latex ,line width=0.5mm,dashed] (4,2) -- (4,3);
			\draw [-> , >=latex ,line width=0.5mm,dashed] (4,2) -- (3,1);
			\draw [-> , >=latex ,line width=0.5mm,dashed] (4,2) -- (3,2);
			\draw [-> , >=latex ,line width=0.5mm,dashed] (4,2) -- (3,3);		
			\draw (4,2) node[TN]{};							
		
			\tikzstyle{Restr}=[thick, dashed, fill=orange!40, opacity=0.5]
			\draw [Restr] (4,2) circle (1) ;	
			\node[text width=3.5cm, color=orange] at(5.85,3.4){$\bar{f}_i^{(1),f}$};	
		
			\draw (7,4) node[IE]{};
			\node[text width=6.5cm] at(10.7,4){Fine nodes without co-located coarse nodes (spatial interpolation)};
			\draw (7,2.75) node[TC]{};
			\node[text width=6.5cm] at(10.7,2.75){Co-located coarse and fine nodes \qquad \qquad (direct transfer from coarse to fine or temporal interpolation)};
			\draw (7,1.5) node[TN]{};
			\node[text width=7.2cm] at(11.1,1.5){Co-located coarse and fine nodes \qquad \qquad (transfer from fine to coarse with filtering step)};
			\node[text width=6.5cm] at(10.7,0.4){Middle fine nodes: fine nodes except \ \ \ ,};
			\draw (13.08,0.4) node[TC]{};
			\draw (13.6,0.4) node[IE]{};

			\node[text width=6.5cm] at(10.7,-0.3){Middle coarse nodes: coarse nodes except};
			\draw (13.8,-0.3) node[TN]{};

		\end{tikzpicture}
		\end{center}
	\caption{\label{fig:Overlap} Two dimensional representation of a plane refinement interface and nodes definition. The overlapping interface is colored in gray. The orange area represents the filter zone to compute $\tilde{f}_i^{(1),f}$.}
\end{figure}

The popular algorithm of Lagrava \textit{et al.}~\cite{Lagrava2012}, that is adopted for this study is based on a cell-vertex formulation and relies on the definition of an overlapping area whose thickness is equal to one coarse cell (Fig.~\ref{fig:Overlap}). Distribution functions from coarse to fine grids are transferred at co-located nodes \TF \ and rescaled using Eq.~(\ref{eq:rescff}). Fine to coarse distributions are exchanged at co-located nodes \TN \ after being filtered with Eq.~(\ref{eq:filterFneq}) and rescaled with Eq.~(\ref{eq:rescfc}).

As there are two fine iterations in one coarse time step, a temporal interpolation of $\rho ^c$, $u^c$  and  $f^{(1),c}$ is mandatory to reconstruct the fine missing populations during the asynchronous iterations on nodes \TF. A third-order polynomial interpolation is used ~\cite{Touil2014}.
	For a quantity $g$, its reads:

\begin{equation} \label{eq:TempInt}
g \left( \TF , t + 1/2 \right) = -\frac{1}{8} g \left( \TF , t-1/2 \right) + \frac{3}{4} g \left( \TF , t\right) + \frac{3}{8} g \left( \TF , t + 1 \right).
\end{equation}

At fine nodes \IE \ which do not have a corresponding coarse node, a third-order spatial interpolation  is used to enforce the second-order accuracy of the LBM at the grid interface~\cite{Lagrava2012}. In the following, an interface with a normal and tangential vector $e_x(1,0)$ and $e_y(0,1)$ respectively is considered. For a quantity \textit{g}, the interpolation scheme reads

\begin{equation} \label{eq:spaceInt}
	g \left( \IE , t \right) = \frac{9}{16} \bigg(g \left( \IE+e_y/2 ,t\right) + g \left( \IE - e_y/2 ,t\right) \bigg) - \frac{1}{16} \bigg(g \left( \IE+3e_y/2 ,t\right) + g \left( \IE-3e_y/2,t\right) \bigg).
\end{equation}

\subsection{Modification of the standard algorithm for the hybrid-recursive regularized collision model}
\label{subseq:modifiedAlgo}
The hybrid-recursive regularized (H-RR) collision model requires the computation of a velocity gradient during the collision step. For this last quantity, the grid refinement algorithm has to be adapted in order to take into account this specificity.
These modifications are not straightforward since, on the \TF \ and \TN \ nodes, the velocity gradient cannot be computed on the fine and coarse mesh respectively. Therefore it has to be transferred using the corresponding velocity gradient computed on the opposite resolution mesh. This computation is particularly delicate for the $\TF ^f$ nodes since it requires knowledge of the velocity on the \TN \ nodes at time $t+1$ which is not known in the classical algorithm. The latter must be computed in advance as follows.

\begin{enumerate}[label=\arabic*)]
	\setlength\itemsep{-0.5em}
	\item \textbf{Reference state} $\rightarrow$ Fine grid $t$ ; Coarse grid $t$
	\begin{enumerate}[label=\alph*.]

		\item All the distribution functions are known on both grids
	\end{enumerate}

	\item \textbf{Asynchronous iteration} $\rightarrow$ Fine grid $t+1/2$ ; Coarse grid $t+1$
	\begin{enumerate}[label=\alph*.]
		\setlength\itemsep{-0.0em}
		\item Propagation step for fine and coarse middle nodes.
		\item Temporal interpolation of $\rho^c,u^c$ on $\TF^f$ nodes (\ref{eq:TempInt}).
		\item Spatial interpolation of $\rho^f,u^f$ on $\IE$ nodes (\ref{eq:spaceInt}).
		\item Collision of fine middle nodes.
		\item Additional fictitious streaming step on $\TN ^f$ nodes over fine neighbors to compute $u(\TN,t+1$). 
		\item Computation of the viscous stress tensor $S^c_{\alpha \beta}$ on  $\TF ^c$ nodes.
		\item Temporal interpolation of $f^c$ on $\TF^f$ nodes (\ref{eq:TempInt}). $f^f$ are reconstructed using (\ref{eq:rescff}). $S^c_{\alpha \beta}$ is also interpolated and rescaled to the fine scale using (\ref{eq:PijResc}) to obtain $S^f_{\alpha \beta}$.
		\item Spatial interpolation of $f^f$ and $S^f_{\alpha \beta}$ on $\IE$ nodes (\ref{eq:spaceInt}).
		\item Collision of $\TF^f$, \IE \ fine nodes.
		\end{enumerate}
	
	\item \textbf{Synchronous iteration} $\rightarrow$ Fine grid $t+1$ ; Coarse grid $t+1$
	\begin{enumerate}[label=\alph*.]
		\setlength\itemsep{-0.em}
		\item Propagation step for fine middle nodes.
		\item Transfer of $\rho^c,u^c,f^c$ and reconstruction of $f^f$ using (\ref{eq:rescff}) on $\TF ^f$ nodes. Transfer and rescaling of $S^c_{\alpha \beta}$ (\ref{eq:PijResc}) to obtain $S^f_{\alpha \beta}$.
		\item Spatial interpolation of $\rho^f,u^f,f^f$ and $S^f_{\alpha \beta}$ on nodes $\IE$ (\ref{eq:spaceInt}).
		\item Filtering step (\ref{eq:filterFneq}) and reconstruction of $f^c$ on \TN \ nodes using (\ref{eq:rescfc}). Transfer and rescaling of $S^f_{\alpha \beta}$ to obtain $S^c_{\alpha \beta}$  (\ref{eq:PijResc}).
		\item Collision of all nodes.
	\end{enumerate}
	\item \textbf{Repetition of steps 2) to 4) until the end of the simulation}.\\
\end{enumerate}

Up to now, grid coupling concepts have been introduced and a cell-vertex algorithm has been described. Grid refinement algorithms are known to produce either spurious vorticity~\cite{Hasert2014} or acoustics~\cite{Gendre2017}. In this study, the spurious artifact's origin is assumed to be tightly linked with non-hydrodynamic phenomena. One useful tool to analyze both hydrodynamic and non-hydrodynamic contributions in the LBM schemes is the spectral analysis tool that is introduced in the following section.


\section{Spectral analysis and non-hydrodynamic modes}
\label{sec:LSA}


\subsection{Von Neumann analyses of LBM schemes}
\label{subsec:LSAlbm}

The von Neumann analysis~\cite{Neumann1950} is a very powerful tool to investigate the behavior of numerical schemes as the LBM, in terms of stability and accuracy properties. This method consists of evaluating the response of a system, which is described by a given set of equations, to small disturbances. It can exhibit the coexistence of physical and non-physical modes~\cite{Adhikari2008} in a computation. 

The standard von Neumann analysis principles can be found in~\cite{Neumann1950}. Sterling and Chen were among the first to apply this method to the LBM scheme~\cite{Sterling1996a}. The first step is to linearize the Eq.~(\ref{eq:LBMequation}) about an equilibrium state~\cite{Sterling1996a}, and to inject perturbations as complex plane monochromatic waves in the following form

\begin{equation} \label{eq:LSA_monowaves}
	f^{'}_i(x,t)=\hat{f_i} \exp \left(i(\omega t - \mathbf{k} \cdot \mathbf{x})\right),
\end{equation}

\noindent where $f_i^{'}$ are the fluctuating populations coming from the decomposition of the global population $f_i$ into a mean part $\bar{f_i}$ plus a fluctuating one. $\hat{f_i} \in \mathbb{C}$, $\omega$ is the complex pulsation of the perturbations and $\mathbf{k}$ is the wavenumber vector. The physical perturbation corresponds to the real part of this complex wave

\begin{equation} \label{eq:LSA_Remonowaves}
	\Re(f_i^{\prime}) = |\hat{f_i}| e^{-\omega_i t} \cos \left( \omega _r t - \mathbf{k \cdot x} + \Phi _i \right),
\end{equation}

\noindent where $\Phi _i = \arg(\hat{f_i})$. The real part $\omega _r = \Re (\omega)$ and the imaginary part $\omega _i = \Im(\omega)$ give respectively the dispersion and the dissipation of the perturbation. The phase velocity $\mathbf{v_{\phi}}$ and the group velocity $\mathbf{v_g}$ of the monochromatic wave are defined as 

\begin{equation} \label{eq:phase_velocity}
	\begin{aligned}
	\mathbf{v_\phi} = \frac{\omega_r}{\mathbf{k}}, \\
	\qquad \mathbf{v_g} = \frac{\mathrm{d}\omega_r}{\mathrm{d}\mathbf{k}}.
		\end{aligned}
\end{equation}

By injecting the complex monochromatic perturbations $f_i^{\prime}$ into Eq.~(\ref{eq:LBMequation}), one can obtain the following eigenvalue problem

\begin{equation} \label{eq:LSA_eigenvalue}
e^{i\omega} \mathbf{F} = \mathbf{M^{LBM}} \mathbf{F},
\end{equation}

\noindent with $\mathbf{M^{LBM}}$ the time-advance matrix which depends on the collision model and $\mathbf{F}=[\hat{f}_i]^T$ the vector of modal fluctuations. Eigenvalues of Eq.~(\ref{eq:LSA_eigenvalue}) are then studied to obtain the dissipation ($\omega _i$) and dispersion ($\omega _r$) properties of the LBM scheme. The time-advance matrices used to analyze the different collision operators presented in Sec.~\ref{sec:LBM} can be found in App.~A.

A similar analysis can be performed on the isothermal Navier-Stokes equations. It gives three linear modes in two dimensions: one shear (or vorticity) mode and two acoustic (one moving upstream $ac-$ and one moving downstream $ac+$) modes. The eigenvalues of these modes are:

\begin{equation} \label{eq:eigenvaluesNS}
	\begin{aligned}
		\omega _{shear} = \mathbf{k} \cdot \bar{\boldsymbol{u}}+i\nu \left\Vert \mathbf{k} \right\Vert^2, \\
		\omega _{ac+} = \mathbf{k} \cdot \bar{\boldsymbol{u}} + \left\Vert \mathbf{k} \right\Vert c_s + i\nu \left\Vert \mathbf{k} \right\Vert^2, \\
		\omega _{ac-} = \mathbf{k} \cdot \bar{\boldsymbol{u}} - \left\Vert \mathbf{k} \right\Vert c_s + i\nu \left\Vert \mathbf{k} \right\Vert^2,	
	\end{aligned}
\end{equation}

\noindent where $\bar{\mathbf{u}}$ is the mean flow velocity. Moreover, following the methodology proposed in~\cite{Wissocq2019}, the LBM eigenvectors $\mathbf{F}$ are used to give a physical interpretation to  modes resulting from the von Neumann analysis. This decomposition is performed by projecting a LBM macroscopic vector $\mathbf{\mathcal{V}}=[\hat{\rho},(\widehat{\rho \boldsymbol{u}})]^T$ composed of moments of $\mathbf{F}$

\begin{equation} \label{eq:momentsF}
	\begin{aligned}
		\hat{\rho} = \sum_{i} \hat{f}_i, \\
		(\widehat{\rho \boldsymbol{u}}) = \sum_{i} \hat{f}_i \boldsymbol{\xi}_i,
	\end{aligned}
\end{equation}

\noindent onto the Navier-Stokes ones. This analysis allows for a systematic identification of the modes carrying a macroscopic information at more than a prescribed ratio $\eta$. In the results presented below, this parameter will be set to $\eta=0.99$.\\
  

\subsection{Von Neumann analysis of the BGK model}
\label{subsec:LSA_BGK}   
  
\begin{figure}[H]
	\begin{center}
		\includegraphics[scale=0.75]{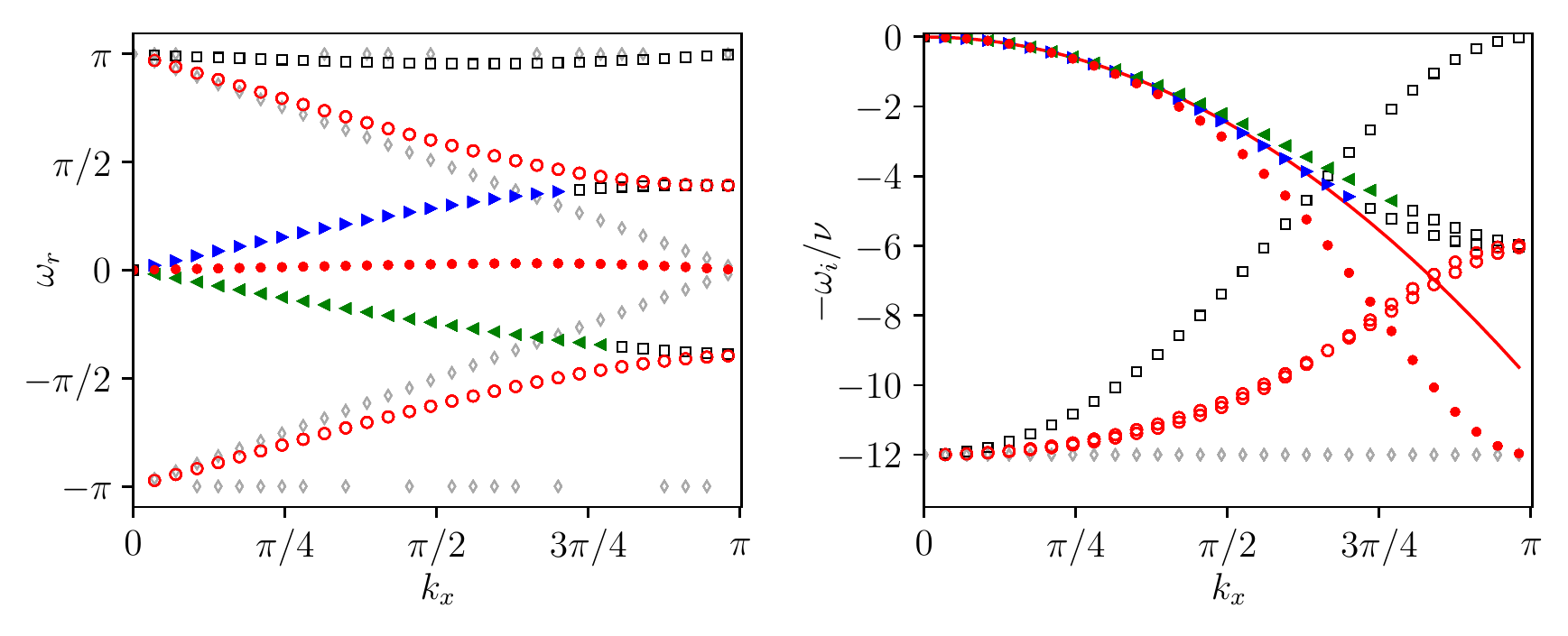}
		\caption{\label{fig:BGKlsa} Propagation and dissipation curves for the BGK collision model. $\nu=10^{-6}$, Ma=0.1. $\left(\protect\acousP\right)$: Ac+ mode, $\left(\protect\acousN\right)$ : Ac- mode, $\left(\protect\shearP\right)$ : Shear mode, $\left(\protect\numS\right)$ : SpuriousS modes, $\left(\protect\numB\right)$ :  SpuriousAc modes, $\left(\protect\numG\right)$ : SpuriousG modes, (\protect\redC) : Isothermal Navier-Stokes.}
		\end{center}
\end{figure}

The von Neumann analysis of the LBM scheme with the BGK model is presented on Fig.~\ref{fig:BGKlsa}. The analysis is performed with a mean flow of $\bar{u} \cdot \vec{e_x}=0.1\, c_s$, corresponding to a Mach number $\mathrm{Ma}=0.1$ along the $x$ axis with an increasing wavenumber $k_x$ ranging from $0$ to $\pi$. 

The dimensionless viscosity is set equal to $\nu=10^{-6}$. This value is retained for the numerical experiments in the following, since it is in the order of magnitude of viscosity of air for a minimal mesh size of $\Delta x=0.01 \mathrm{m}$. It is likely that the mesh will be more resolved, leading to a larger value of dimensionless kinematic viscosity. Here, this very low value is retained to put the emphasis on the spurious phenomena that will arise, knowing that the value of $\tau$ will not affect the results and the explanations provided below.

Since the analysis has been performed on a D2Q9 lattice, nine modes are observed. Furthermore, the imaginary part of the Navier-Stokes eigenvalues (\textit{cf.} Eq.~\ref{eq:eigenvaluesNS}), is, in the following, taken as the reference for the dissipation curves (\redC) that physically takes place in real flows.

The eigenvector analysis proposed in~\cite{Wissocq2019} allows for the identification of modes carrying more than $99\%$ of an acoustic information identified with (\acousP )  and (\acousN )  on Fig.~\ref{fig:BGKlsa}, and modes carrying a shear information displayed with (\shearP ) and (\numS ). Three modes are found for the latter category, while only one shear wave is expected by the NS equations. In the present article, and for a sake of clarity, these shear modes will be further distinguished thanks to their propagation speed. Indeed, on the case illustrated here, only one mode has a propagation speed close the expected one and can be identified as a physical shear mode. Hence, the nine modes observed on Fig.~\ref{fig:BGKlsa} can be classified into six categories:
\newline

\noindent The physical modes:

\begin{itemize}
\setlength\itemsep{-1.em}
\item (\shearP) The \textbf{shear} modes.\\
\item (\acousP) The acoustic \textbf{Ac+} modes.\\
\item (\acousN) The acoustic \textbf{Ac-} modes.
\end{itemize}

\noindent The non-hydrodynamic modes:

\begin{itemize}
\setlength\itemsep{-1.em}
\item (\numS) The \textbf{SpuriousS} modes that carry a shear quantity (transverse velocity fluctuation) at an incorrect phase velocity.\\
\item (\numB) The \textbf{SpuriousAc} modes thats carry a combination of the two acoustics waves. These modes do not propagate at the sound celerity. \\
\item (\hspace*{0.036cm}\numG \hspace*{0.036cm}) The \textbf{SpuriousG} modes thats does not carry any physical quantities. These modes are invisible at the macroscopic level.\end{itemize}

The ``Spurious" terms are related to the non-hydrodynamic modes since they are unexpected in a simulation. Furthermore, as it will be shown in Sec.~\ref{sec:NHeffect}, they can have troublesome effects, especially in the presence of non-uniform grids. The spuriousAc (\numB) modes have a projection on both the density and the longitudinal velocity, but do not have any contribution on the transversal velocity. As a matter of fact, they can be considered as spurious acoustic modes since their projections on the Navier-Stokes acoustic waves are not null. Moreover, a tight link between the spuriousAc (\numB) modes and acoustics (\acousP,\acousN) will be shown in Sec.~\ref{subsec:acousPulse}.\\

In this article, collision models are chosen with regards to the properties of their non-hydrodynamic modes. For the BGK collision model, three non-hydrodynamic modes have an impact from the physical point of view: two spuriousS (\numS) modes that carry shear quantities and one spuriousAc (\numB) that carry acoustics. Furthermore, the dissipation of these modes is in the same order of magnitude of that of the physical ones, especially for under resolved waves. This specificity is particularly emphasized in~\cite{Dellar2002}.

\subsection{Von Neumann analysis of the RR model}
\label{subsec:LSA_RR}

\begin{figure}[H]
	\begin{center}
		\includegraphics[scale=0.75]{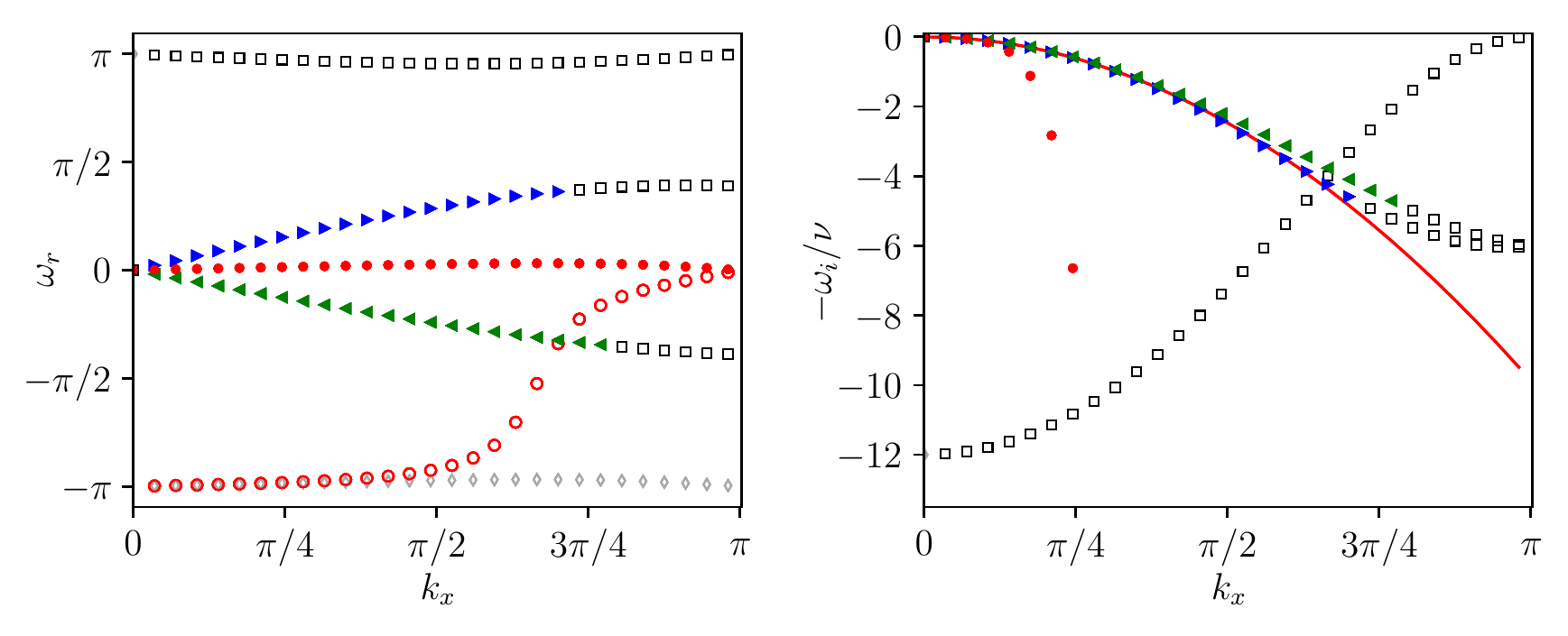} 
		\caption{Propagation and dissipation curves for the RR collision model. $\nu = 10^{-6}$, Ma=0.1. $\left(\protect\acousP\right)$: Ac+ mode, $\left(\protect\acousN\right)$ : Ac- mode, $\left(\protect\shearP\right)$ : Shear mode, $\left(\protect\numS\right)$ : SpuriousS modes, $\left(\protect\numB\right)$ :  SpuriousAc modes, $\left(\protect\numG\right)$ : SpuriousG modes, (\protect\redC) : Isothermal Navier-Stokes.} 
		\label{fig:RRlsa}
	\end{center}
\end{figure}

The results of the spectral analysis of the RR model are displayed on Fig.~\ref{fig:RRlsa} and one can observe many differences with the BGK model. In the present case, only six modes are present instead of nine, the eigenvalues associated with the three other ones having a null modulus. Compared to the BGK model, the two spuriousG modes (\numG) and one spuriousS mode (\numS) have been filtered out. Furthermore, the physical shear mode (\shearP) is more attenuated than in the BGK case. However, two spurious modes that carry physical quantities have remained: the spuriousAc modes (\numB), with identical dissipation and dispersion properties to those seen with the BGK model, and one spuriousS mode (\numS) that is much more attenuated, and has a lower group velocity $v_g$ for low wavenumbers.

The spectral behavior of this model with respect to non-hydrodynamic modes is interesting since some modes have disappeared and the others that carry shear are much more dissipated. On the contrary, the spuriousAc (\numB) one is not influenced  by this model. That is one of the reasons why the H-RR model is studied in the following.

\subsection{Von Neumann analysis of the H-RR model}
\label{subsec:LSA_HRR}

\begin{figure}[H]
	\begin{center}
		\includegraphics[scale=0.75]{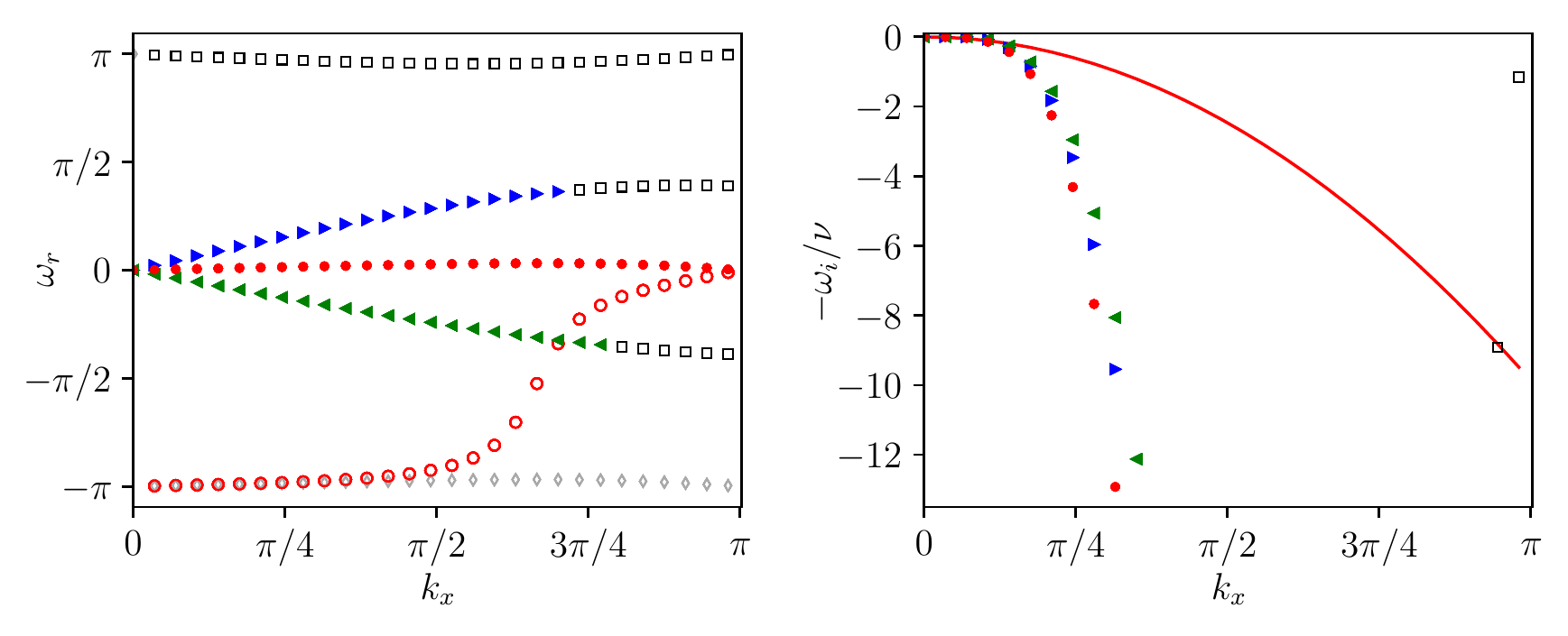}
		\caption{\label{fig:HRRlsa} Propagation and dissipation curves for the H-RR collision model with $\sigma=0.995$. $\nu = 10^{-6} $, Ma=0.1. $\left(\protect\acousP\right)$: Ac+ mode, $\left(\protect\acousN\right)$ : Ac- mode, $\left(\protect\shearP\right)$ : Shear mode, $\left(\protect\numS\right)$ : SpuriousS modes, $\left(\protect\numB\right)$ :  SpuriousAc modes, $\left(\protect\numG\right)$ : SpuriousG modes, (\protect\redC) : Isothermal Navier-Stokes.}
	\end{center}
\end{figure}

The spectral analysis of the H-RR collision model is presented on Fig.~\ref{fig:HRRlsa} with the parameter $\sigma$ set to $\sigma=0.995$. This implies that the viscous stress computation is computed at 99.5\% with standard approaches~(\ref{eq:a1Proj}) and with 0.5\% with a finite difference estimation of the stress tensor~(\ref{eq:stressT}).

\noindent The finite difference reconstruction of the viscous stress tensor is the key element to vanish the non-hydrodynamic contibutions as shown in Sec.~\ref{sec:improvement}. At the same time, it also increases the dissipation of physical modes as shown in Fig.~\ref{fig:HRRlsa}. For these reasons, a hybridation is mandatory to benefit of both advantages of the two methods.

\noindent The value of 0.995 is retained for all the numerical experiments of this article since it induces a strong dissipation of non-hydrodynamic modes while keeping a satisfactory dissipation of the physical ones. 

\begin{figure}[H]
	\begin{center}
		\includegraphics[scale=0.6]{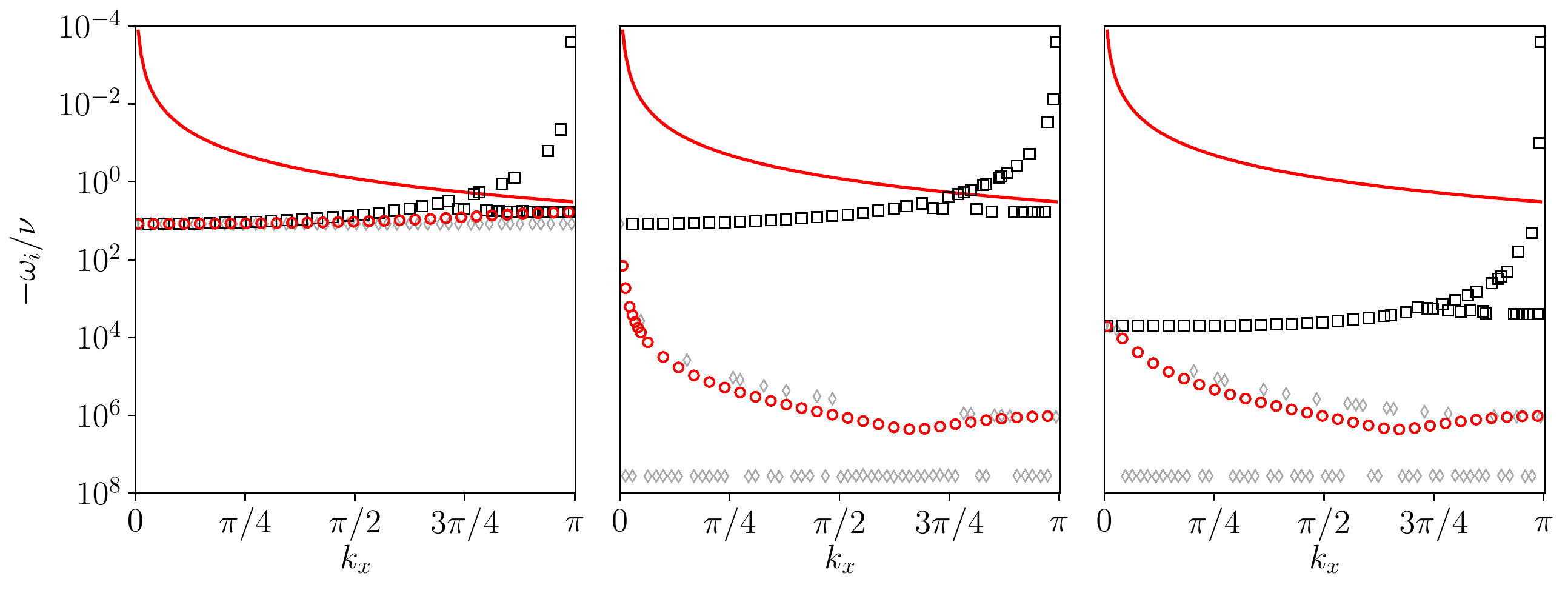}
		\caption{\label{fig:compDissipNH} Comparison of the non-hydrodynamic modes dissipation for the three collision models. Left: BGK, Middle: RR, Right: H-RR ($\sigma=0.995$). $\nu = 10^{-6} $, Ma=0.1. $\left(\protect\numS\right)$ : SpuriousS modes, $\left(\protect\numB\right)$ :  SpuriousAc modes, $\left(\protect\numG\right)$ : SpuriousG modes, (\protect\redC)~:~Isothermal Navier-Stokes.}
	\end{center}
\end{figure}

One can observe that dispersion properties are rigorously identical to that of the RR model. On the other hand, discrepancies appear on the dissipation curves. A better overview is available on Fig.~\ref{fig:compDissipNH} where the dissipation rate $-\omega_i/\nu$ is  displayed on a logarithmic scale, and compared to the other collision models. At first, the remained spuriousS mode (\numS) is much more attenuated than for the RR model for low wavenumbers. The spuriousG modes (\numG) have equivalent dissipation properties, but the main differences appear on the spuriousAc modes (\numB). The dissipation of this mode is rigorously equivalent for the BGK or RR models, but for the H-RR model it is a lot more dissipated.\\

For the H-RR collision model, an estimation of the spuriousAc mode (\numB) dissipation based on the spectral analysis can be proposed:

\begin{equation} \label{eq:tauSpuriousAc}
	\nu^{\left(\numB \right)} \simeq \nu + \frac{1-\sigma}{4}c_s^2.
\end{equation}

This estimation seems valid for a dimensionless viscosity $\nu<0.1$ and is qualitatively assessed on Fig.~\ref{fig:compDissipNH} where, among other things, the dissipation of the spuriousAc (\numB) mode is compared for the RR and H-RR collision models. For $\nu=10^{-6}$ and $\sigma=0.995$, the spuriousAc mode is damped at a rate $\nu^{\left(\numB \right)} \simeq 418 \nu$.\\

Since the spuriousAc mode (\numB) carries macroscopic information linked with acoustics, it can also be attenuated by increasing the bulk viscosity. This can be easily done by using a multiple relaxation time collision operator~\cite{Lallemand2000a,DHumieres2002a} for example. However, any modification of the bulk viscosity will also modify the damping of physical acoustic waves. In the present case, the bulk viscosity should be increased by several orders of magnitude to reach the same level of dissipation of the SpuriousAc mode (\numB) than the one obtained with the H-RR model (Fig.~\ref{fig:compDissipNH}).

\noindent A summary of the spectral properties of the aforementioned collision models is given in Table.~\ref{tab:SummaryCM}

\begin{table}[H]
	\centering
	\begin{tabu}{ c|[1.2pt]c|[1.2pt]c|[1.2pt]c|[1.2pt]}
		 & \textbf{SpuriousAc (\protect\numB)} & \textbf{SpuriousS (\protect\numS)} & \textbf{SpuriousG (\protect\numG)}\\\tabucline[1.2pt]{-}
		\textbf{BGK} &  $\boldsymbol{\times}$ & $\boldsymbol{\times}$ & $\boldsymbol{\times}$ \\
		\textbf{RR} &  $\boldsymbol{\times}$ & $\boldsymbol{\checkmark}$ for high $k_x$ & $\boldsymbol{\checkmark}$ \\
		\textbf{H-RR} &  $\boldsymbol{\checkmark}$ & $\boldsymbol{\checkmark}$ & $\boldsymbol{\checkmark}$ \\
	\end{tabu}
	\caption{\label{tab:SummaryCM}Summary of the dissipation of non-hydrodynamic modes for the three collision models. ($\boldsymbol{\checkmark}$) high dissipation, ($\boldsymbol{\times}$) low dissipation.}
\end{table}

This table justifies the choice of these collision models: the dissipation rates of the spurious modes are very different for the three models. A subsequent comparison of their behavior across a mesh refinement interface will be used to highlight the role of each spurious mode. Before performing such analysis, it is therefore interesting to quantify the effect of a change of resolution onto these modes, independently of the grid refinement algorithm used. This study is proposed in the next section.

\section{Energy transfer induced by a change of resolution}
\label{sec:modal_interactions}

The aim of this section is to study the effect of a resolution change on the LBM modes, regardless the grid coupling algorithm. Since the spectral properties of the LBM schemes strongly depend on the dimensionless wavenumber vector $\boldsymbol{k}$, then on the mesh resolution, it is interesting to wonder how a given mode may be affected by a resolution change. To address this question, it is proposed here to study the passage matrix $\mathbf{P}$ between modes with a wavenumber $k_x^f$ and those with a wavenumber $k_x^c=2k_x^f$. \newline

Let us denote $\mathbf{P}_f$ (respectively $\mathbf{P}_c$) the passage matrix composed of the eigenvectors obtained by the von Neumann analysis at $k_x^f$ (resp. $k_x^c$) written in the basis of the distribution functions. A given vector $\mathbf{F}$, written in the basis of the distribution functions, can equivalently be represented either by a vector $\mathbf{V}^f$ in the basis of the eigenmodes at $k_x^f$, or $\mathbf{V}^c$ in the basis of the eigenmodes at $k_x^c$, where:
\begin{equation} \label{eq:matrix_Pf_Pc}
	\mathbf{V}^f = \mathbf{P_f^{-1}} \mathrm{\mathbf{F}}, \qquad \mathbf{V}^c = \mathbf{P_c^{-1}} \mathrm{\mathbf{F}}.
\end{equation}
Each component of $\mathbf{V}^f= \big[(\acousP)^f,(\acousN)^f,(\numB)^f,(\shearP)^f,(\numS)^f,(\numS)^f,(\numG)^f,(\numG)^f,(\numG)^f \big]^{T}$ (resp. $\mathbf{V}^c$) represents the decomposition of $\mathbf{F}$ in the LBM modes of the fine mesh (resp. the coarse mesh). For example, $\mathbf{V}^f= \big[1, 0, .., 0\big]^{T}$ denotes a pure downstream acoustic wave at $k_x^f$. The link between $\mathbf{V}^f$ and $\mathbf{V}^c$ is then straightforward as
\begin{equation} \label{eq:matrix_Pij}
	\mathbf{V}^f = \mathbf{P} \mathbf{V}^c \qquad \mathrm{with} \qquad \mathbf{P}=\mathbf{P_f^{-1}P_c}.
\end{equation}

%
%
%

Coefficient $P_{ij}$ of $\mathbf{P}$ provide the decomposition of the fine modes expressed in the coarse modes basis. For example, the linear decomposition of a fine acoustic Ac+\ mode $(\acousP)^f$ onto the coarse basis reads

\begin{equation} \label{eq:Proj_Acmode}
	(\acousP)^f = P_{11}(\acousP)^c + P_{12}(\acousN)^c + P_{13}(\numB)^c + P_{14}(\shearP)^c 
	 + P_{15}(\numS)^c + P_{16}(\numS)^c + P_{17}(\numG)^c + P_{18}(\numG)^c + P_{19}(\numG)^c.
\end{equation}
Each component of $\mathbf{P}$ is \textit{a priori} complex, whose argument is linked with the phase shift between the modes. Here, only their modulus $|P_{i,j}|$ will be of interest. Moreover, they will be normalized as 
\begin{equation} \label{eq:Pij_norm}
P^*_{ij}=|P_{ij}|^2/\sum_{k}|P_{ik}|^2,
\end{equation}
so that $\sum_{j} P_{i,j}^* = 1$.\newline

\begin{figure}[H]
 	\begin{center}
		\includegraphics[scale=0.55]{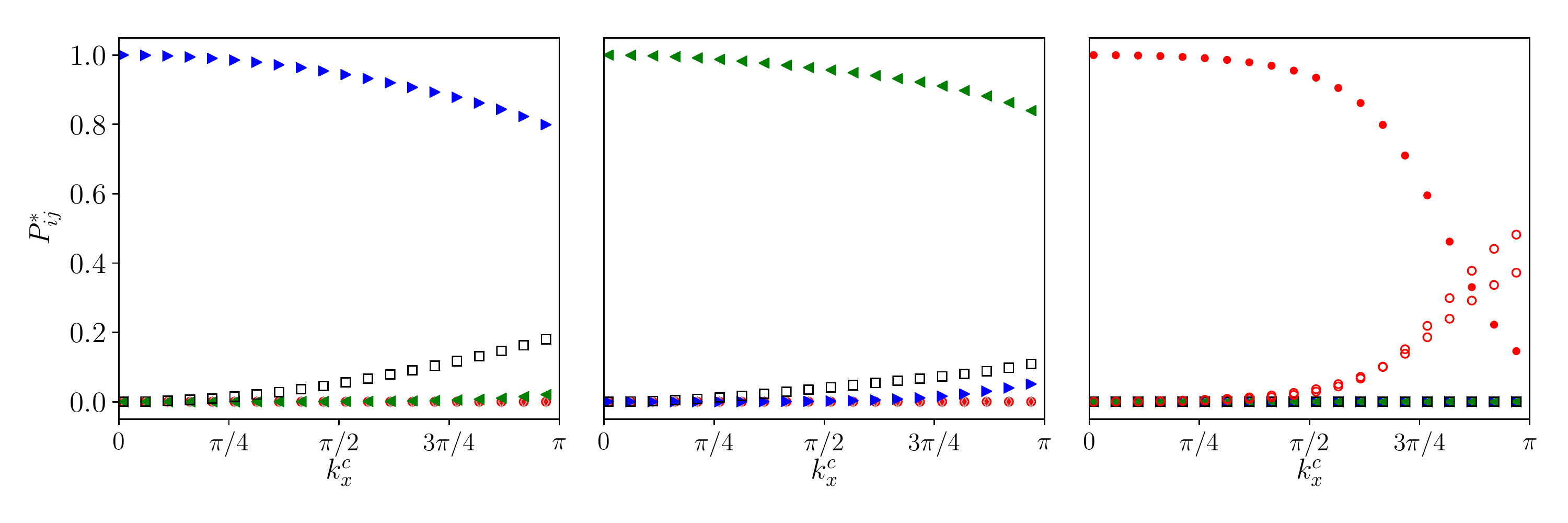}
		\caption{\label{fig:proj_phymode} Normalized coefficients $P_{i,j}^*$ for the three physical modes with the BGK collision operator with $\nu = 10^{-6}$ and Ma=0.1. Left: Components $P_{1j}^*$ of the decomposition of the fine Ac+ mode in the coarse eigenbasis, middle: $P_{2j}^*$ (\textbf{Ac-}), right: $P_{4j}^*$ \textbf{Shear}. $\left(\protect\acousP\right)$: Ac+ mode ($P_{i1}^*$), $\left(\protect\acousN\right)$: Ac- mode ($P_{i2}^*$), $\left(\protect\shearP\right)$: Shear mode ($P_{i4}^*$), $\left(\protect\numS\right)$: SpuriousS modes ($P_{i5}^*$ and $P_{i6}^*$), $\left(\protect\numB\right)$: SpuriousAc mode ($P_{i3}^*$), $\left(\protect\numG\right)$: SpuriousG modes ($P_{i7}^*$ to $P_{i9}^*$).}
	\end{center}
\end{figure}

Normalized coefficients $P_{ij}^*$ are displayed on Fig.~\ref{fig:proj_phymode} for the three physical modes (\acousP, \acousN, \shearP) obtained with the BGK collision operator. This analysis is performed for several $k_x^c \in [0,\pi]$ and  $k_x^f \in [\min(k_x^{c})/2,\pi/2]$.

It is noticeable that the physical mode of a fine mesh at wavenumber $k_x^f$ is not preserved when changing the mesh resolution to the wavenumber $k_x^c$. It is indeed decomposed into a superposition of coarse modes, hydrodynamic and non-hydrodynamic ones, that carry a quantity of the same nature. Moreover, this phenomenon is amplified as $k_x^c$ increases, \textit{i.e.} as the mode is less resolved. For instance, for the two physical acoustic modes (\acousP,\acousN), the decomposition onto the SpuriousAc mode (\numB) is favored, while shear and ghost modes are not excited. Concerning the shear mode (\shearP), its decomposition is distributed over both SpuriousS modes (\numS) to reach less than 20\% of the projection onto a coarse physical shear mode (\shearP) for high wavenumbers.

\begin{figure}[H]
 	\begin{center}
		\includegraphics[scale=0.55]{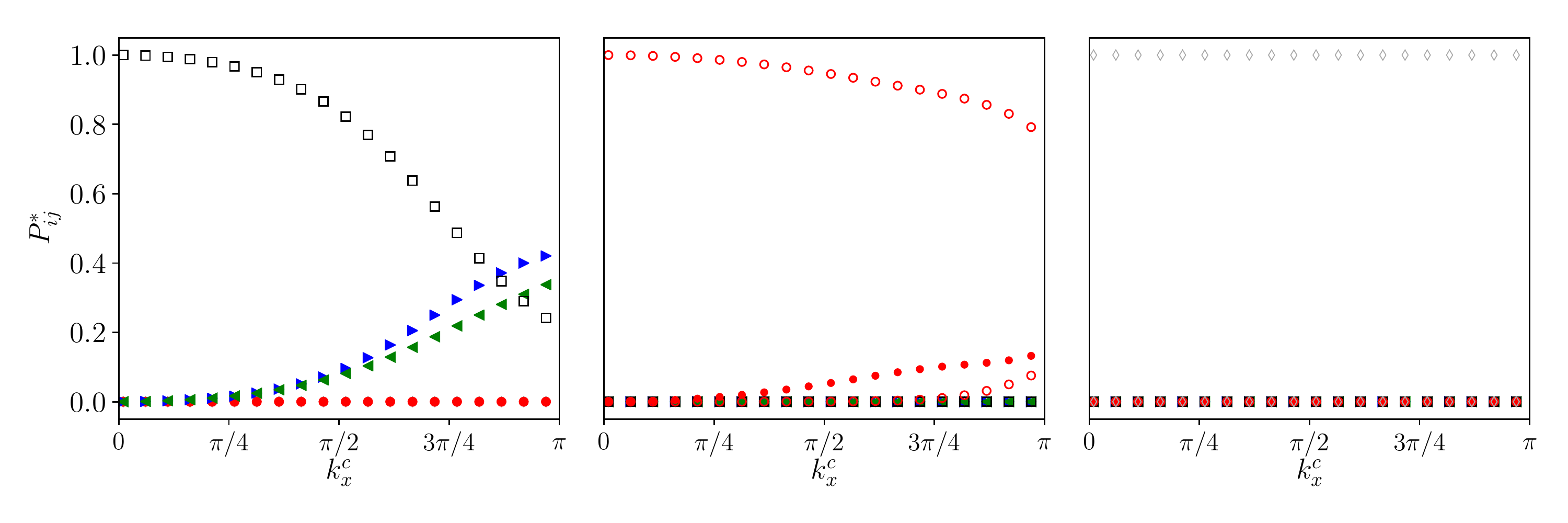}
		\caption{\label{fig:proj_NHmode}
		 Normalized coefficients $P_{i,j}^*$ for the the non-hydrodynamic modes with the BGK collision operator with $\nu = 10^{-6}$ and Ma=0.1. Left: Components $P_{3j}^*$ of the decomposition of the fine SpuriousAc mode in the coarse eigenbasis, middle: $P_{5j}^*$ \textbf{SpuriousS}, right: $P_{7j}^*$ \textbf{SpuriousG}. $\left(\protect\acousP\right)$: Ac+ mode ($P_{i1}^*$), $\left(\protect\acousN\right)$: Ac- mode ($P_{i2}^*$), $\left(\protect\shearP\right)$: Shear mode ($P_{i4}^*$), $\left(\protect\numS\right)$: SpuriousS modes ($P_{i5}^*$ and $P_{i6}^*$), $\left(\protect\numB\right)$: SpuriousAc mode ($P_{i3}^*$), $\left(\protect\numG\right)$: SpuriousG modes ($P_{i7}^*$ to $P_{i9}^*$).}
	\end{center}
\end{figure}

The same analysis is performed on Fig.~\ref{fig:proj_NHmode} for the non-hydrodynamic modes that are projected from a fine resolution ($k_x^f$) to a coarser one ($k_x^c$).

Here again, the non-hydrodynamic modes are composed only of a combination of modes that carry information of the same nature. The spuriousAc mode (\numB) generates acoustics in a significant way. Only one spuriousS mode (\numS) and one spuriousG mode (\numG) are plotted since a similar observation is provided for the other ones. The spuriousG modes are preserved by the change of resolution.

With these results, it is then possible to rewrite $\mathbf{P}$ as a block diagonal matrix, with acoustic modes (\acousP,\acousN,\numB) and shear modes (\shearP,\numS) that form two separated blocks and with the ghost modes (\numG) preserved by the transformation:

\begin{figure}[H]
	\begin{eqnarray}
	\nonumber
    	\qquad \qquad \qquad \quad
    	\underbrace{
    	\begin{pmatrix} \protect\acousP \\ \protect\acousN \\ \protect\numB \\ \protect\shearP \\ \protect\numS \\ \protect\numS \\ \protect\numG \\ \protect\numG \\ \protect\numG \end{pmatrix}^f
    	}_{\mathbf{V}^f}
		=
		\underbrace{
		\begin{bmatrix}P_{11} & P_{12} & P_{13} & 0 & 0 & 0 & 0 & 0 & 0\\
P_{21} & P_{22} & P_{23} & 0 & 0 & 0 & 0 & 0 & 0\\
P_{31} & P_{32} & P_{33} & 0 & 0 & 0 & 0 & 0 & 0\\
0 & 0 & 0 & P_{44} & P_{45} & P_{46} & 0 & 0 & 0\\
0 & 0 & 0 & P_{54} & P_{55} & P_{56}&  0 & 0 & 0\\
0 & 0 & 0 & P_{64} & P_{65} & P_{66} & 0 & 0 & 0\\
0 & 0 & 0 & 0 & 0 & 0 & 1 & 0 & 0\\
0 & 0 & 0 & 0 & 0 & 0 & 0 & 1 & 0\\
0 & 0 & 0 & 0 & 0 & 0 & 0 & 0 & 1
		\end{bmatrix}
		}_{\mathbf{P}}
		\cdot
		\underbrace{
 		\begin{pmatrix} \protect\acousP \\ \protect\acousN \\ \protect\numB \\ \protect\shearP \\ \protect\numS \\ \protect\numS \\ \protect\numG \\ \protect\numG \\ \protect\numG \end{pmatrix}^c
 		}_{\mathbf{V}^c}.
	\end{eqnarray}
	\caption{\label{fig:proj_matrix}Block diagonal projection matrix $\mathbf{P}$ of fine modes $\mathbf{V}^f$ with resolution $k_x^f$ on a coarser resolution $k_x^c=2k_x^f$ to obtain $\mathbf{V}^c$. $P_{1-3}$: bloc of acoustic modes (\protect\acousP, \protect\acousN, \protect\numB), $P_{4-6}$: bloc of shear modes (\protect\shearP, \protect\numS), $P_{7-9}$: ghost modes (\protect\numG).}
\end{figure}

This short analysis highlights the transfers occurring between both hydrodynamic and non-hydrodynamic modes that carry information of the same nature. This consideration is extremely important and will be further discussed in the numerical experiments of Sec.~\ref{sec:NHeffect}. Moreover, it is important to notice that this study has been carried out through von Neumann analyses performed in the fluid core only, regardless of the grid refinement algorithm. These results are therefore generic, and represent the ideal case of mesh transitions that do not introduce any more numerical errors. Obviously, it is never the case in practice, where the algorithm may affect the mode redistribution, such as by introducing non-linear effects and high-frequency waves that cannot be predicted by this analysis. In any case, the aforementioned mode transfer seems unavoidable given the spectral properties of the BGK model. In this context, since the $\mathbf{P}$ matrix is dependent of the collision operator, changing the latter can thus be used to act on the $P_{ij}$ coefficients. This observation will be the key point in the solution proposed in Sec.~\ref{sec:improvement} to improve the behavior of the mesh transition.

Up to now, spectral analysis tools have been used to emphasize some non-hydrodynamic modes with given properties depending on the collision model. In addition, a projection of physical modes onto non-hydrodynamic ones of the same nature and \textit{vice-versa} is very likely to happen at grid interface as described just above. However, the spectral analysis tools are not systematically applicable in a real simulation, for which the linear hypothesis with plane monochromatic perturbations may not be valid. For this reason, in the next section, a derivation of sensors is proposed to make a systematic link between spectral analysis tool outcomes, and non-hydrodynamic modes observation during a real simulation.


\section{Introduction of sensors to locate non-hydrodynamic modes}
\label{sec:NHsensor}

This section aims at proposing different kinds of sensors in order to detect the presence of non-physical modes in a simulation. The objective is to make the link between modes exhibited by the von Neumann analysis and phenomena observed during simulations.\\

Currently in the literature, the entropic lattice Boltzmann models are based on a similar attempt to systematically identify non-hydrodynamic content~\cite{Karlin2014}. More precisely, it is proposed to decompose the populations $f_i$ into three parts

\begin{equation} \label{eq:fEntropic}
	f_i = k_i + s_i + h_i,
\end{equation}
 
\noindent where $k_i$, $s_i$ and $h_i$ respectively refer to a kinematic part, a shear part and the remaining higher-order moments. The kinematic part $k_i$ relies only on conserved variables ($\rho,\mathbf{u}$). The shear part $s_i$ includes second-order moments of $f_i$, and $h_i$ includes higher-order moments only. Based on this decomposition, Karlin \textit{et al.}~\cite{Karlin2014} proposed to dynamically modify the relaxation time of $h_i$, especially when interactions appears with the shear moments $s_i$ in order to damp them. This dynamic model is performed using an entropic sensor (referred to as entropic stabilizer), that highlights interactions between the off-equilibrium part of both shear and higher-order moments. 

\noindent Usually, only the deviatoric stress tensor contribution is included in $s_i$ and the trace of the second-order moment can frequently be found in the $h_i$ part and, thus, its relaxation rate can be a free parameter. The bulk viscosity is then modified allowing an enhanced behavior in presence of non-uniform grids as it will be shown Sec.~\ref{sec:improvement}.\\

However, this entropic sensor allows for detecting only non-hydrodynamic moments of $f_i$. Yet, the present study focuses on the effects of non-hydrodynamic modes, rather than moments. Before introducing the sensors that will be used below, it is therefore important to emphasize the differences between moments and modes of a LBM scheme:

\begin{itemize}
	\item Moments are macroscopic variables that can be: hydrodynamics and conserved variables ($\rho$,$\mathbf{u}$) during collisions, non-conserved but hydrodynamic variables (stress tensor components), or non-conserved non-hydrodynamic variables for moments of order greater than two. The latter are usually referred to as ``ghost variables"~\cite{Benzi1990}. For many multiple-relaxation-time collision models, the relaxation parameters of these ghost variables are set free regarding the physics~\cite{Lallemand2000a,DHumieres2002a}.
	\item The modes are built as the eigenvectors of the linear stability analysis applied to the LB scheme. As it has been shown by Wissocq \textit{et al.}~\cite{Wissocq2019}, they can be ``observable" if they carry kinetic variables ($\rho,\mathbf{u}$) or not. In other cases they are ``ghost" and are invisible at the macroscopic level. For the purposes of this study, ``observable" modes are classified depending on the quantity their carry, as well as on their velocity as proposed in Sec.~\ref{subsec:LSAlbm}.
\end{itemize} 

Knowing that the spuriousAc (\numB) and spuriousS (\numS) modes can have a projection on the macroscopic moments ($\rho$,$\mathbf{u}$), and since the spuriousG mode (\numG) is linked to non-observable variables, these non-hydrodynamic modes can thus be located during a simulation by building sensors.\\

\textbf{SpuriousAc mode (\protect\numB) sensor:}\\
The first sensor introduced here aims at detecting the spuriousAc modes (\numB). These modes carry acoustic disturbances, i.e. compressive waves, thus they can be visualized thanks to the velocity divergence. Furthermore, these modes have a real pulsation $\omega_r$ very close to $\omega_r=\pi$ so that their amplitude is inverted at each time step. By using these two properties, one can build a sensor based on the velocity divergence product between two iterations which has to be negative.

\begin{equation} \label{eq:sensorSpuriousAc}
	\vec{\nabla} \cdot \boldsymbol{u} (t-1) * \vec{\nabla} \cdot \boldsymbol{u} (t)
		\quad
	    \begin{cases}
    	<0 \ \Rightarrow \mathrm{NHsensor}^{\left(\protect\numB\right)} = 1,\\
    	\ge 0 \ \Rightarrow \mathrm{NHsensor}^{\left(\protect\numB\right)} = 0.
	    \end{cases}
\end{equation}

\textbf{SpuriousS mode (\protect\numS) sensor:}\\
The second sensor aims to detect the spuriousS modes (\numS). These modes carry shear quantity, therefore they can be detected by looking at the vorticity field. As previously, these modes have also a real pulsation $\omega_r$ very close to $\omega_r=\pi$ so that their amplitude is inverted at each iteration. Here, it is possible to build a sensor based on the vorticity product between two iterations, which has to be negative. However, the inversion does not occur in strongly sheared areas. Consequently, this sensor enables the detection of spuriousS mode (\numS) outside regions with strong hydrodynamic variations. 

\begin{equation} \label{eq:sensorSpuriousS}
	\vec{\nabla} \times \boldsymbol{u} (t-1) *  \vec{\nabla} \times \boldsymbol{u}(t)
		\quad
	    \begin{cases}
    	<0 \ \Rightarrow \mathrm{NHsensor}^{\left(\protect\numS\right)} = 1,\\
    	\ge 0 \ \Rightarrow \mathrm{NHsensor}^{\left(\protect\numS\right)} = 0.
	    \end{cases}
\end{equation}

\textbf{SpuriousG mode (\protect\numG) sensor:}\\
The third sensor aims to detect the spuriousG modes (\numG). As stated above, they cannot be detected by looking at the macroscopic quantities. Through the von Neumann analysis, it is possible to show that these ghost modes are linked with ``ghost variable"~\cite{Adhikari2008}. Therefore, similarly to the entropic LBM sensor, a decomposition of the off-equilibrium distribution functions is proposed in a shear part and a ghost part corresponding to higher-order contributions, as

\begin{equation} \label{eq:fneqSplit}
	f_i^{(1)} = f_i^{(1),S} + f_i^{(1),G},
\end{equation}
 
\noindent where $f_i^{(1),S}$ is computed by projection of the off-equilibrium populations onto the second-order Hermite polynomials~\cite{Latt2006}

\begin{equation} \label{eq:reguProceduref1S}
  f^{(1),S}_i = \frac{w_i}{2c_s^4} \herm_i^{(2)}:\textit{\textbf{a}}^{(2)}_{1}.
\end{equation}

Using this decomposition, one can detect a spuriousG mode (\numG) when the norm of $f^{(1),G}$ is not null. However, the two previous non-hydrodynamic modes can also be detected with this definition, so it is proposed to withdraw their contributions. The remaining part is, de facto, the spuriousG mode.

\begin{equation} \label{eq:sensorSpuriousG}
		\quad
	    \begin{cases}
	    \ \mathrm{if} \qquad \left\Vert f^{(1),G} \right\Vert > 0 \quad \mathrm{and} \quad \left(\mathrm{NHsensor}^{\left(\protect\numB \right)} = 0 \quad \mathrm{and} \quad \mathrm{NHsensor}^{\left(\protect\numS \right)} = 0\right) \quad \Rightarrow \mathrm{NHsensor}^{\left(\protect\numG\right)} = 1,\\
    	\ \mathrm{else} \ \quad \mathrm{NHsensor}^{\left(\protect\numG\right)} = 0,
	    \end{cases}
\end{equation}

\noindent with $\left\Vert f^{(1),G} \right\Vert=\sqrt{\sum \left( f_i^{(1),G} \right)^2}$ the norm of the vector $f^{(1),G}$. This sensor allows to detect the spuriousG mode (\numG) outside the influence of the spuriousAc (\numB) and the spuriousS (\numS) ones. Since the group velocity of this mode is larger than the other ones, one can expect that they will be well separated with each other, allowing an easier identification.\\

Sensors to detect non-hydrodynamic modes have been proposed in this section. They allow to make the link between the spectral properties of collision models described in Sec.~\ref{sec:LSA} and unexpected phenomena occurring in simulations. They will be very useful to understand interactions that can appear between physical and spurious modes induced by the drastic change of spectral properties at grid refinement interface.


\section{Non-hydrodynamic modes effects on grid refinement algorithms}
\label{sec:NHeffect}

The aim of this section is to highlight some issues of grid refinement algorithms on very simple cases. First of all, a convected shear wave will be introduced to characterize the effect of the spuriousS (\numS) modes. Then an upstream acoustic wave is studied to look at the influence of the spuriousAc (\numB) modes.


\subsection{Convected shear wave}
\label{subsec:shearWave}

The first test case introduced is a plane convected shear wave, which is one of the most simple cases allowing to have a look at a shear flow across a grid refinement interface.

\begin{figure}[H]
	\includegraphics[scale=0.8]{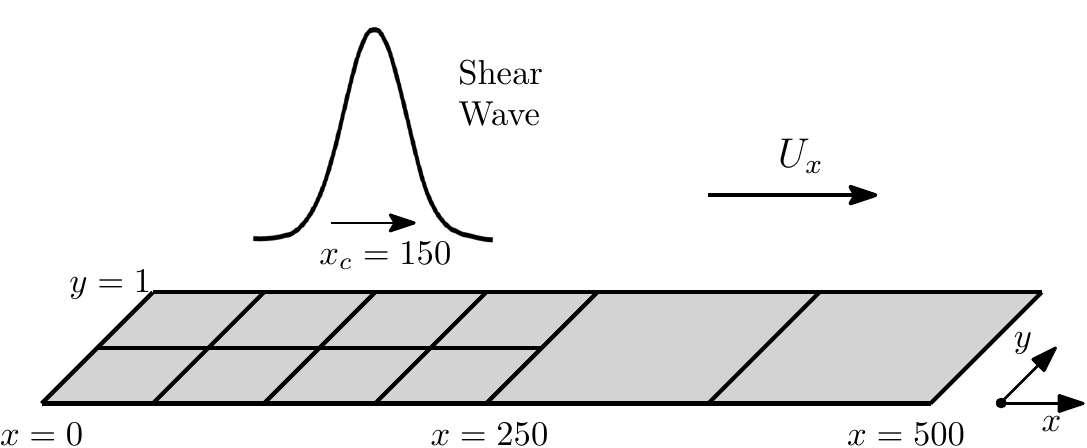}
\caption{\label{fig:Shear_wave_BGK} Schematic representation of the convected shear wave test case.}
\end{figure}

\noindent The convected plane shear wave is initialized on the fine grid as follow:
\begin{equation}
\label{eq:init_ShearW}
    \begin{cases}
		\quad \rho \left(x,y \right) = \rho_0, \\
	 	\quad u_x\left(x,y\right) = U_{x},  \\
	 	\quad u_y\left(x,y\right) = U_{y} \exp \left(- \frac{(x-x_c)^2}{2R_c^2} \right),     
	 	\end{cases}
\end{equation}

\noindent with
\begin{equation}
	\label{eq:init_ShearW2}
	\quad \rho_0=1,
   	\quad U_x = U_y = 0.1 c_s,
	\quad R_c = 5,
    \quad x_c = 150,
    \quad \nu = 10^{-6}.
\end{equation}

\noindent Every quantities are given in the coarse dimensionless unit.\\

The refinement interface is located at $x=250$ with the fine domain defined between $0 < x < 250$ and the coarse domain for $x>250$. The simulation domain is extended below $x<0$ and above $x>500$ to avoid any  reflection of waves on the domain boundary. Since the case is invariant along the $y$ axis, the domain is defined with a thickness of one coarse cell.

\subsubsection{Convected shear wave with the BGK collision model}
\label{subsubsec:shearBGK}

The first numerical experiment on the convected shear wave is performed using the BGK collision model. The results are presented on Fig.~\ref{fig:Shear_wave_BGK} where, for each plot, the two fine iterations and one corresponding coarse iteration are represented. This allows for the visualization of the amplitude inversion of non-hydrodynamic modes in the fine domain.

\begin{figure}[H]
	\includegraphics[scale=0.54]{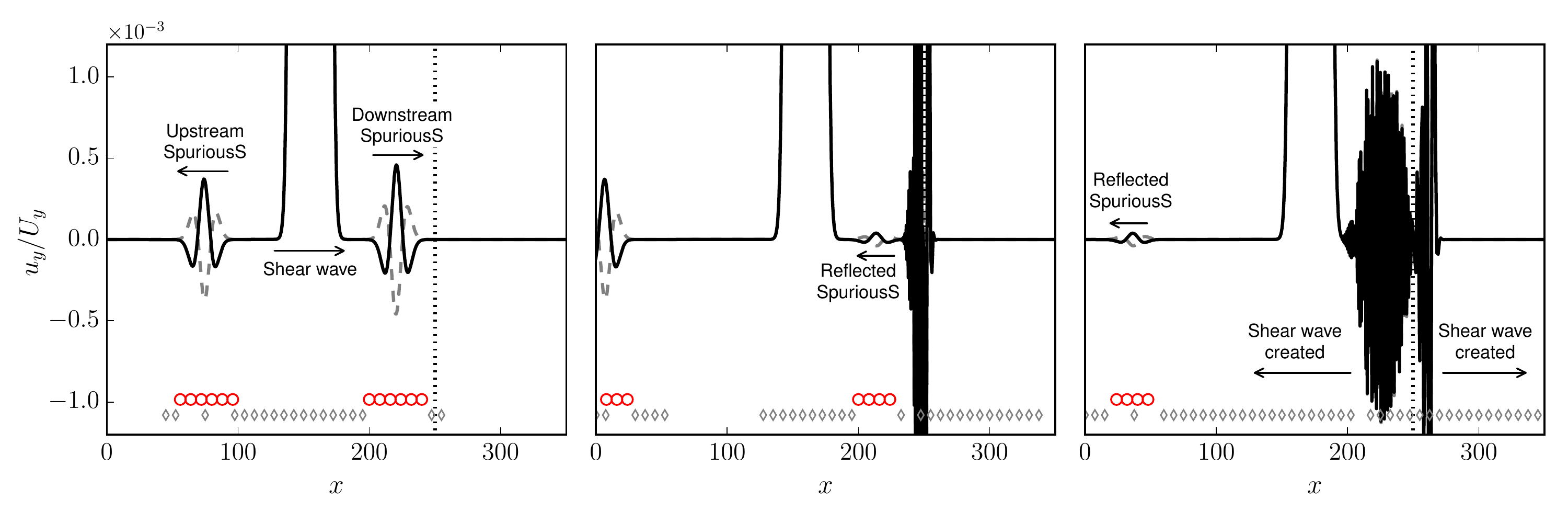}
\caption{\label{fig:Shear_wave_BGK} Convected shear wave with the BGK collision model. The sub-fine iteration (\protect\grayCdash) as well as the coarse iteration (\protect\blackC) are plotted to visualize the amplitude inversion of non-hydrodynamic modes. The (\protect\numS) symbols allow detecting the spuriousS modes and (\protect\numG) the spuriousG ones, by the $\mathrm{NHsensor^{\left(\protect\numS\right)}}$ and $\mathrm{NHsensor^{\left(\protect\numG\right)}}$ respectively.  Left: $t=90$, Middle: $t=170$, Right: $t=380$ coarse iterations.}
\end{figure}

As predicted by the von Neumann analysis (\textit{cf.} Fig.~\ref{fig:BGKlsa}), after the initialization state, two spuriousS (\numS) modes are excited, one going upstream and the other one downstream. The amplitude of these modes is reversed at each iteration, due to the high value of their pulsation $\omega_r$. Furthermore, the wavelength of these modes is about $2R_c$. 

At $t=170$, the spuriousS (\numS) mode travelling downstream does not properly cross the interface. A spuriousS mode detected by the corresponding sensor is reflected on the interface and is advected upstream. However, most of the energy is converted into physical waves going both upstream and downstream as shown at $t=380$. Indeed, for wavenumbers close to $k_x=\pi$, physical shear modes can be advected upstream due to a negative group velocity.\\

\begin{figure}[H]
 	\begin{center}
		\includegraphics[scale=0.53]{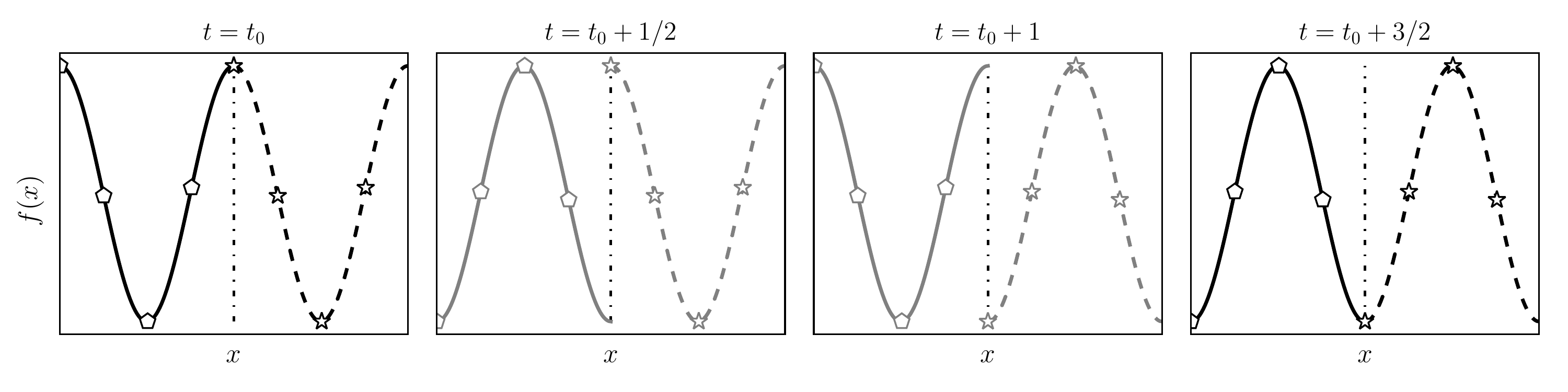}
		\caption{\label{fig:discontinuity} Schematic representation of a spurious mode at a grid refinement interface during two consecutive coarse iterations. Due to asynchronous time evolution between the two meshes, modes could be in phase or in phase opposition. Phased modes (\protect\blackPolyg):~fine mesh, (\protect\blackStar):~coarse mesh. Phase opposed modes 	(\protect\grayPolyg):~fine mesh. (\protect\grayStar):~coarse mesh.}
	\end{center}
\end{figure}

The shear waves created by the spuriousS modes are amplified at the interface. An explanation for this amplification may be found on Fig.~\ref{fig:discontinuity} where two coarse iterations are schematically decomposed. Indeed, the spuriousS modes are reversed at each iteration, and, due to the acoustic scaling of the LBM, fine cells are updated twice as frequently as coarse ones. As a matter of fact, the spuriousS modes are often in phase opposition from one side to the other of the interface. This leads to a huge error during the temporal interpolation for one dimensional cases, and for both space and time interpolations for two dimensional phenomena. Furthermore, significant errors appear also on $\left(\protect\TN\right)$ nodes during the fine to coarse reconstruction.

One can also notice that at $t=90$, the spuriousG (\numG) modes going upstream cross the interface without having any impact at the macroscopic level. It is confirmed at $t=170$ where no variation of transversal velocity is observed downstream the interface (and no variation of density, which remained null every time for this test case). Still, the grid refinement has created upstream and downstream spuriousG modes, each detected by the corresponding sensor.\\

The effect of the grid refinement on the convected shear wave is, with the BGK collision model, to redistribute the energy on all shear modes. This behavior is in agreement with the Sec.~\ref{sec:modal_interactions} study but some non-linear effects and amplifications are also observed. High frequency physical shear waves are created with a significant amplitude at the interface. Hence, with this collision model, a proper advection of a shear wave through a mesh transition is unlikely to be possible, given the redistribution of the energy between physical and non-hydrodynamic modes.

\subsubsection{Convected shear wave with the RR collision model}
\label{subsubsec:shearRR}
The same test case is performed using the recursive regularized collision model introduced in Sec.~\ref{subsec:RR}. For the sake of clarity, the spuriousG modes (\numG) are not studied in the following since they do not have any effect at the macroscopic level.

\begin{figure}[H]
	\includegraphics[scale=0.54]{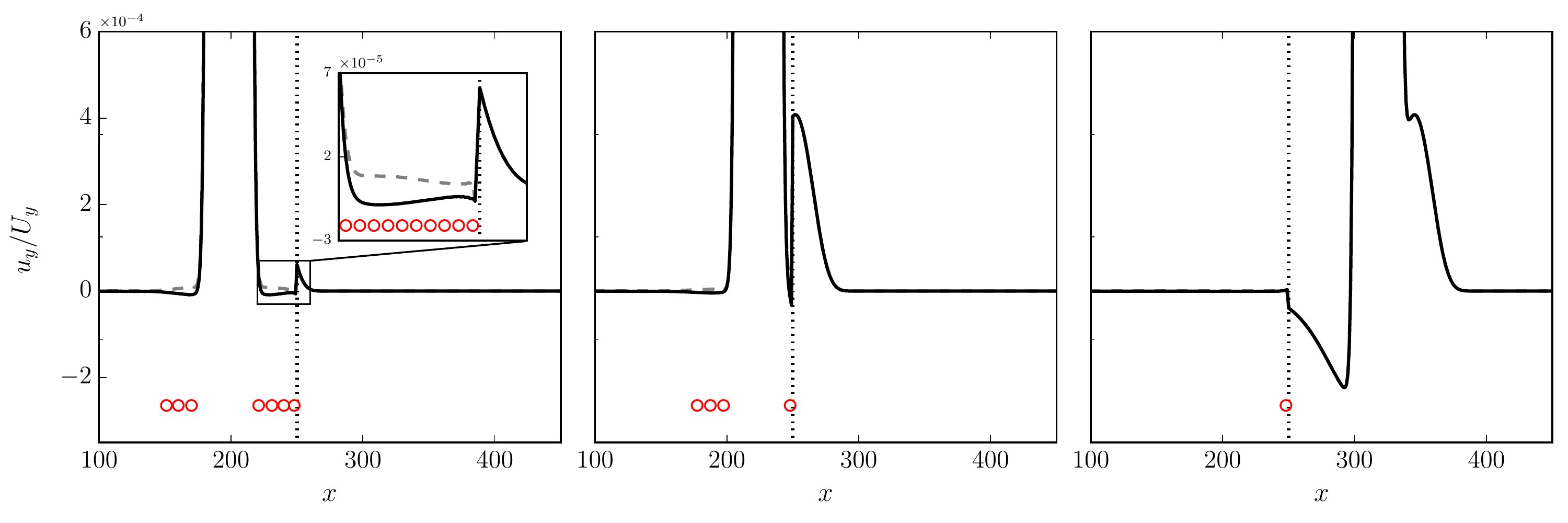}
\caption{\label{fig:Shear_wave_RR} Convected shear wave with the RR collision model. The sub-fine iteration (\protect\grayCdash) as well as the coarse iteration (\protect\blackC) are plotted to visualize the amplitude inversion of non-hydrodynamic modes. The (\protect\numS) symbols allow detecting the spuriousS modes by the $\mathrm{NHsensor^{\left(\protect\numS\right)}}$. Left: $t=840$, Middle: $t=1280$, Right: $t=2900$ coarse iterations.}
\end{figure}

The von Neumann analysis predictions are, once again, verified on Fig.~\ref{fig:Shear_wave_RR}. After the initialization state, the spuriousS (\numS) mode going upstream is excited, especially for low wavenumbers where it is not strongly attenuated (\textit{cf.} Fig.~\ref{fig:compDissipNH}). Its amplitude is reversed at each iteration, and it is detected before and after the physical shear wave due to the large range of its group velocity. Indeed, since its group velocity $v_g$ is always positive, $v_g$ is lower than that of the physical shear mode (\shearP) for very low wavenumbers, and higher afterwards.

The spuriousS (\numS) mode reaches the refinement interface at $t\simeq 840$ coarse iterations, and a unphysical peak appears on the transversal velocity $u_y$. The peak grows up to $t=1280$. Since no mode that carry shear exists with a negative group velocity using the RR collision model, no shear wave is reflected upstream. The spuriousS (\numS) mode fails to go over the interface. This last is fully converted into a physical shear mode (\shearP), and after crossing the interface, the convected shear wave is distorted with a positive and a negative transversal velocity component.\\
The behavior is quite different than for the BGK collision model. No more high frequency waves are generated at the interface. A possible reason might be that these waves are very attenuated by the RR collision model.\\

With the RR collision model for a fine to coarse grid crossing, the energy is fully converted from spuriousS modes (\numS) to physical shear ones (\shearP). This collision model allows for the shear wave to cross a grid refinement interface with a significant deformation. Note that no spurious noise is observed in this case.\\

The effects of shear flows over a grid interface have been studied with a convected shear wave. To complete the study, the effect of the acoustics is investigated in the following.

\subsection{Convected acoustic wave}
\label{subsec:acousPulse}

The second test case introduced is a convected one dimensional acoustic wave. This test case has the advantage of exciting only the modes carrying the acoustics, \textit{i.e.} the physical acoustics Ac+ (\acousP) and Ac- (\acousN) modes and the spuriousAc (\numB) ones.

Similarly, with the previous shear Gaussian excitation, one expects to generate a well-resolved spuriousAc mode. According to the spectral analysis of Fig.\ref{fig:BGKlsa}, this mode does not have any positive group velocity for low wavenumbers, and the resolved spuriousAc mode may be advected upstream. Hence, to investigate the effect of this mode across a transition, it is chosen to initialize the wave on the right of the transition. A upstream acoustic wave is then initialized as follows:

\begin{figure}[H]
	\includegraphics[scale=0.8]{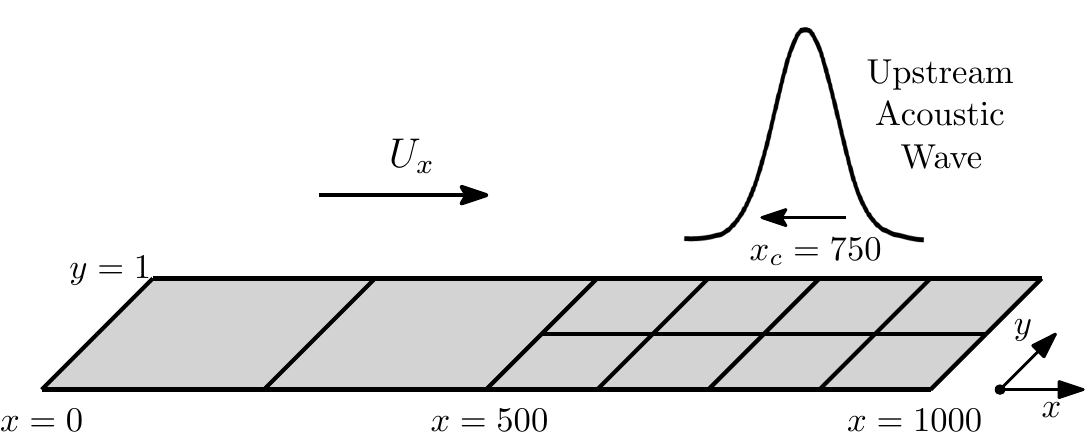}
	\caption{\label{fig:Acous_wave_case} Schematic representation of the convected acoustic wave test case.}
\end{figure}

\begin{equation}
\label{eq:init_Pulse}
    \begin{cases}
		\quad \rho \left(x\right) = \rho_0\left( 1 + A \exp  \left(- \frac{(x-x_c)^2}{2R_c^2}\right)\right), \\
	 	\quad u_x\left(x\right) = U_x - \frac{\rho'}{\rho_0}c_s,
    \end{cases}
\end{equation}
with
\begin{equation}
	\label{eq:init_Pulse2}
	\begin{split}
		\quad \rho_0 = 1 \ ,
		\quad U_x = 0.1 \cdot c_s,
   		\quad A = 10^{-4},
		\quad R_c = 5,
    	\quad x_c = 750,  
	    \quad \nu = 1.10^{-6}.    	
    \end{split}
\end{equation}

The refinement interface is located at $x=500$ with the fine domain located between $500 < x < 1000$ and the coarse domain for $x<500$. The simulation domain is extended below $x<0$ and above $x>1000$ to avoid any wave reflection on the domain boundary. Since the case is invariant along the $y$ axis, the domain is defined with a thickness of one coarse cell.

\begin{figure}[H]
	\includegraphics[scale=0.576]{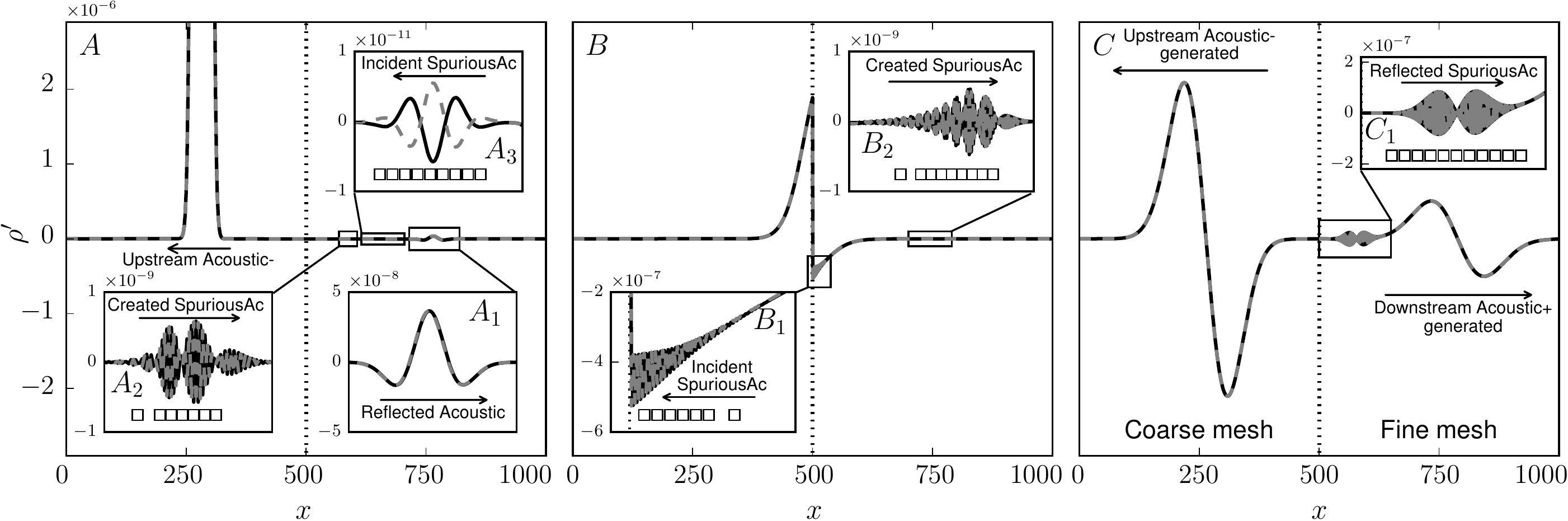}
	\caption{\label{fig:Acous_wave_BGK} Relative density evolution $\rho'=\rho-\rho_0$ of the upstream acoustic wave with the BGK collision model. The sub-fine iteration (\protect\grayCdash) as well as the coarse iteration (\protect\blackC) are plotted to visualize the amplitude inversion of non-hydrodynamic modes. The (\protect\numB) symbol allow detecting the spuriousAc modes by the $\mathrm{NHsensor^{\left(\protect\numB\right)}}$. Left: $t=900$, Middle: $t=2050$, Right: $t=2620$ coarse iterations.}
\end{figure}

The numerical experiment results are presented on Fig.~\ref{fig:Acous_wave_BGK} where, for each plot,  the coarse iterations are decomposed with the two fine corresponding sub-iterations. Many phenomena appear, and, for the sake of clarity, they are described one by one as numerated on the Fig.~\ref{fig:Acous_wave_BGK}.\\

Fig.~\ref{fig:Acous_wave_BGK}-$A$ presents the effect of the upstream acoustic wave that has been advected from $x=750$ to $x=270$, and has already passed through the grid interface. As expected ~\cite{Marie2008}, a reflected acoustic wave (Fig.~\ref{fig:Acous_wave_BGK}-$A_1$) appears. Its amplitude is 2000 times less than its corresponding incident acoustic wave. Furthermore, a high wavenumber ($k_x \simeq \pi$) spuriousAc mode (\numB) is generated at the interface and advected upstream (\textit{cf.} Fig.~\ref{fig:Acous_wave_BGK}-$A_2$). Moreover, as intended, spuriousAc modes (\numB) have been excited at the initialization with a negative group velocity. Their amplitudes are $10^7$ times smaller than that of the upstream acoustic wave (\textit{cf.} Fig.~\ref{fig:Acous_wave_BGK}-$A_3$). Both spuriousAc modes are succesfully detected by the $\mathrm{NHsensor^{\left(\protect\numB\right)}}$. 

Fig.~\ref{fig:Acous_wave_BGK}-$B$ shows the instant when the incident SpuriousAc mode (\numB), that was previously described on Fig.~\ref{fig:Acous_wave_BGK}-$A_3$, has impinged the interface (\textit{cf.} Fig.~\ref{fig:Acous_wave_BGK}-$B_1$). A huge amplification appears on the density field and a high frequency spurious wave is generated.

In the end, two acoustic waves are generated by the incident spuriousAc ones, as shown on Fig.~\ref{fig:Acous_wave_BGK}-$C$: one is going downstream on the fine mesh with an amplitude 100 times smaller than the initial acoustic wave, and the other one is advected upstream with an amplitude 25 times smaller. The incident SpuriousAc modes have been amplified with a factor up to $4.10^5$ to reach no less than $4\%$ of the initial acoustic wave amplitude.

This test case highlights the transfer of energy between physical acoustic modes (\acousP,\acousN) and the spuriousAc modes (\numB) that can appear when a change of mesh resolution occurs. It is in agreement with the analysis performed on Sec.~\ref{sec:modal_interactions} where these three modes have non-nul off-diagonal $P_{ij}$ coefficients. This kind of transfer has not been studied until here, as the classical benchmark for acoustic propagation across a grid interface is commonly performed at a null Mach number~\cite{Gendre2017}. In such case, the spuriousAc (\numB) mode are static and never cross the interface. This transfer is at the core of the spurious noise emission that can appear when vortices pass through a grid refinement interface~\cite{Gendre2017}.\\

Both the BGK and the RR collision models have rigorously the same spectral properties with regards to acoustics for a one-dimensional case (\textit{cf.} Fig.~\ref{fig:BGKlsa} \& Fig.~\ref{fig:RRlsa}). As a consequence, exactly the same emission appears with the RR collision model. Since the aim of this article is to perform proper aeroacoustic simulations, the BGK collision model will not be studied in the following, as it is less convincing for shear flows than the RR model.\\

Finally, it has been shown that transfers occur between modes that carry the same kind of quantity. Acoustic (\acousP,\acousN) modes can exchange energy with the spuriousAc (\numB) modes and the physical shear mode (\shearP) with the spuriousS (\numS) ones as anticipate in Sec.~\ref{sec:modal_interactions}.


\section{Improvement of fluid modeling for grid refinement algorithms}
\label{sec:improvement}

The previous section has highlighted the undesirable effects of non-hydrodynamic modes at grid refinement interfaces. Since these lasts can exchange energy with physical modes, and as their amplitude is inverted at each iteration, they are very difficult to handle properly with classical grid refinement algorithms.
In light of this, any lattice Boltzmann scheme that can effectively attenuate non-hydrodynamic modes seems to be a good candidate to avoid these exchanges. 

To this end, many ways exist. One can add low-pass filters~\cite{Ricot2009,Qian1997b}, or modify the collision model. The first solution allows  for dissipating both physical and non-physical modes at high-wavenumbers only. Still, it has been shown in Sec.~\ref{subsubsec:shearRR} that dissipating high-wavenumbers is not sufficient to avoid spurious phenomena occurring. To the authors' knowledge and following von Neumann analyses of many collision models, generally the second solution may allow for the enhancement of the dissipation of the spuriousS (\numS) modes. However, its does not allow to increase that of the spuriousAc (\numB) ones without increasing the bulk viscosity, which is not acceptable for aeroacoustic simulations will also affect the physical acoustic waves damping.

\begin{figure}[H]
	\begin{center}
		\includegraphics[scale=0.75]{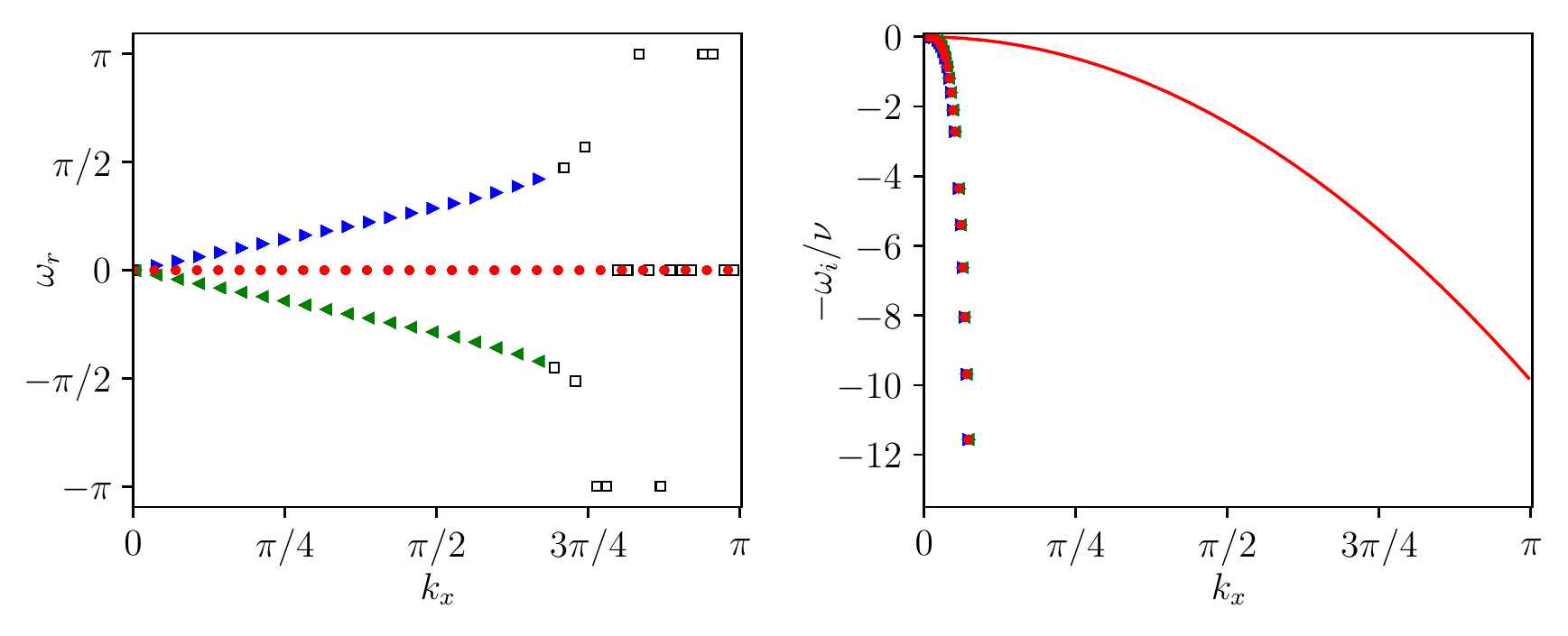}
		\caption{\label{fig:DFlsa} Propagation and dissipation curves for H-RR collision model with $\sigma=0$. $\nu=10^{-6}$, Ma=0. $\left(\protect\acousP\right)$: Ac+ mode, $\left(\protect\acousN\right)$ : Ac- mode, $\left(\protect\shearP\right)$ : Shear mode, $\left(\protect\numB\right)$ :  SpuriousAc modes, (\protect\redC) : Isothermal Navier-Stokes.}
	\end{center}
\end{figure}

An efficient way to dissipate both spuriousS (\numS) and spuriousAc (\numB) modes has been introduced in Sec.~\ref{subsec:LSA_HRR}: the reconstruction of the viscous stress tensor using finite differences. The von Neumann analysis of the H-RR collision model with a $\sigma=0$ parameter is presented on Fig.~\ref{fig:DFlsa}. This is equivalent to evaluating the viscous stress tensor with finite differences only.

The reconstruction of the viscous stress tensor has the main advantage making all the spurious modes fully vanish for $k_x < 3\pi /4$. Indeed, whatever the number of velocity of the lattice, the number of remaining modes corresponds to the number of physical modes, as would be the case when solving the isothermal Navier-Stokes equations. In two dimensions, three modes remain, while a fourth mode is observed in three dimensions, corresponding to an additional physical shear wave.

 The counterpart of this beneficial effect is the huge dissipation rate of physical modes introduced, which is unbearable for acoustics purpose.

\noindent That is why the hybridization of the viscous stress tensor computation allows the best possible compromise between the two models: the remarkable stability properties and the low dissipation of physical modes induced by the recursive regularized procedure, altogether with the massive dissipation of non-hydrodynamic modes resulting from the finite difference reconstruction.\\

In Sec.~\ref{subsec:validShear}, the H-RR collision model will be assessed for both  convected shear and acoustic waves, and a more complex case, a convected vortex with a value of $\sigma=0.995$ corresponding to the spectral properties shown on Fig.~\ref{fig:RRlsa}.

\subsection{Assessment of the H-RR collision model on a convected shear wave}
\label{subsec:validShear}

\begin{figure}[H]
	\includegraphics[scale=0.54]{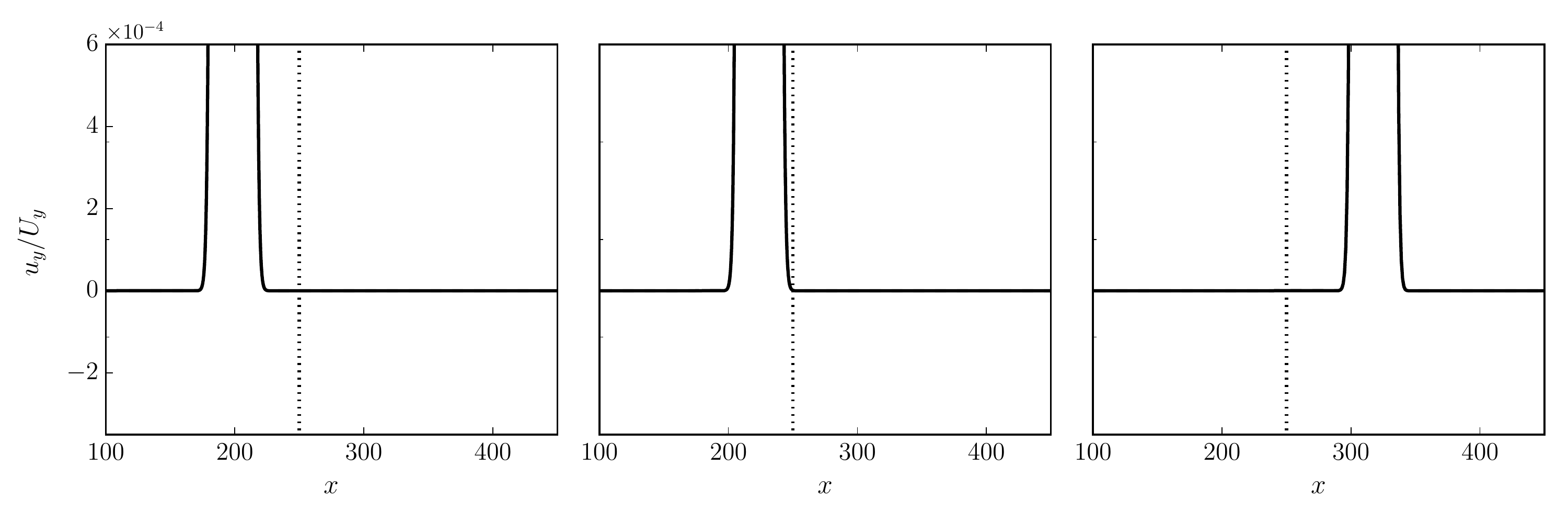}
\caption{\label{fig:Shear_wave_AMU} Convected shear wave with the H-RR collision model with $\sigma=0.995$. No spuriousS modes (\protect\numS) are detected by the $\mathrm{NHsensor^{\left(\protect\numS\right)}}$. Left: $t=840$, Middle: $t=1280$, Right: $t=2900$.}
\end{figure}

The aforementioned convected shear wave test case is now performed using the H-RR collision model. As displayed on Fig.~\ref{fig:Shear_wave_AMU}, at $t=840$, no spuriousS modes (\numS) are  detected by the $\mathrm{NHsensor^{\left(\protect\numS\right)}}$, since they have been totally damped by the H-RR model. As a consequence, the convected shear wave perfectly crosses the interface without being deformed. 

The H-RR collision model thus seems to be perfectly capable of dealing with shear flows in the presence of mesh refinement.

\subsection{Assessment of the H-RR collision model on a convected acoustic wave}

The convected acoustic wave is assessed with the H-RR collision model, and compared to the results yielded by the RR model, which yields results similar to those produced by the BGK model.

\begin{figure}[H]
	\includegraphics[scale=0.576]{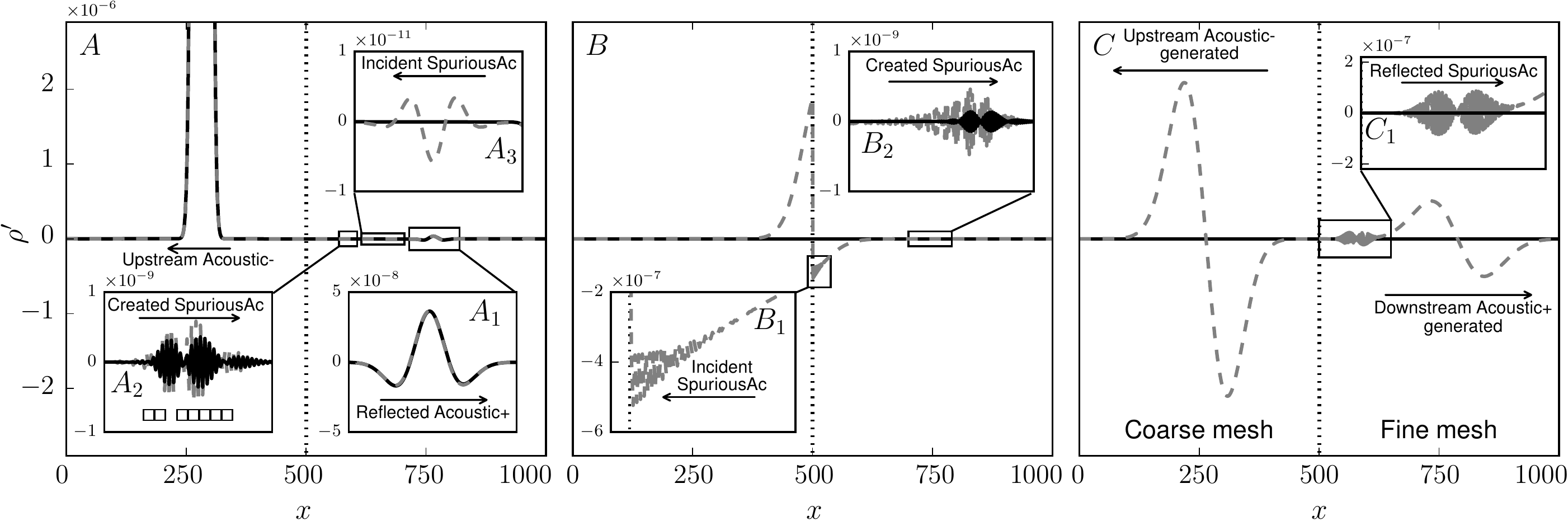}
	\caption{\label{fig:Acous_wave_HRR} Comparison of the RR (\protect\grayCdash) and H-RR (\protect\blackC) collision models with $\sigma=0.995$. Relative density evolution $\rho'=\rho-\rho_0$ for the upstream acoustic wave. The (\protect\numB) symbol allow detecting the spuriousAc modes by the $\mathrm{NHsensor^{\left(\protect\numB\right)}}$. Left: $t=900$, Middle: $t=2050$, Right: $t=2620$.}
\end{figure}

Fig.~\ref{fig:Acous_wave_HRR}-$A$ shows the effects of the upstream acoustic wave that has been advected from $x=750$ to $x=270$ and has already passed through the grid interface, in comparison with the results obtained with the RR collision model. The reflected acoustic wave (Fig.~\ref{fig:Acous_wave_HRR}-$A_1$) appears in the same manner as with the RR collision model. This result was expected, since this phenomenon is independent of the non-hydrodynamic contribution. Furthermore, high wavenumber ($k_x=\pi$) spuriousAc modes (\numB) are also generated at the interface and advected upstream (\textit{cf.} Fig.~\ref{fig:Acous_wave_HRR}-$A_2$). These modes are more strongly attenuated with the H-RR model. Moreover, the spuriousAc modes (\numB) generated at the initialization have been fully damped with this last model (\textit{cf.} Fig.~\ref{fig:Acous_wave_HRR}-$A_3$). This is confirmed by the $\mathrm{NHsensor^{\left(\protect\numB\right)}}$, which does not detect any spuriousAc mode.

Afterwards, the spuriousAc modes (\numB) that have been generated by the upstream acoustic wave is more and more damped with the H-RR collision model (\textit{cf.} Fig.~\ref{fig:Acous_wave_HRR}-$B_1$). These high frequency modes are not well attenuated by both the RR and H-RR collision model as shown on Fig.~\ref{fig:compDissipNH}. To enhance this dissipation, the use of low-pass filters might be required.

No more acoustic emission appears with the H-RR model, since the incident spuriousAc mode (\numB) has been fully damped.\\

The H-RR collision model seems perfectly able to deal with both shear flows and acoustics on non-uniform grids. This is seen in the subsequent more realistic case: a convected vortex.

\subsection{Assessment of the H-RR collision model on a convected Vortex}
\label{subsec:validCOVO}

The improvements induced by filtering out non-hydrodynamic modes using the H-RR collision model are now validated considering a convected vortex that crosses a grid refinement interface. This test case is of interest for most aeroacoustic applications. Indeed, this situation appears in many configurations like for the prediction of noise produced by turbulent jet noise~\cite{Brogi2017}, or landing gears~\cite{Sengissen2015}.

One should notice that, in order to avoid a transient adaptation which might generate spurious waves that would impact the mesh transition, the convected vortex cannot be initialized here by the common analytical expression of the well-known isentropic Lamb-Oseen vortex~\cite{Lamb1932,Oseen1927}. Indeed, these expressions have been derived from the isentropic Euler equations, whereas the notion of `isentropic' has no meaning in a standard athermal LBM solver. In order to avoid a spurious transient adaptation, the vortex is initialized with the more suited batrotropic vortex derived in~\cite{Wissocq2019a}, as follows

\begin{equation}
\label{eq:init_COVO}
    \begin{cases}
		\quad \rho \left(x,y \right) = \rho _0 \exp \left[ -\frac{\epsilon ^2}{2 c_s ^2} \exp\left( -\frac{(x-x_c)^2 + (y-y_c)^2}{R_c ^2} \right) \right], \\
		\quad u_x\left(x,y\right) = U_x - \epsilon \left( \frac{y-y_c}{R_c} \right) \exp \left( - \frac{(x-x_c)^2 + (y-y_c)^2}{2 R_c ^2} \right),  \\
	 	\quad u_y\left(x,y\right) = \epsilon \left( \frac{x-x_c}{R_c} \right) \exp \left( - \frac{(x-x_c)^2 + (y-y_c)^2}{2 R_c ^2} \right),  \\
    \end{cases}
\end{equation}

with

\begin{equation}
	\label{eq:init_COVO2}
	\quad \rho _0 = 1.,
	\quad U_x = 0.1 c_s,
	\quad \epsilon = 0.15U_x,
    \quad R_c = 5,
    \quad (x_c,y_c) = (100,75),  
    \quad \nu = 10^{-6}.
\end{equation}

The refinement interface is located at $x=150$ with the fine domain defined between $0 < x < 150$ and the coarse domain for $x>150$.

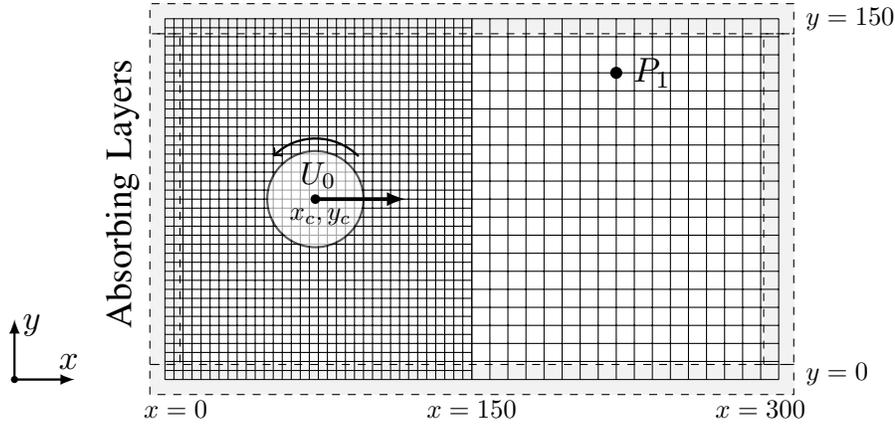
\begin{figure}[H]
	\begin{center}
		\begin{tikzpicture}[scale=0.8]
			\draw (0,0) grid[step=0.15] (5.1,6.001);
			\draw (5.1,0) grid[step=0.3] (10.2,6);
			\fill[gray,opacity=0.1] (-0.25,-0.25) rectangle (10.45,0.25) ;
			\draw [dashed] (-0.25,-0.25) rectangle (10.45,0.25) ;
			\fill[gray,opacity=0.1] (-0.25,5.75) rectangle (10.45,6.25) ;
			\draw [dashed] (-0.25,5.75) rectangle (10.45,6.25) ;
			\fill[gray,opacity=0.1] (-0.25,0.25) rectangle (0.25,5.75) ;
			\draw [dashed] (-0.25,0.25) rectangle (0.25,5.75) ;
			\fill[gray,opacity=0.1] (9.95,0.25) rectangle (10.45,5.75) ;
			\draw [dashed] (9.95,0.25) rectangle (10.45,5.75) ;
			\node[text width=3cm,rotate=90,scale=1.4] at(-0.7,3.5){Absorbing Layers};
			\draw  [thick,fill=white,opacity=0.65] (2.5,3) circle (0.8) ;	
			\draw [-> , >=latex ,line width=0.5mm] (2.5,3) -- (4,3);	
			\node[text width=1cm,scale=1.2] at(3.,3.4){$U_0$};
			\draw [-> , >=latex ,line width=0.3mm] (-2.5,0) -- (-1.5,0);	
			\node[text width=0.2cm,scale=1.3] at(-1.6,0.3){$x$};
			\draw [-> , >=latex ,line width=0.3mm] (-2.5,0) -- (-2.5,1.);
			\node[text width=0.2cm,scale=1.3] at(-2.2,0.9){$y$};
			\draw node[circle,draw,thick,fill=black, scale=0.2] at (-2.5,0.)  {};
			\draw node[circle,draw,thick,fill=black, scale=0.3] at (2.5,3.) {};
			\node[text width=1cm,scale=1.] at(2.7,2.7){$x_c,y_c$};
			\draw node[circle,draw,thick,fill=black, scale=0.4] at (7.5,5.1) {};
			\node[text width=1cm,scale=1.3] at(8.6,5.1){$P_1$};
			\node[text width=2cm,scale=1.] at(5.6,-0.5){$x=150$};
			\node[text width=2cm,scale=1.] at(0.9,-0.5){$x=0$};
			\node[text width=2cm,scale=1.] at(10.4,-0.5){$x=300$};
			\node[text width=2cm,scale=1.] at(11.9,0.1){$y=0$};
			\node[text width=2cm,scale=1.] at(11.9,6.){$y=150$};

			\draw [<-,line width=0.3mm,rotate=45](3.9,1.35) arc (90:0:1) ;
		\end{tikzpicture}
		\caption{\label{fig:COVO_mesh} Schematic representation of the simulation domain for the convected vortex. $\mathbf{P_1}$ probe is added for acoustic recording. Absorbing layers map the domain boundaries to avoid reflection of spurious acoustic emission.}
	\end{center}
\end{figure}

The convected vortex combines a perturbation on the density field $\rho$, and the two velocity components $u_x,u_y$. This leads to an excitation of all kinds of modes that we can find in a LBM scheme. Both modes that carry shear quantities (\shearP,\numS) and acoustics (\acousP,\acousN,\numB) are supposed to be excited. Furthermore, in order to avoid any reflection of spurious waves, Neumann boundary conditions and explicit absorbing layers are added at the domain boundaries in the same way as in \cite{Chevillotte2016}. A density probe is also inserted in the simulation domain far from hydrodynamic fluctuations at location ($x=200$, $y=125$) as shown on Fig.~\ref{fig:COVO_mesh}. An estimation of the spurious noise emitted is performed by computing the power spectral density (PSD) of the density fluctuations recorded on this probe.

Before discussing the results obtained with the H-RR collision model, the two spurious phenomena highlighted in Sec.~\ref{sec:NHeffect} are decomposed for the convected vortex using the RR model. Indeed, since the main property of the H-RR is to damp the spurious modes, it is first interesting to show their negative effect on this case.

\begin{figure}[H]
	\begin{center}
		\includegraphics[scale=0.56]{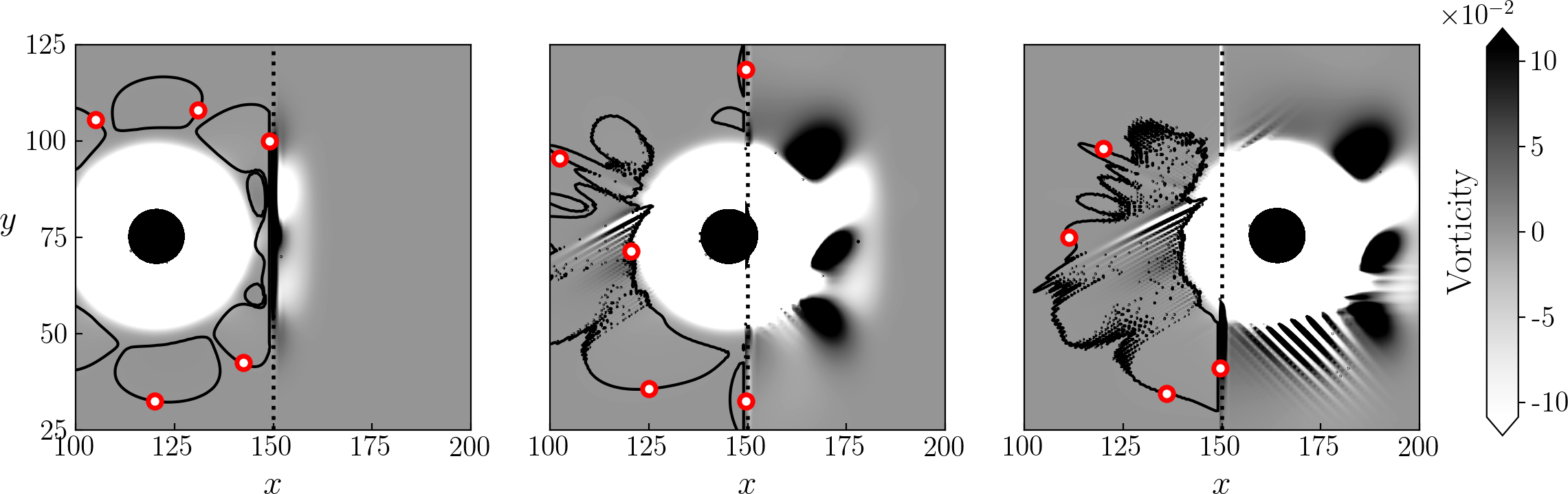}
		\caption{\label{fig:COVO_SpuriousS}  Vorticity field of the convected vortex that crosses a grid refinement using the RR collision model. Iso-contours denote the presence of spuriouS (\protect\numS) modes detected by the $\mathrm{NHsensor^{\left(\protect\numS\right)}}$.\ Left: $t=350$,\ Middle: $t=780$,\  Right: $t=1100$.}
	\end{center}
\end{figure}

The emphasis is first focused on the shear quantity. Fig.~\ref{fig:COVO_SpuriousS} shows the vorticity and the spuriousS modes (\numS) that are detected by the $\mathrm{NHsensor^{\left(\protect\numS\right)}}$ inside the isocontours. The sensor allows for the detection of the spuriousS modes only outside the vortex as explained in Sec.~\ref{sec:NHsensor}.

The vortex approaches the transition at $t=350$ iterations. As it was previously shown for the convected shear wave using this collision model, the spuriousS modes (\numS) are completely converted into physical ones for fine to coarse mesh crossing. Exactly the same behavior occurs here. At $t=780$, the $\mathrm{NHsensor^{\left(\protect\numS\right)}}$ detects a spreading of the spuriousS mode (\numS) along the grid interface, and a quantity of spurious vorticity is generated around the vortex with the same thickness as the previous spuriousS mode (\numS) location. In the end, a large deformation of the convected vortex appears after crossing the interface. Previous results obtained on the convected shear wave are then fully recovered on this more complex case.

\begin{figure}[H]
	\begin{center}
		\includegraphics[scale=0.56]{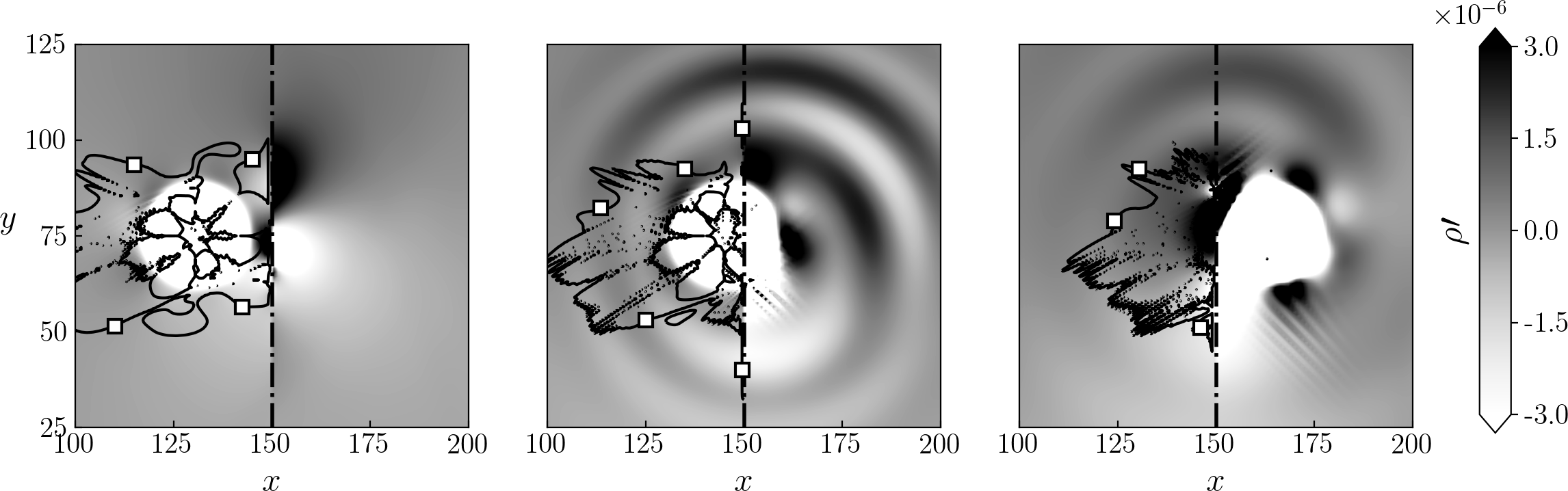}
		\caption{\label{fig:COVO_SpuriousAc} Density field for the convected vortex test case that crosses a grid refinement using the RR collision model. Iso-contours denote the presence of spuriouAc (\protect\numB) modes detected by the $\mathrm{NHsensor^{\left(\protect\numB\right)}}$.  Left: $t=530$,\ Middle: $t=780$,\   Right: $t=1100$.}
	\end{center}
\end{figure}

Next, the focus is put on acoustic modes (\acousP,\acousN,\numB) on Fig.~\ref{fig:COVO_SpuriousAc}. The spuriousAc modes (\numB) are detected inside the isocontours with the $\mathrm{NHsensor^{\left(\protect\numB\right)}}$. Note that, as predicted by the linear stability analysis, the only spuriousAc modes advected downstream are high frequency fluctuations in the range $k_x \in \left[\pi/2,\pi \right]$. Here, the sensor allows for the detection of the presence of these modes, both inside and outside the vortex. Firstly, when the vortex approaches the refinement interface ($t=530$), unwanted pressure spots arise on both sides of the interface. These spots appear far from the influence of hydrodynamic fluctuation areas. Their locations correspond to that of the spuriousAc modes (\numB) which are converted into physical acoustics. 

For a fine to coarse crossing, the modes which are resolved on the fine mesh in the range $k_x \in [\pi /2,\pi]$  cannot exist in the coarse one, due to the resolution change. Therefore, the spuriousAc modes (\numB) that do not have a positive group velocity $v_g$ in the range $k_x \in [0,\pi /2]$, cannot cross the interface, and are fully converted into physical acoustics.
 
At $t=780$, a huge spurious emission comes out due to the intensity of the spuriousAc (\numB) mode which is higher inside the vortex. At the end, spurious emissions appear on a large frequency range depending on the spuriousAc mode's wavenumbers.

\begin{figure}[H]
	\begin{center}
		\includegraphics[scale=0.56]{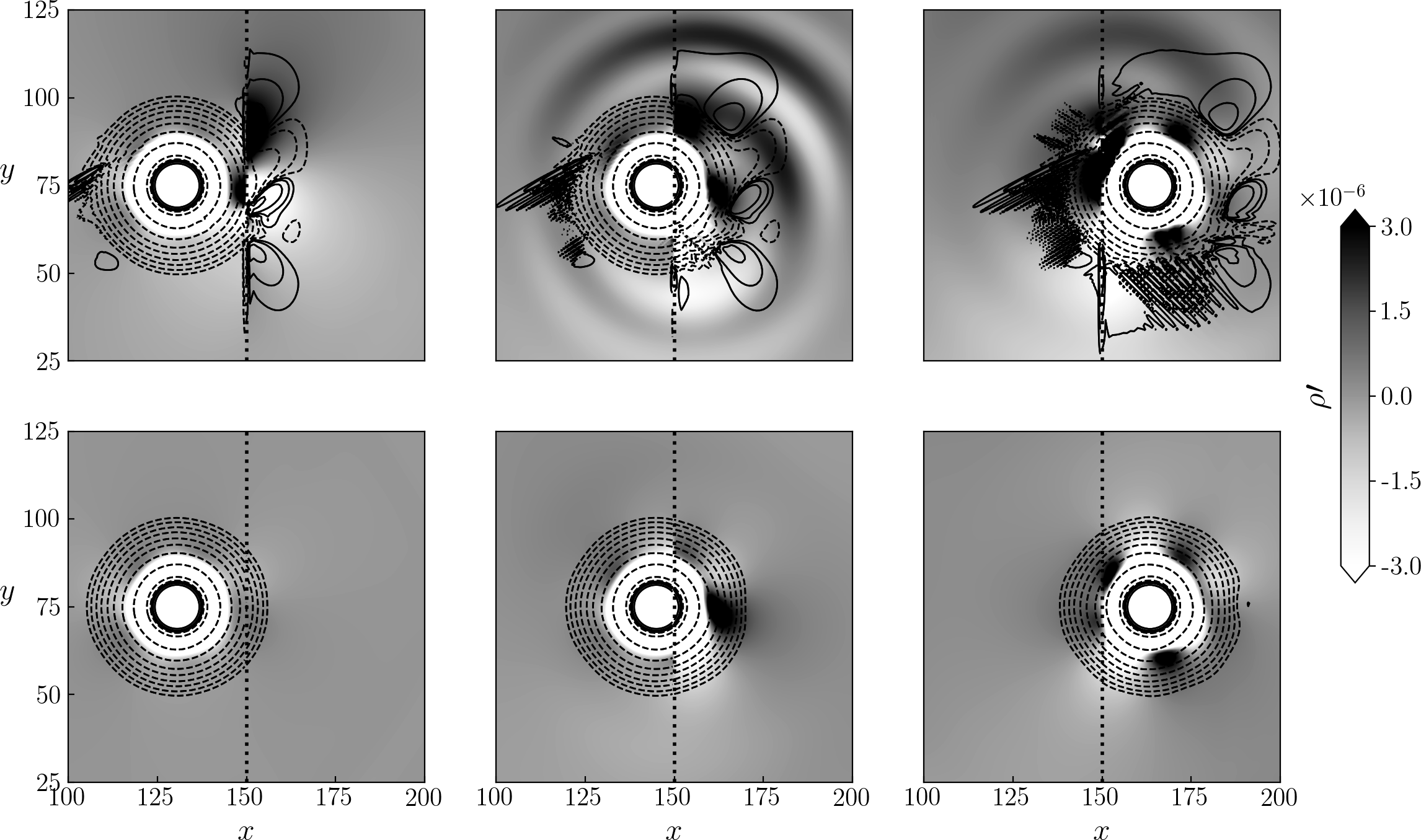}%
		\caption{\label{fig:COVO_AMU} Maps of relative density fields $\rho'=\rho-\rho_0$ with isocontours of vorticity for the vortex convected across a mesh interface, from a fine mesh to a coarse one. Top: RR collision model, bottom: H-RR ($\sigma=0.995$). Left: $t=530$,\ Middle: $t=780$,\   Right: $t=1100$.}
	\end{center}
\end{figure}

The RR and H-RR collision models are now compared on this test case on Fig.~\ref{fig:COVO_AMU}. This time, only physical quantities are displayed on the figure, since for the H-RR model, no more spurious modes are detected by the non-hydrodynamic mode sensors, which proves that they have been fully damped by the collision model. The two displayed quantities are the density, so as to observe the spurious acoustic emission, and the vorticity, to examine the vortex deformation. 

With the H-RR collision model, no more deformation or pressure spots appear on the grid refinement at $t=530$. After that, a tiny deformation is observed, and the spurious emission has been drastically reduced. At $t=530$, a small discontinuity in the pressure field can be noticed on both vortex sides. This deformation is now only due to the ability of the grid refinement algorithm to properly deal with physical waves, which is obviously of crucial importance. Gendre \textit{et al.}~\cite{Gendre2017} have actually described the lack of accuracy of the algorithm adopted here in such conditions. Without going into such considerations, the phenomena explained in the present study come from the change of mesh resolution, and are independent of the grid refinement algorithm.\\

\begin{figure}[H]
	\begin{center}
		\includegraphics[scale=0.6]{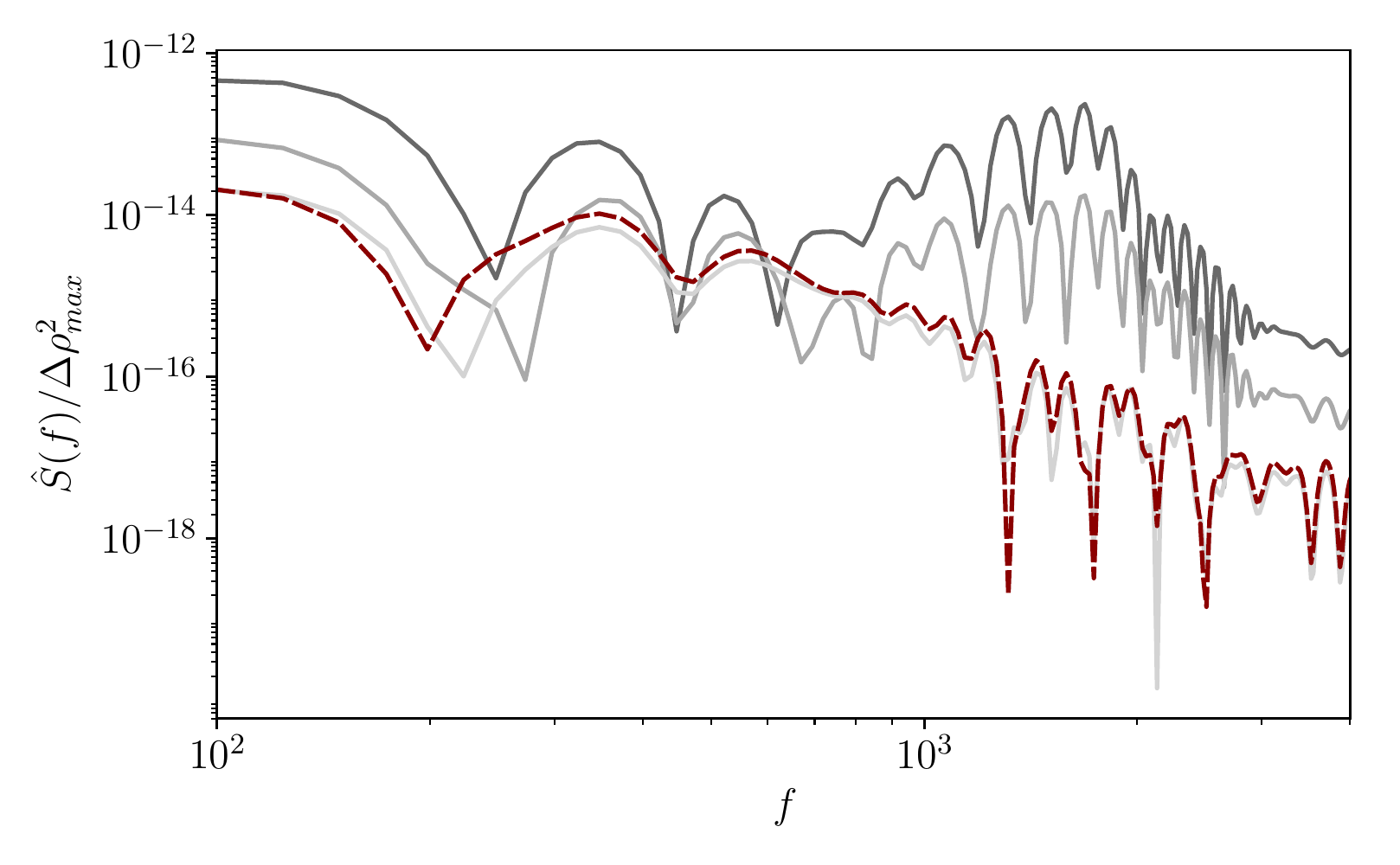}
		\caption{\label{fig:psd_COVO} PSD of spurious density fluctuations recorded at Probe $\mathbf{P_1}$ with $\Delta\rho^2_{max}=\rho_0-\rho(x_c,y_c,t=0)$. \ (1) -- Comparison of RR and H-RR collision model at iso dimensionless kinetic viscosity $\nu=10^{-6}$ (\protect\PSDred): H-RR ($\sigma=0.995$), (\protect\PSDgrayA): RR.\ (2) -- Influence of the dimensionless kinetic viscosity on the spurious noise emitted for the RR collision model.  (\protect\PSDgrayB) :\ $\nu=10^{-4}$, (\protect\PSDgrayC) :\ $\nu=4.18.10^{-4}$.}
	\end{center}
\end{figure}

More quantitative results are displayed on Fig.~\ref{fig:psd_COVO} where PSD of density fluctuations are displayed. This figure highlights two points.

The first one is a comparison of the spurious noise emitted for the two collision models (\PSDred , \PSDgrayA) at a dimensionless viscosity $\nu=10^{-6}$. One can notice an improvement by up to four orders of magnitude on the spurious noise emission on a large frequency range. 

The second comparison is achieved to highlight the increase of dimensionless kinetic viscosity that is required to reach the same spurious emission. Indeed, as most of the spurious emission is directly linked to the dissipation of the spuriousAc (\numB) mode, by increasing the kinetic viscosity using the RR collision model, one can obtain similar results to the one obtained by using the H-RR model with $\sigma=0.995$. As discussed in Sec.~\ref{subsec:LSA_HRR}, the strength of the H-RR model is to add an equivalent of the viscosity focused on non-hydrodynamic modes, especially on the spuriousAc (\numB) one. An estimation of the dissipation of the spuriousAc mode (\numB) was given in Eq.~(\ref{eq:tauSpuriousAc}). Using this relation in the present case, to reach the same level of spurious emission, the kinetic viscosity has to be increased by a factor  418.

It has been shown above that the acoustic emission comes from the spuriousAc modes (\numB) and the deformation from the spuriousS (\numS) ones. As a matter of fact, the intensities of the spurious artifacts are, to a lower degree, linked with the initial position of the vortex. Actually, the greater the distance between the vortex initial position and the refinement interface, the more the spurious modes will be dissipated by the LBM scheme. However, as shown on Fig.~\ref{fig:RRlsa}, the dissipation of the spuriousAc modes (\numB) for the RR collision model is of the same order of magnitude of that of the physical modes. A very long distance is then required to observe a significant effect of the choice of the initial location.\\

This remark is of great importance for industrial applications. The same behaviors are expected to appear in wakes where vortices are convected and, often, at cross grid refinement interfaces. In such situations, it is crucial to dissipate non-hydrodynamic modes before they impact it. Hopefully, in large eddy simulations (LES), the subgrid scale model has a beneficial effect, because it adds turbulent viscosity inside vortices, and helps to dissipate non-hydrodynamic modes. Nevertheless, the range of the turbulent viscosity for LES is about $0$ to $50\mathrm{m^2.s^{-1}}$, which is very far from the order of dissipation obtainable by computing the stress tensor using finite differences as in the H-RR collision model.

Finally, a nuance should be noted. The dissipation of physical modes is slightly increased in high wavenumbers, notably for acoustics (\acousP,\acousN) using the H-RR collision model (\textit{cf.} Fig.~\ref{fig:HRRlsa}). To overcome this issue and allow a better farfield acoustic propagation in presence of non-uniform grids, the parameter $\sigma$ can be associated with non-hydrodynamic sensors and can be modified dynamically to dissipate non-hydrodynamic modes only where they are presents.\\

\section{Validation on a high Reynolds number turbulent flow around a cylinder}
\label{sec:cylinder}

In this section, a validation is carried out on a three-dimensional high Reynolds turbulent flow around a circular cylinder. The flow physics is not examined here, since it depends mainly on parietal modeling, which is not the subject of this paper. The objective is to simulate a low-viscosity turbulent flow across refinement interfaces, minimizing parasitic vorticity and spurious noise.

\noindent This test case is purely qualitative and intends to highlight specific problems that may occur when adding solid walls. Quantitative results on the use of H-RR model on a turbulent cylinder can be found in \cite{Jacob2018}. Furthermore, a quantitative aeroacoustic application on a landing gear using the H-RR collision model can also be found in \cite{Hou2019}.\\

\noindent This test case aims at:
\begin{itemize}
\item evidencing the impact of non-hydrodynamic modes on mesh transitions in a more realistic case,
\item highlighting the parasitic acoustic waves emitted from the wake generated by the cylinder through the grid refinement interface.\\
\end{itemize}

The LBM solver used in this section is based on a D3Q19 lattice. The subgrid scale viscosity $\nu_{SGS}$ is modeled using a Shear improved Smagorinsky model from~\cite{Leveque2010} and computed from the strain-rate tensor. 
The latter is calculated by the same gradients as those used for the H-RR model, which minimizes the cost of this collision model. A wall-law taking into account an adverse pressure gradient and curvature effects~\cite{Afzal1999} is imposed on the cylinder walls. The no-slip condition is implemented using a full reconstruction of the distribution functions using finite differences \cite{Verschaeve2010}. This LBM code adopts the refinement algorithm described in Sec.~\ref{sec:refinement}. \\

\begin{figure}[H]
	\begin{center}
		\begin{tikzpicture}[scale=0.6]
			\draw (2.4,2.4) grid[step=0.15] (6.,4.8);
			\draw (1.8,1.8) grid[step=0.3] (9.6,5.4);
			\draw (0.,0) grid[step=0.6] (12.6,7.2);

			\fill[gray,opacity=0.1] (-0.25,-0.25) rectangle (12.85,0.25) ;
			\draw [dashed] (-0.25,-0.25) rectangle (12.85,0.25) ;
			\fill[gray,opacity=0.1] (-0.25,6.9) rectangle (12.85,7.45) ;
			\draw [dashed] (-0.25,6.9) rectangle (12.85,7.45) ;
			\fill[gray,opacity=0.5] (-0.25,0.25) rectangle (0.25,6.9) ;
			\draw [dashed] (-0.25,0.25) rectangle (0.25,6.9) ;
			\fill[gray,opacity=0.1] (12.35,0.25) rectangle (12.85,7.45) ;
			\draw [dashed] (12.35,0.25) rectangle (12.85,7.45) ;
			\node[text width=4.5cm,rotate=90,scale=1.,darkgray] at(-0.65,4.){Absorbing Layers - Velocity};
			\node[text width=4.5cm,rotate=-90,scale=1.,gray] at(13.28,3.2){Absorbing Layers - Pressure};

			\draw  [thick,fill=white,opacity=1] (3.5,3.6) circle (0.8) ;	
			\draw [<->] (2.75,3.6) -- (3.5,3.6) -- (4.25,3.6);
			\node[text width=0cm,rotate=0,scale=0.8] at(3.35,3.95){D};

			\draw  [fill=white,opacity=1] (4.5,4.2) rectangle (6.,4.8) ;	
			\draw  [fill=white,opacity=1] (8.1,4.8) rectangle (9.6,5.4) ;	
			\draw  [fill=white,opacity=1] (10.2,6.) rectangle (11.4,6.6) ;	

			\node[text width=0cm,rotate=0,scale=1.] at(4.7,4.5){RD1};
			\node[text width=0cm,rotate=0,scale=1.] at(8.3,5.1){RD2};
			\node[text width=0cm,rotate=0,scale=1.] at(10.23,6.3){RD3};

			\draw node[circle,draw,thick,fill=black, scale=0.4] at (3.6,6) {};
			\draw  [fill=white,opacity=1,white] (3.9,5.43) rectangle (4.77,6.57) ;	
			\node[text width=1cm,scale=1.3] at(5,6){$\mathrm{P_2}$};

			\draw [-> , >=latex ,line width=0.3mm] (-2.5,3.6) -- (-1.5,3.6);	
			\node[text width=0.2cm,scale=1.3] at(-1.6,4.){$x$};
			\draw [-> , >=latex ,line width=0.3mm] (-2.5,3.6) -- (-2.5,2.6);
			\node[text width=0.2cm,scale=1.3] at(-2.1,2.6){$y$};
			\draw node[circle,draw,thick,fill=black, scale=0.2] at (-2.5,3.6)  {};
		\end{tikzpicture}
		\caption{\label{fig:cylinder_domain} Schematic representation of the simulation domain for the cylinder test case. Three refinement domains (RD) are used. A $\mathbf{P_2}$ probe is added for acoustic recording. Absorbing layers map the domain boundaries to avoid reflection of spurious acoustic emission.}
	\end{center}
\end{figure}
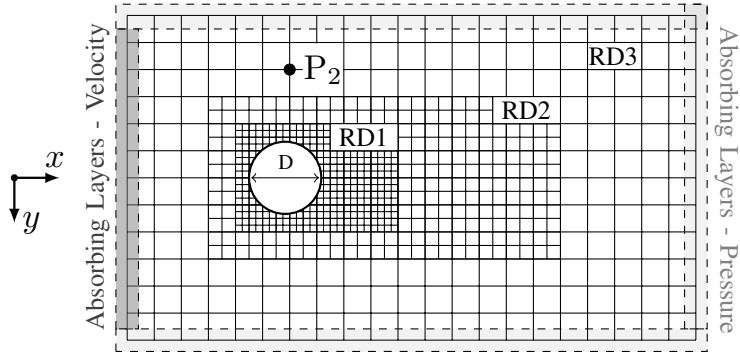

A sketch of the simulation domain is displayed on Fig.~\ref{fig:cylinder_domain}. Three levels of refinement are placed in order to evidence the effects of, firstly, the non-hydrodynamic modes generated by the cylinder in the RD1 zone, and secondly, the effect of the wake impinging the RD2 to RD3 interface once it has already passed through the first refinement area. A probe $\mathrm{P_2}$ is added in the domain, far from hydrodynamic fluctuations at location (0, -160) to record acoustic fluctuations.

\noindent The operating conditions of the simulation, given in dimensionless units based on the coarser mesh resolution (RD3), are as follows:
\begin{equation}
	\label{eq:cylinder}
	\quad U_{\infty} = 0.1c_s,
	\quad \rho_\infty = 1,
    \quad \mathrm{D} = 75,
    \quad \nu = 5.77.10^{-6},
    \quad \mathrm{Re}=750\ 000,
    \quad \mathrm{T} = 320\ 000,
\end{equation}

where $U_\infty$ is the inflow velocity upstream the cylinder, $D$ is its diameter (expressed in number of coarse voxels RD3. It means that there are 300 fine cells over a diameter), $\nu$ is the dimensionless viscosity (expressed in number of coarse voxels RD3) and $T$ is the number of iterations of the simulation.\\

\noindent The boundary conditions are the following: a velocity Dirichlet $U_\infty$ is imposed at the inlet and a density Dirichlet $\rho_\infty$ is imposed on the domain outlets. Both conditions are implemented using a full reconstruction of distribution functions with finite differences as in \cite{Latt2008,Verschaeve2010}. The domain is periodic in the spanwise direction ($z$ axis) with a domain span equal to $1.6\mathrm{D}$ to allow the turbulence to be fully developed. 
The boundaries of the domain are covered with absorbing layers \cite{Chevillotte2016} to avoid parasitic wave reflections and so that the turbulent wake generated by the cylinder can be properly evacuated.

Following the same approach as in Sec.~\ref{sec:improvement}, the vorticity and then the acoustic waves emitted are analyzed separately. Only the RR and H-RR ($\sigma=0.98$) models are compared in this section because the BGK model is unstable at this high Reynolds number. 

\begin{figure}[H]
	\begin{center}
		\includegraphics[scale=0.55]{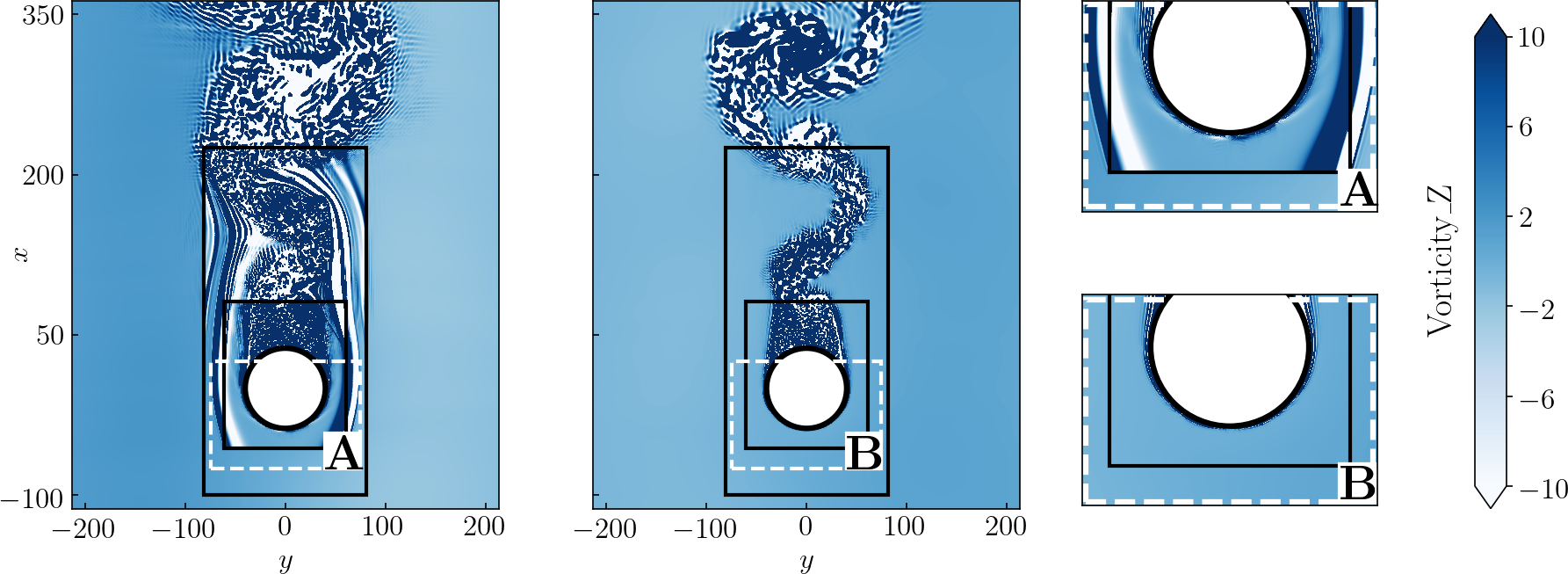}
		\caption{\label{fig:cylinder_vortZ} Field of Z component of the vorticity generated by the cylinder with two different collision models. Left: RR, Right: H-RR.}
	\end{center}
\end{figure}

Fig.~\ref{fig:cylinder_vortZ} shows that parasitic vorticity fluctuations appear on the Z-component of the vorticity upstream the cylinder. 
This vorticity is clearly generated at the refinement interface due to SpuriousS modes (\numS) generated by the walls. 
Then, it is convected by the flow through the rest of the domain. This observation is consistent with the results obtained on the convected vortex (Fig.~{\ref{fig:COVO_SpuriousS}}): the SpuriousS (\numS) modes do not succeed in properly crossing the mesh transitions and are converted into physical shear. This vorticity strongly disturbs the flow. It surrounds the cylinder, may interferes with the boundary layer that develops on it and and interacts with the wake in ways that are not physical. Finally, strips of parasitic vorticity induced by the RD3 transition are observed. They are the consequence of parasitic waves generated when the spurious vorticity created upstream of the cylinder intersects the RD1. All of these glitches are greatly improved with the H-RR model which dissipates non-hydrodynamic modes that parasitize the vorticity field. Neither parasitic vorticity upstream, nor any striations around the wake are visible with the H-RR model.

\begin{figure}[H]
	\begin{center}
		\includegraphics[scale=0.55]{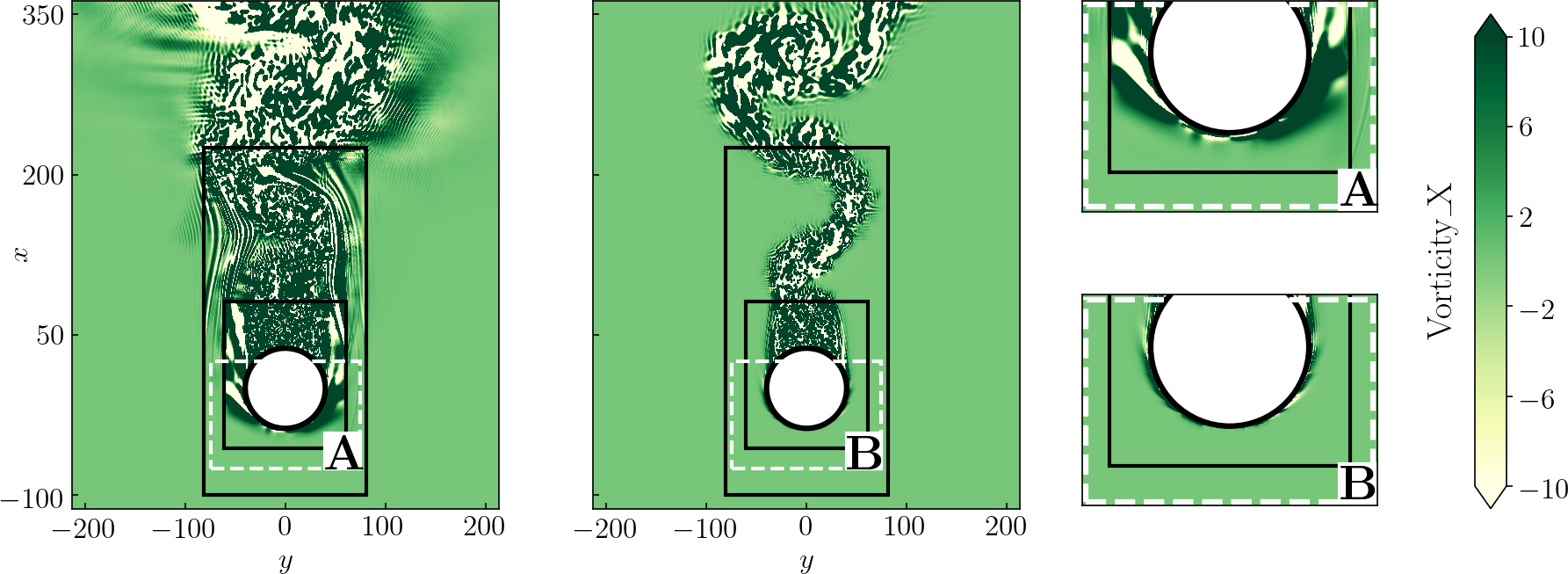}
		\caption{\label{fig:cylinder_vortY} Field of X component of the vorticity generated by the cylinder with the two collision models. Left: RR, Right: H-RR.}
	\end{center}
\end{figure}

It is also noteworthy to focus on the $x$-component of the vorticity (\textit{cf.} Fig.~\ref{fig:cylinder_vortY}). For this component, there are no more parasitic modes located at the transition upstream of the cylinder with the RR model. However, spurious waves appear around the cylinder and everywhere in the downstream region of RD1. In this resolution domain, parasitic waves propagated in the direction normal to shedding vortices are also observable. These are SpuriousS (\numS) waves that are continuously generated in the fluid core by the vortices. The same phenomenon has been observed with a uniform mesh simulation and is also filtered by the H-RR model. It is important to filter these waves, especially before they impact another resolution area, as they would create parasitic vorticity again. The fields obtained on the $y$-component of the vorticity are very similar to those on $x$-component and are not presented in this study.

\begin{figure}[H]
	\begin{center}
		\includegraphics[scale=0.55]{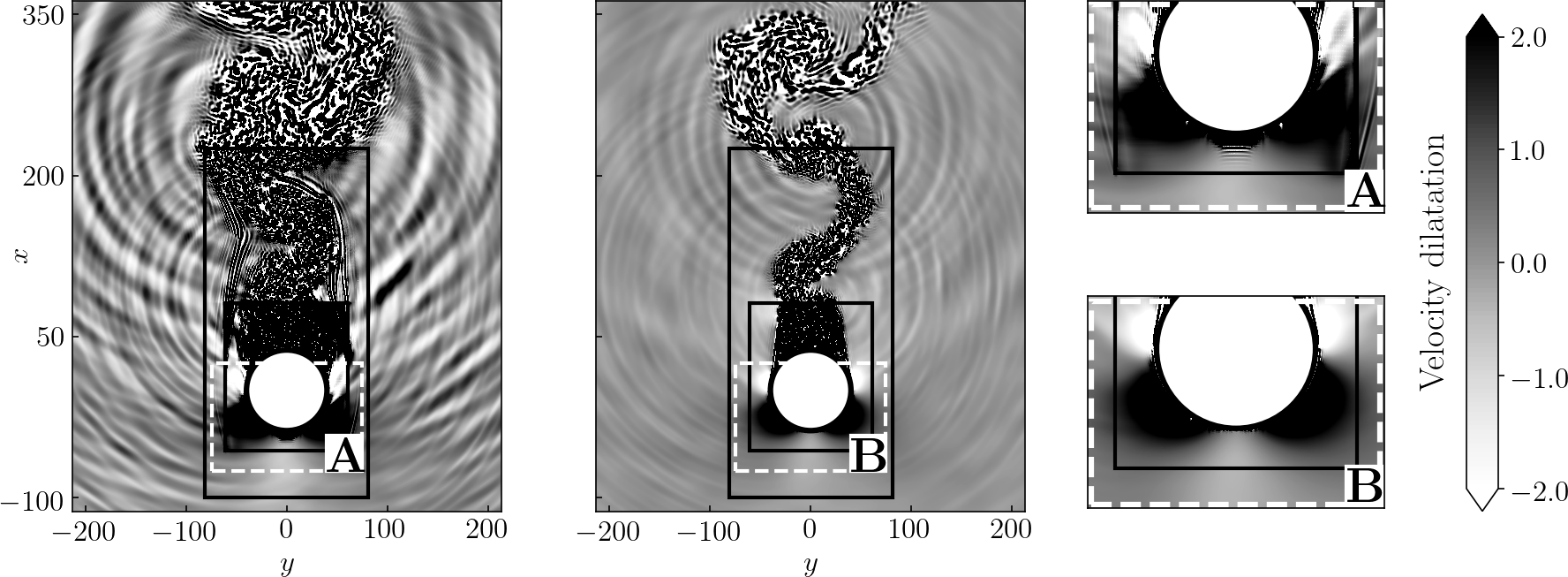}
		\caption{\label{fig:cylinder_divU} Velocity divergence field of the flow around the turbulent cylinder with the two collision models. Left: RR, Right: H-RR.}
	\end{center}
\end{figure}

Let us now look at the velocity divergence (Fig.~\ref{fig:cylinder_divU}) in order to highlight the parasitic acoustic waves emitted by the mesh refinement interface. 
Up to now, it has been shown that two sources of parasitic noise exist at the mesh interface. \\

The first one is the conversion of SpuriousAc (\numB) modes into acoustic waves. These modes are present in the vortices generated by the cylinder and contribute significantly to the large parasitic emission that is centered downstream of RD1. These modes are also displayed on zoom (A) of Fig.~\ref{fig:cylinder_divU}, where high-frequency waves are visible.\\

The second emission is attributed to the lack of accuracy of the grid refinement algorithm, where the slightest discontinuity in the transfer of information from one resolution level to another one may result in the emission of acoustic waves. This issue has already been highlighted on the convected vortex test case. To date, the only two refinement algorithms whose acoustic emission due to the passage of a vortex has been quantified are the present one and the \textit{Directional Splitting algorithm} presented in~\cite{Gendre2017}. To act on this second source of noise, the flow that impinges the refinement areas must be as clean as possible since every vortex that impacts the interface will ineluctably generate spurious acoustics.\\

\begin{figure}[H]
	\begin{center}
		\includegraphics[scale=0.279]{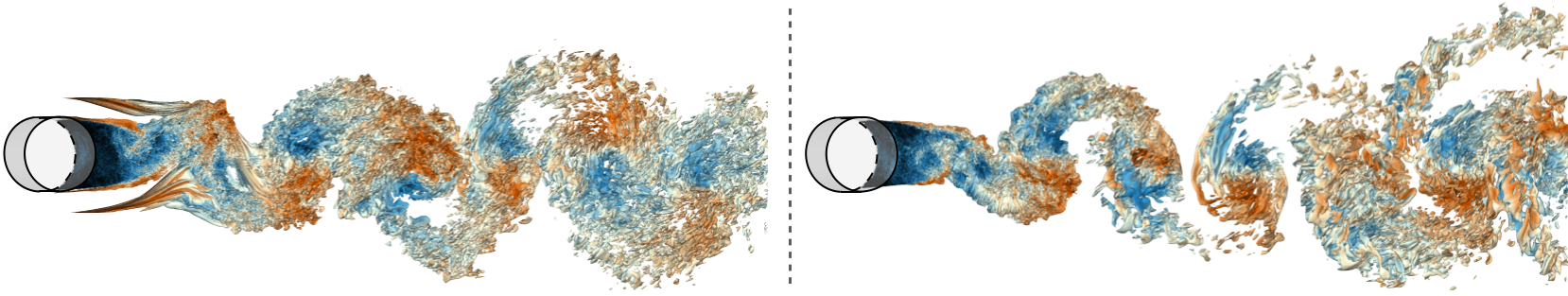}
		\caption{\label{fig:cylinder_critQ} Isosurface of Q criterion $Q=10^{5}$ colored by velocity magnitude for the two collision models. Left: RR, Right: H-RR.}
	\end{center}
\end{figure}

As illustrated on the Q criterion isosurface of Fig.~\ref{fig:cylinder_critQ}, the parasitic vorticity created by the RD3 are of an intensity comparable to the vortices generated in the wake of the cylinder. These vortices interact with the wake and enrich the wake turbulence that crosses the grid interface in an unphysical way.

\noindent All of these parasitic phenomena contribute to an increase in the quantity and intensity of the vortices. Thus, they greatly increase the parasitic acoustic emissions induced by grid interfaces positioned in the wake. This is clearly evidenced on Fig.~\ref{fig:cylinder_divU}. Moreover, the parasitic vorticity is, in most cases, under-resolved and will thus be unlikely to transmit properly from one mesh to another. 

\noindent Nevertheless, it is also clearly shown in Fig.~\ref{fig:cylinder_divU} that the wake that crosses the RD3 also emits a parasitic acoustic wave with the H-RR model. This last is therefore strongly attenuated.

\noindent The quality of the algorithm remains of paramount importance. This was also shown in Sec.~\ref{sec:acous_context} where the acoustic emission of the cell-vertex algorithm used was considerably lower than for the cell-centered one.

\begin{figure}[H]
	\begin{center}
		\includegraphics[scale=0.6]{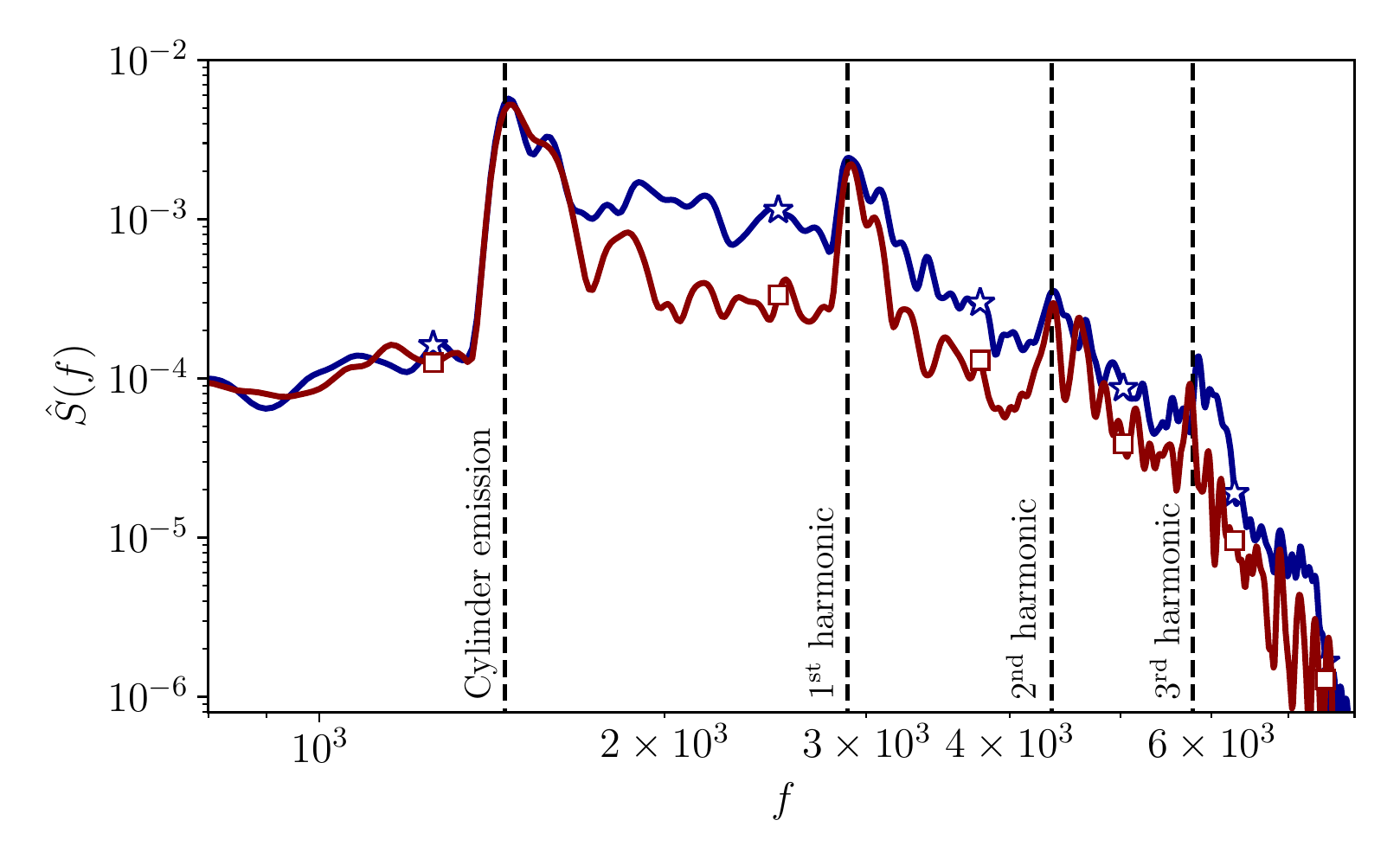}
		\caption{\label{fig:psd_cylinder} Power Spectral Density of pressure fluctuations recorded at Probe $\mathbf{P_2}$. Comparison of RR~(\protect\blueStar) and H-RR~(\protect\redSquare) collision model.}
	\end{center}
\end{figure}

Although this test case is qualitative and mainly selected to break down the diverse spurious phenomena, a Power Spectral Density (PSD) of pressure fluctuations is displayed on Fig.~\ref{fig:psd_cylinder}. This PSD reveals the tremendous impact of the collision model on the spurious noise emitted over a wide frequency range. Indeed, the two types of acoustic sources mentioned in the previous paragraph have a strong impact on the entire spectrum. The noise induced by the aliasing of SpuriousAc (\numB) modes is completely suppressed with the H-RR model, and the vortices crossing the transitions are cleaned of parasitic vorticity that also affects the whole spectrum. On this plot, four peaks are also noticeable. They are all physical and correspond to the dipole noise emitted by the cylinder with the associated harmonics.

This turbulent three-dimensional cylinder test case allows for corroborating the observations made on the elementary test cases on a more realistic configuration. It has been demonstrated that solid boundaries can be responsible for the generation of non-hydrodynamic modes that have to be properly handled at the mesh transitions.
More generally, it has been observed by the authors that any boundary condition, mesh refinement, and even any vortex produces non-hydrodynamic modes that are likely to pollute hydrodynamic and acoustic fields. It is therefore necessary to filter them out by different means through the whole domain, to increase the stability of high Reynolds number computations, as well as the accuracy of simulations.


\section{Conclusion}
\label{sec:conclusion}

This paper has investigated the transfer of energy between non-hydrodynamic and physical modes occurring at a grid refinement interface. More precisely, by clearly sorting the modes by their carried macroscopic information, referred to above as shear or acoustic modes, and by systematically identifying them in a simulation thanks to adequate newly proposed sensors, it has been shown that the energy of a non-hydrodynamic mode can be redistributed on every mode carrying a quantity of the same nature when a change of grid resolution occurs. 
This observation was anticipated by spectrally analyzing the projection of a fine mode into a coarser resolution. Furthermore, these exchanges can be harmful and the undesired effects are amplified by the high pulsation $\omega_r$ of some modes which leads to a huge discontinuity at the interface, due to the asynchronous evolution of both meshes. Moreover, these non-hydrodynamic modes might suffer from a severe dispersion, which makes their group velocity $v_g$ strongly dependent on the wavenumber $k$. It can even be reversed depending on the mesh resolution, so that a wave packet travelling in a given direction has no equivalent after the mesh transition: the only solution is then to convert energy into a physical mode. For all these reasons, non-hydrodynamic modes that cross a mesh transition are likely to be the main source of spurious noise emission, whereas the physical waves seem to have a correct behavior in all the simulations presented in this article. To this extent, major improvements have been observed by changing the collision model in the fluid core in order to damp any non-hydrodynamic mode.

Issues observed in this paper are commonly found in industrial simulations. Even when focusing on pure aerodynamic outcomes, it has been proven that the spurious modes carrying the shear quantity can hugely distort vortices, which may have an important impact on wakes shape. Boundary layers are also likely to be strongly affected by the spurious vorticity as in most industrial applications, refinement areas are placed close to walls. Furthermore, spurious vorticity may be created at grid interface, even far from the hydrodynamic areas~\cite{Hasert2014}. It is even more crucial to handle carefully these phenomena for aeroacoustic simulations where, in addition to accurate aerodynamic predictions, the spurious acoustic sources must be avoided as much as possible. To this end, the turbulent cylinder test case has highlighted the necessity to damp both the spuriousAc (\numB) modes that are converted into spurious acoustics and the spuriousS (\numS) modes to avoid parasitic vorticity that increases the turbulence in a non-physical manner.

A solution to this issue has been introduced by finding a way to strongly dissipate these modes: the reconstruction of the stress tensor using finite differences. As this solution is very dissipative, a hybridization of this reconstruction with the recursive regularized model~\cite{Jacob2018} was chosen to strongly dissipate non-hydrodynamic modes while keeping correct dissipation properties for the physical ones. A good alternative to properly handle these modes while avoiding a systematic over-dissipation could also be to dynamically damp them with a variable $\sigma$ parameter in the H-RR model, linked with the non-hydrodynamic sensors like those introduced in this paper.

Compatibility with large-eddy simulations has been addressed. It appears that although the subgrid scale model works in favor of dissipating non-hydrodynamic modes in sheared areas, the order of dissipation added is far from that introduced by the H-RR model. Furthermore, the non-hydrodynamic modes responsible for spurious acoustic emission are not often directly impacted by the subgrid scale model since they may not be detected by sensors based on turbulent quantities. The H-RR model, or another one capable of dissipating spurious modes, is then strongly advised, especially for aeroacoustic simulations. Furthermore, the additional CPU cost of the H-RR model is reduced when a subgrid scale model based on velocity gradient is used since these quantity are already computed.

A brief comparison of grid coupling algorithms has been proposed. It has been shown that parasitic behaviors appear regardless of the algorithm structure used. Nevertheless, once the influence of non-hydrodynamic modes have been ruled out, the choice of such an algorithm remains essential \cite{Gendre2017}.

\noindent Finite difference algorithms such as proposed in~\cite{Fakhari2014,Fakhari2015} have not been investigated in this paper. These algorithms do not seem well adapted to an industrial context. Indeed, the Courant–Friedrichs–Lewy number (CFL) ~\cite{Courant1967} is constant in all refinement domains. Thus, when many RD are used, it significantly increases the CPU cost of the simulation. Furthermore, the spectral properties are degraded when the mesh is coarsened making them unattractive for aero-acoustic simulations. Nevertheless, since the timestep is constant in both meshes, it may help reducing the discontinuities in non-hydrodynamic modes imposed by the acoustic scaling (\textit{cf.}~Fig.~\ref{fig:discontinuity}).

Finally, the H-RR model, proposed as an efficient candidate for solving spurious artifacts at grid refinement, is probably not the only one to have these kinds of spectral properties. To this extent, the von Neumann analysis of advanced collision models like the cumulant model~\cite{Geier2015}, cascaded model~\cite{Geier2006,Dubois2015} or some variants of entropic lattice Boltzmann models~\cite{Karlin2014} might be of interest. This article mainly focused on a methodology for efficiently selecting a collision model adapted to perform accurate simulations in the presence of non-uniform grids.

\section*{Acknowledgments}
The authors would like to gratefully acknowledge Felix Gendre for the fruitful discussions on grid refinement algorithms. Acknowledgements are also expressed to Airbus Operations and ANRT/CIFRE for the financial support.

\section*{Appendix 1: Expressions of the time-advance matrices for the von Neumann analyses}
\label{AppA}

The aim of this appendix is to give the expression of the time-advance matrices required to performed the linear stability analyses. The linearization and the eigenvalue problems are solved thanks to a Python code, using the NumPy library \cite{oliphant2006guide}.

\noindent Starting from the discrete LBM equation rewritten into the following form

\begin{equation}
\label{eq:lsa_lbm}
	f_i(\mathbf{x}+\mathbf{\xi_i},t+1) = f_i^{(0)}(\mathbf{x},t) + \left( 1 - \frac{1}{\tau} \right) f_i^{(1)}(\mathbf{x},t) = \Phi_i(\mathbf{x},t).
\end{equation}

\noindent The distribution functions are expanded into the sum of a stationary part $\overline{f_i}$ and a fluctuating part $f_i^{'}$

\begin{equation}
\label{eq:lsa_f}
	f_i = \overline{f_i} + f_i^{'}.
\end{equation}

\noindent Both the equilibrium part $f_i^{(0)}$ and the non-equilibrium part $f_i^{(1)}$ are non linear in $f$. A linearization of $\Phi_i$ is then required for the linear stability analysis.

\begin{equation}
\label{eq:lsa_PhiLin}
\Phi_i(f_j) = \Phi_i(\overline{f_j}+f_j^{'}) = \Phi_i(\overline{f_j}) +  \underbrace{\left. {\frac{\partial \Phi_i}{\partial f_j}}%
_{\stackunder[1pt]{}{}}%
 \right|_{%
 \stackon[1pt]{$\scriptscriptstyle f_j=\overline{f_j}$}{$\scriptscriptstyle \ $}}}_{J_{ij}} f_j^{'} + O(f_j^{'2}),
\end{equation}

\noindent where the Jacobian $J_{ij}$ of the $\Phi_i$ operator appears. Keeping only the first-order in distribution fluctuation, the LBM equations becomes

\begin{equation}
\label{eq:lsa_systF}
	f^{'}_i(\mathbf{x}+\boldsymbol{\xi_i},t+1) =  J_{ij} f^{'}_j(\mathbf{x},t).
\end{equation}

\noindent By injecting complex plane wave perturbations given by Eq.~(\ref{eq:LSA_monowaves}) into Eq.~(\ref{eq:lsa_systF}), the linear system to solve reads

\begin{equation}
\label{eq:lsa_lbm}
	e^{i \omega} \mathbf{F} = \underbrace{e^{i \mathbf{k} \cdot \boldsymbol{\xi_i}} \mathbf{J}}_{\mathbf{M^{LBM}}} \mathbf{F},
\end{equation}

\noindent with $\mathbf{F}=[\hat{f}_i]^T$ the vector of modal fluctuations. For the collision models introduced in Sec.~\ref{sec:LBM}, only the $\Phi_i$ operator, and thus the jacobian $J_{ij}$ has to be adapted.\\

\textbf{Time-advance matrix for the BGK collision model}\\

\noindent For the BGK collision model, the non-equilibrium distribution reads $f_i^{(1)}=f_i-f_i^{(0)}$. Thus the following relation is obtained 

\begin{equation}
	\Phi _i^{BGK}(\mathbf{x},t) = f_i^{(0)}(\mathbf{x},t) + \left( 1 - \frac{1}{\tau} \right) \left( f_i(\mathbf{x},t) - f_i^{(0)}(\mathbf{x},t) \right).
\end{equation}

\textbf{Time-advance matrix for the RR collision model}\\

\noindent The non-equilibrium part $f_i^{(1),RR}$ for the RR collision model is given in Eq.~(\ref{eq:f1RR}). The $\Phi_i$ operator reads

\begin{equation}
    \label{eq:lsa_PhiRR}
	\Phi _i^{RR}(\mathbf{x},t) = f_i^{(0)}(\mathbf{x},t) + \left( 1 - \frac{1}{\tau} \right) f_i(\mathbf{x},t)^{(1),RR},
\end{equation}

\noindent The third order off-equilibrium coefficients needed to calculate $f_i^{(1)}$ depend on the velocities (Eq.~\ref{eq:a13RR}). The macroscopic quantities can be decomposed into a steady and a fluctuating part of $f$

\begin{equation}
	\rho  = \sum_i f_i = \underbrace{\sum_i \overline{f_i}}_{\overline{\rho}} + \underbrace{\sum_i f_i^{'}}_{\rho^{'}} ,
\end{equation}

\begin{equation}
	\rho \boldsymbol{u} = \sum_i \boldsymbol{\xi}_i f_i = \underbrace{\sum_i \boldsymbol{\xi}_i \overline{f_i}}_{(\overline{\rho \boldsymbol{u}'})} + \underbrace{\sum_i \boldsymbol{\xi}_i f_i^{'}}_{(\rho \boldsymbol{u})^{'}} ,
\end{equation}

\noindent Since $(\rho \boldsymbol{u})^{'} = \overline{\rho}\boldsymbol{u}^{'} + \rho^{'}\overline{\boldsymbol{u}}$,

\begin{equation}
	\label{eq:lsa_uprime}
	\boldsymbol{u}^{'} = \frac{(\rho \boldsymbol{u})^{'} - \rho ^{'}\overline{\boldsymbol{u}}}{\overline{\rho}} = \frac{\sum_i \xi _i f_i^{'} - \overline{\boldsymbol{u}} \sum_i f_i^{'}}{\overline{\rho}}.
\end{equation}

\noindent Thus, $\boldsymbol{u}^{'}$ is first order in distribution function as the $\boldsymbol{a}_1^{(2)}$ coefficients. The third order off-equilibrium coefficients are therefore computed using the steady part of the velocity $\boldsymbol{\overline{u}}$

\begin{equation}
	\label{eq:lsa_a13}
	\textit{a}^{(3)}_{1,xxy} = 2 \overline{u_x} a_{1,xy}^{(2)}+\overline{u_y} a_{1,xx}^{(2)}, \qquad \textit{a}^{(3)}_{1,xyy} = 2\overline{u_y} a_{1,xy}^{(2)}+\overline{u_x} a_{1,yy}^{(2)}.\\
\end{equation}

\textbf{Time-advance matrix for the HRR collision model}\\

\noindent The $\Phi _i^{HRR}$ expression is identical to Eq.~(\ref{eq:lsa_PhiRR}). Specificity comes up when calculating the $\boldsymbol{a}_1^{(2),FD}$ coefficients, since they depend on spatial derivatives (Eq.~(\ref{eq:a1ij_HRR})). The velocity gradients of the steady part vanish and the coefficients $\boldsymbol{a}_1^{(2)}$ read

\begin{equation}
	a_{1,ii}^{(2),FD} = -2 \tau \overline{\rho} c_s^2 \frac{\partial u_i^{'}}{\partial x_i}, \qquad a_{1,ij}^{(2),FD} = - \tau \overline{\rho} c_s^2  \left( \frac{\partial u_i^{'}}{\partial x_j} + \frac{\partial u_j^{'}}{\partial x_i} \right),
\end{equation}

\noindent Estimated using second order finite difference schemes, in can be expressed in the Fourrier space as follow

\begin{equation}
	a_{1,ii}^{(2),FD} = -2 \tau \overline{\rho} c_s^2 i u_i^{'} \sin(k_i), \qquad a_{1,ij}^{(2),FD} = - \tau \overline{\rho} c_s^2  i \left( u_i^{'} \mathrm{sin}(k_j) + u_j^{'} \sin(k_i) \right),
\end{equation}

\noindent where $u_i^{'}$ are non linear in $f$ and are linearized using Eq.~(\ref{eq:lsa_uprime}). The third order coefficients $\boldsymbol{a}_1^{(3)}$ are then computed following Eq.~(\ref{eq:lsa_a13}).


\bibliographystyle{unsrt}
\bibliography{refs}

\end{document}